\documentclass[acmsmall,authorversion,nonacm]{acmart}

\AtBeginDocument{%
  \providecommand\BibTeX{{%
    \normalfont B\kern-0.5em{\scshape i\kern-0.25em b}\kern-0.8em\TeX}}}

\setcopyright{acmcopyright}
\copyrightyear{2023}
\acmYear{2023}
\acmDOI{XXXXXXX.XXXXXXX}

\acmJournal{TOIS}
\acmVolume{0}
\acmNumber{0}
\acmArticle{111}
\acmMonth{0}

\usepackage{multirow}
\usepackage{booktabs}
\usepackage[subrefformat=parens,labelformat=parens]{subfig}
\usepackage{bm}
\usepackage{balance}
\usepackage{tikz}
\usepackage{colortbl}
\usepackage{subfig}
\usepackage{graphicx}
\usetikzlibrary{positioning}
\usepackage{enumitem}
\usepackage{amsthm}
\usepackage{amsmath}
\usepackage{amsfonts}
\usepackage{bbm}
\usepackage{apptools}
\usepackage{xcolor}
\usepackage[ruled]{algorithm2e}
\usepackage{algorithmic}
\usepackage{placeins}

\usepackage[show]{chato-notes}

\SetKwComment{Comment}{$\triangleright$\ }{}

\usepackage{xcolor}
\definecolor{red}  {rgb}{0.9,0.0,0.0}
\definecolor{green}{rgb}{0.0,0.7,0.0}
\definecolor{blue} {rgb}{0.0,0.0,0.9}

\DeclareMathOperator*{\argmax}{arg\,max}

\newcommand{\sinnamon}{\textsc{Sinnamon}}
\newcommand{\weaksinnamon}{\textsc{Weak Sinnamon}}
\newcommand{\linscan}{\textsc{LinScan}}
\newcommand{\splade}{\textsc{Splade}}
\newcommand{\esplade}{\textsc{Efficient Splade}}

\usepackage{soul}
\usepackage{xcolor}

\begin{document}

\title[Bridging Dense and Sparse Maximum Inner Product Search]{Bridging Dense and Sparse Maximum Inner Product Search}

\author{Sebastian Bruch}
\affiliation{%
  \institution{Pinecone}
  \city{New York}
  \state{NY}
  \country{USA}
}
\email{sbruch@acm.org}
\orcid{0000-0002-2469-8242}

\author{Franco Maria Nardini}
\affiliation{%
  \institution{ISTI-CNR}
  \city{Pisa}
  \country{Italy}
}
\email{francomaria.nardini@isti.cnr.it}
\orcid{0000-0003-3183-334X}

\author{Amir Ingber}
\affiliation{%
  \institution{Pinecone}
  \city{Tel Aviv}
  \country{Israel}
}
\email{ingber@pinecone.io}
\orcid{0000-0001-6639-8240}

\author{Edo Liberty}
\affiliation{%
  \institution{Pinecone}
  \city{New York}
  \state{NY}
  \country{USA}
}
\email{edo@pinecone.io}
\orcid{0000-0003-3132-2785}



\begin{abstract}
Maximum inner product search (MIPS) over dense and sparse
vectors have progressed independently in a bifurcated literature for decades;
the latter is better known as top-$k$ retrieval in Information Retrieval.
This duality exists because sparse and dense vectors serve different end goals.
That is despite the fact that they are manifestations of the same mathematical problem.
In this work, we ask if algorithms for dense vectors
could be applied effectively to sparse vectors, particularly those
that violate the assumptions underlying top-$k$ retrieval methods.
We study IVF-based retrieval
where vectors are partitioned into clusters and
only a fraction of clusters are searched during retrieval.
We conduct a comprehensive analysis of dimensionality reduction for sparse vectors,
and examine standard and spherical KMeans for partitioning.
Our experiments demonstrate that IVF serves as an efficient solution for sparse MIPS.
As byproducts, we identify two research opportunities
and demonstrate their potential.
First, we cast the IVF paradigm as a dynamic pruning technique
and turn that insight into a novel organization of the inverted index for
approximate MIPS for general sparse vectors. Second, we offer a unified regime for MIPS
over vectors that have dense and sparse subspaces,
and show its robustness to query distributions.
\end{abstract}
\maketitle              

\section{Introduction}
\label{section:introduction}

Retrieval is one of the most fundamental questions in Information Retrieval (IR),
as the name of the discipline itself reflects. Simply put, given a large number of objects,
we wish to find, in an efficient manner, the closest subset of those objects to a query according
to some notion of closeness. The data structure and algorithmic inventions~\cite{zobel2006invertedindex,tonellotto2018survey}
that have emerged from the IR literature to address this deceptively simple
question have had enormous impact on the field and birthed major research directions.
They provide the machinery to scale ranking to massive datasets within
multi-stage ranking systems~\cite{asadi2013phd,lin2021pretrained,asadi2013efficiency,bruch2023fntir},
for instance, or power large-scale applications, of which search is a notable and ubiquitous example.

Much of the IR research on retrieval targets textual data,
where documents and queries are texts in natural languages.
Unsurprisingly, then, the retrieval machinery that exists today is highly optimized for data that is
governed by the laws of natural languages (such as Zipf's law)
and the way users interact with retrieval and search systems (e.g., by means of short, keyword queries).
The inverted index~\cite{zobel2006invertedindex}, for example, is inspired by how we historically organized and found
information in a book or at a library. Our measures of closeness, such as TF-IDF and BM25~\cite{bm25original},
rely on statistics that reflect our understanding of the relevance between two pieces of text.
The dynamic pruning algorithms that help us traverse inverted indexes
efficiently~\cite{mackenzie2021anytime,broder2003wand,ding2011bmwand,mallia2017blockmaxwand_variableBlocks,tonellotto2018survey,crane2017wsdm,jass,petri2019accelerated}
to find the top $k$ most relevant documents to a query, too, rely on the statistical properties of
language and relevance measures.

While the form of retrieval above is the bedrock
of flurry of other research and applications in IR,
the rise of deep learning in recent years brought a different
form of retrieval into the IR spotlight: Approximate Nearest Neighbor (ANN) search~\cite{malkov2016hnsw,wang2021graphanns,Johnson2021faiss,pq2011,pcpq,scann} in dense vector spaces.

ANN search has for decades played an outsize role in
research problems that are adjacent to text retrieval such as image and multimedia
retrieval~\cite{Zhou2017imageRetrieval,Peng2018crossMedia}.
Its machinery is optimized for objects and queries that are
real vectors in some high-dimensional space, and where closeness is determined by inner product or
proper metrics such as Euclidean distance. Today, efficient and effective data structures and algorithms
for this problem are often critical components in, among other applications, \emph{semantic} search,
where, using deep learning, we learn a vector representation of documents and queries
in a space where closeness of vectors implies semantic similarity of their corresponding texts~\cite{lin2021pretrained}.

\subsection{Maximum Inner Product Search as the Unifying Problem}
The fact that these two branches of retrieval have historically progressed independently
makes a great deal of sense: they have targeted quite different applications.
Today's reality driven by the burgeoning role of deep learning in IR and
the effectiveness of learnt representations in many related domains, however,
begins to challenge the status quo. Let us illustrate our point by considering
joint lexical-semantic
search~\cite{bruch2023fusion,chen2022ecir,wang2021bert,Kuzi2020LeveragingSA,karpukhin-etal-2020-dense,Ma2021ARS,Ma2020HybridFR,Wu2019EfficientIP}
as an example. In that setup, documents and queries are represented as learnt
vectors \emph{and} as bags of words. Retrieval is then performed over both representations
to find the documents that are both lexically and semantically close to a query.
This application is at the confluence of (inverted index-based) top-$k$ retrieval and
ANN search. The challenge presented by the historical dichotomy
is that researchers and practitioners alike must study and develop two disparate systems
that are characteristically different.

At the same time, we are witnessing the success of methods that \emph{learn} term importance weights from
texts~\cite{sparterm,formal2021splade,formal2022splade,zhuang2022reneuir,dai2020sigir,coil,mallia2021learning,zamani2018cikm,unicoil},
rather than compute it based on term frequency and propensity.
It has been shown that the weights learnt this way exhibit distributional
properties that do not conform to the expectations of inverted-index based retrieval algorithms~\cite{mackenzie2022tois,bruch2023sinnamon}.
This challenges some of the assumptions underlying dynamic pruning algorithms
and thus the efficacy of inverted index-based
retrieval in the face of arbitrarily-distributed term weights~\cite{bruch2023sinnamon,mackenzie2021wacky}.

The existing literature gives effective solutions of various degrees of complexity
to each and every one of the shortcomings
above~\cite{mackenzie2022tois,mallia2022sigir,Wu2019EfficientIP,yang2021sparsifying,mackenzie-etal-2022-accelerating}.
In this work, we wish to investigate a more general question that
arises if we returned to the principles and re-examined the most
glaring fact: It should come as no surprise that both branches of retrieval
operate on vectors and, often, attempt to solve Maximum Inner Product Search (MIPS).
It just so happens that in one branch the vectors are
\emph{dense} (i.e., all coordinates are almost surely non-zero) and in the other
\emph{sparse} (i.e., where, relative to the dimensionality of the space,
very few coordinates are non-zero). We call the former ``dense MIPS'' and the latter ``sparse MIPS'' for brevity.

\subsection{Sparse MIPS as a Subclass of Dense MIPS}
It is clear that solutions devised for sparse MIPS are not immediately applicable to dense MIPS.
That is because sparse MIPS algorithms operate under stricter distributional assumptions than
dense MIPS algorithms do; in other words, the class of sparse vectors for which MIPS solutions
exist is a subset of the class of dense vectors.
For example, inverted index-based solutions are only efficient if the vectors are
sparse\footnote{In fact, query vectors are often required to be
much more sparse than document vectors for a sparse MIPS solution to remain reasonably efficient.}
and non-negative, and if their sparsity pattern takes on a Zipfian shape.
Dense MIPS algorithms, on the other hand, have fewer inherent limitations.
A natural question that arises given the observation above is whether dense MIPS algorithms
remain effective and efficient when applied to sparse vectors.
That is the primary motivation behind this study.

While conceptually simple and admittedly pedestrian,
applying dense MIPS solutions to sparse vectors faces many challenges.
And therein lies our technical contribution:
We present, as a proof of concept, the machinery that enables such a formulation.

We start by foregoing \emph{exactness} and instead developing ideas on the principle of
\emph{probably approximately correctness} (PAC). In other words, instead of insisting on finding
the exact set of top $k$ documents, we settle with an approximate set that may erroneously contain some
farther-afield documents and mistakenly miss other close-by documents.
In the IR literature, this is the familiar notion of \emph{rank-unsafe} retrieval~\cite{tonellotto2018survey}.

Having accepted some (quantifiable) error in the retrieval outcome,
we are faced with the next, rather debilitating challenge of
working with often extremely high dimensional sparse vectors.
It is here that we appeal to results from related disciplines that study
data-oblivious $\ell 2$-subspace embedding~\cite{woodruff2014sketching} and
non-linear sketching\footnote{We use ``sketch'' to describe a compressed representation of
a high-dimensional vector, and ``to sketch'' to describe the act of compressing a vector into a sketch.}
(itself sparse) of sparse vectors~\cite{bruch2023sinnamon}.
These dimensionality reduction techniques use the elegant yet simple idea of
random projections to preserve Euclidean distance or inner product between vectors.
To understand the ramifications of reducing dimensions (and thereby losing information)
for sparse MIPS, we study the behavior of two particular random projection
techniques when applied to sparse vectors:
the linear Johnson-Lindenstrauss (JL)~\cite{fjlt,JLLemma1984ExtensionsOL,ailon2006fjlt,edo-fastjk-soda,edo-fastjl-acm}
transform and the non-linear \sinnamon{}~\cite{bruch2023sinnamon} transform.
We study this particular topic in depth in Section~\ref{section:sketching}.

By projecting sparse high-dimensional vectors into a (possibly dense) low-dimensional
subspace, we have removed the main barrier to applying dense MIPS solutions to sparse
vectors and are therefore prepared to investigate our main research question above.
We are particularly interested in a method commonly known as Inverted File-based (IVF)
retrieval: It begins by clustering vectors into partitions in an unsupervised manner.
When it receives a query vector, it identifies a subset of the more ``promising'' partitions,
and conducts (exact or approximate) retrieval only over the subset of documents assigned to them.
The search over the sub-collection can be delegated to another MIPS algorithm, the most na\"ive of
which is an exhaustive, exact search.
To understand how (sketches of) sparse vectors behave in an IVF retrieval system,
we empirically evaluate standard and spherical KMeans~\cite{sphericalKMeans} on
a range of datasets. This analysis is the main topic of Section~\ref{section:clustering}.

\begin{algorithm}[!t]
\SetAlgoLined
{\bf Input: }{Collection $\mathcal{X}$ of sparse vectors in $\mathbb{R}^{N}$;
Number of clusters, $P$;
Random projector,  $\phi: \mathbb{R}^N \rightarrow \mathbb{R}^n$ where $n \ll N$;
Clustering algorithm $\textsc{Cluster}$ that returns partitions of input data and their representatives.}\\
\KwResult{Cluster assignments $\mathcal{P}_i = \{ j \;|\; x^{(j)} \in \text{ Partition } i \}$
and cluster representatives $\mathcal{C}_i$'s.}

\begin{algorithmic}[1]
    \STATE $\tilde{\mathcal{X}} \leftarrow \{ \phi(x) \;|\; x \in \mathcal{X} \}$ \label{algorithm:indexing:sketching}
    \STATE $\textsc{Partitions}, \textsc{Representatives} \leftarrow \textsc{Cluster}(\tilde{\mathcal{X}}; P)$
    \STATE $\mathcal{P}_i \leftarrow \{ j \;|\; \tilde{x}^{(j)} \in \textsc{Partitions}[i] \}, \quad \forall 1 \leq i \leq P$
    \STATE $\mathcal{C}_i \leftarrow \textsc{Representatives}[i], \quad \forall 1 \leq i \leq P$
    \RETURN $\mathcal{P}$ and $\mathcal{C}$
 \end{algorithmic}
 \caption{Indexing}
\label{algorithm:indexing}
\end{algorithm}

\begin{algorithm}[!t]
\SetAlgoLined
{\bf Input: }{Sparse query vector, $q \in \mathbb{R}^{N}$;
Clusters and representatives,
$\mathcal{P}$, $\mathcal{C}$ obtained from Algorithm~\ref{algorithm:indexing};
Random projector $\phi: \mathbb{R}^N \rightarrow \mathbb{R}^n$ where $n \ll N$;
Number of data points to examine, $\ell \leq |\mathcal{X}|$, where $\lvert \mathcal{X} \rvert$
denotes the size of the collection;
MIPS sub-algorithm $\mathcal{R}$.}\\
\KwResult{Approximate set of top $k$ vectors that maximize inner product with $q$.}

\begin{algorithmic}[1]
    \STATE $\tilde{q} \leftarrow \phi(q)$
    \STATE \textsc{SortedClusters} $\leftarrow \textbf{SortDescending}(\mathcal{P} \text{ by } \langle \tilde{q}, \mathcal{C}_i \rangle)$

    \STATE $\textsc{TotalSize} \leftarrow 0$
    \STATE $\mathcal{I} \leftarrow \emptyset$ \Comment*[r]{Records the index of the partitions $\mathcal{R}$ should probe.}
    \FOR{$\mathcal{P}_{\pi_i} \in $ \textsc{SortedClusters}}
    \STATE $\mathcal{I} \leftarrow \pi_i$
    \STATE $\textsc{TotalSize} \leftarrow \textsc{TotalSize} + \lvert \mathcal{P}_{\pi_i} \rvert$
    \STATE \textbf{break if} $\textsc{TotalSize} \geq \ell$
    \ENDFOR
    
    \RETURN Top $k$ vectors from partitions $\mathcal{P}_\mathcal{I} \triangleq \{ \mathcal{P}_i \;|\; i \in \mathcal{I} \}$
    w.r.t $\langle q, \cdot \rangle$ using $\mathcal{R}$ \label{algorithm:retrieval:restricted-mips}
 \end{algorithmic}
 \caption{Retrieval}
\label{algorithm:retrieval}
\end{algorithm}

Together, dimensionality reduction via random projections and clustering, enable
the IVF paradigm for sparse vectors. Algorithm~\ref{algorithm:indexing} describes the
end-to-end indexing procedure, and Algorithm~\ref{algorithm:retrieval} gives details
of the retrieval logic. We encourage the reader to refer to Section~\ref{section:setup}
for an overview of our adopted notation.

\subsection{Research Byproducts}

As we demonstrate, it is certainly feasible and---given an appropriate tolerance for
error---often effective, to apply Algorithms~\ref{algorithm:indexing} and~\ref{algorithm:retrieval} to sparse vectors.
That possibility immediately leads to two important observations that we explore
later in this work.

First, we remark that, in effect, clustering a document collection
and performing search over only a fraction of the resulting clusters,
constitutes a dynamic pruning method---albeit a rank-unsafe one.
We use this insight to propose an organization of the inverted index
where inverted lists comprise of blocks, with each block containing documents
that fall into the same partition, and sorted by partition identifier.
We show that, appropriately using skip pointers over inverted lists facilitates
fast approximate top-$k$ retrieval for \emph{general} sparse vectors---vectors that need not
conform to any distributional requirements.
Experiments confirm the efficiency and effectiveness of our proposal.

Secondly, we offer a fresh but natural perspective to unify the
two worlds of dense and sparse MIPS into a single, elegant framework at the systems level.
In particular, we consider \emph{hybrid} vectors (i.e., vectors that may contain dense
\emph{and} sparse subspaces) in an IVF retrieval system.
We demonstrate empirically that the clusters formed by our proposal
are effective, and, regardless of how the $\ell 2$ mass is split between
the dense and sparse subspaces, retrieval can be arbitrarily accurate.

\subsection{Contributions}

We summarize our contributions as follows:
\begin{itemize}
    \item We analyze the effect of linear and non-linear random projection algorithms
    on the inner product approximation of sparse vectors;
    \item We extend the clustering-based IVF method of dense MIPS to (sketches of) sparse vectors,
    and, in that context, empirically evaluate standard and spherical KMeans clustering algorithms;
    \item We use our findings to propose a novel organization of the inverted index
    that facilitates approximate MIPS over general sparse vectors, thereby freeing sparse MIPS
    from strict distributional requirements of traditional top-$k$ retrieval algorithms in IR; and,
    \item We propose a unification of dense and sparse MIPS using IVF,
    and present a preliminary empirical evaluation of the proposal.
\end{itemize}

Throughout our presentation, we hope to convey the simplicity that our proposals provide in working with vectors,
regardless of their density or sparsity, for both researchers and practitioners.
But we are more excited by what this new perspective enables and the major research questions it inspires.
To start, we believe our framework and the retrieval machinery it offers
provide substantial flexibility to researchers who wish to study learnt term weights
without the constraints imposed by traditional inverted index-based retrieval algorithms.
We are equally encouraged by our initial findings on hybrid vector retrieval
and hope our framework enables further research on lexical-semantic search,
multi-modal retrieval, multimedia retrieval, and other domains.

We additionally claim, as we argue later, that our proposed view opens the door to new and exciting research directions in IR,
while, as a meta-algorithm, still allowing the incorporation of decades of research.
From principled distributed system design, to the mathematics of alternative sparse vector sketching,
to improved clustering or partitioning algorithms, our conceptual framework motivates a number of research questions to
pursue. Moreover, our proposal gives a new flavor to the important research on efficient and effective
systems in IR~\cite{bruch2022reneuir,bruch2023reneuir}: the PAC nature of the framework
offers intrinsic levers to trade off efficiency for
effectiveness that deserve a thorough theoretical and empirical examination.

\subsection{Structure}

The remainder of this manuscript is organized as follows.
We review the relevant parts of the literature in Section~\ref{section:related-work}.
We then describe our notation and setup in Section~\ref{section:setup}. That
will let us put in context our analysis and discussion of the behavior of linear and non-linear
random projections for sparse vectors in Section~\ref{section:sketching}, and
subsequently clustering in Section~\ref{section:clustering}.
In Section~\ref{section:dynamic-pruning}, we show that clustering for IVF
and dynamic pruning for inverted indexes are intimately connected,
and describe a natural organization of the inverted index through clustering.
We philosophize on a unified, density-agnostic framework for MIPS
in Section~\ref{section:unified}. We conclude this manuscript in
Section~\ref{section:conclusion}.

\section{Related Work}
\label{section:related-work}

This section sets the stage by briefly reviewing the literature on sparse and dense MIPS.

\subsection{Sparse MIPS}

Numerous sparse MIPS algorithms exist in the IR literature that are specifically tailored to text data
and that are behind the success of the field in scaling to massive text collections.
We refrain from reviewing this vast literature here and, instead, refer the reader
to excellent existing surveys~\cite{zobel2006invertedindex,tonellotto2018survey} on the topic.
But to give context to our work, we quickly make note of key algorithms and explain what makes
them less than ideal for the setup we consider in this work.

\subsubsection{Sparse MIPS for Text Collections}
MaxScore~\cite{maxscore} and WAND~\cite{broder2003wand}, along with their intellectual
descendants~\cite{ding2011bmwand,topk_bmindexes,mallia2019faster-blockmaxwand,mallia2017blockmaxwand_variableBlocks}
are the \emph{de facto} sparse MIPS algorithms, applied typically to vectors obtained
obtained from a BM25-encoding~\cite{bm25original} of text. This family of algorithms augment
a document identifier-sorted inverted index with upper-bounds on the partial
score contribution of each coordinate to the final inner product. With that
additional statistic, it is possible to traverse the inverted lists
one document at a time and decide if a document may possibly end up in the top $k$
set: if the document appears in enough inverted lists whose collective score upper-bound
exceeds the current threshold (i.e., minimum of scores in the current top-$k$ set),
then that document should be fully evaluated; otherwise, it has no prospect of ever
making it to the top-$k$ set and can therefore be safely rejected.

As articulated elsewhere~\cite{bruch2023sinnamon}, the logic above is effective
when vectors have very specific properties: non-negativity, asymmetricly higher
sparsity rate in queries, and a Zipfian distribution of the length of inverted lists.
It should be noted that these assumptions are true of relevance measures
such as BM25~\cite{bm25original}; sparse MIPS algorithms were designed for text
distributions after all.

The limitations of existing algorithms render them inefficient for the general
case of sparse MIPS, where vectors may be real-valued and whose sparsity rate is closer to
uniform across dimensions. That is because, coordinate upper-bounds become more uniform,
leading to less effective pruning of the inverted lists. That, among other problems~\cite{crane2017wsdm,bruch2023sinnamon},
renders the particular dynamic pruning strategy in MaxScore and WAND ineffective,
as demonstrated empirically in the past~\cite{bruch2023sinnamon,mackenzie2021wacky}.

\subsubsection{Signatures for Logical Queries}
There are alternatives to the inverted index, however, such as the use of signatures for retrieval
and sketches for inner product approximation~\cite{bitfunnel,binsketch,cat_binsketch}.
In this class of algorithms, Goodwin et al.~\cite{bitfunnel} describe the BitFunnel indexing machinery.
BitFunnel stores a bit signature for every document vector in the index using Bloom filters.
These signatures are scanned during retrieval to deduce if a document contains the terms of
a conjunctive query. While it is encouraging that a signature-based replacement to inverted
indexes appears not only viable but very much practical, the query logic BitFunnel supports is
limited to logical ANDs and does not generalize to the setup we are considering in this work.

Pratap et al. considered a simple algorithm~\cite{binsketch} to
sketch sparse \emph{binary} vectors so that the inner product of sketches
approximates the inner product of original vectors. They do so by randomly projecting each coordinate in the original
space to coordinates in the sketch. When two or more non-zero coordinates collide, the sketch records their logical OR.
While a later work extends this idea to categorical-valued vectors~\cite{cat_binsketch}, it is not obvious
how the proposed sketching mechanisms may be extended to real-valued vectors.

\subsubsection{General Sparse MIPS}
The most relevant work to ours is the recent study of general sparse MIPS by Bruch et al.~\cite{bruch2023sinnamon}.
Building on random projections, the authors proposed a sketching algorithm, dubbed \sinnamon{},
that embeds sparse vectors into a low-dimensional sparse subspace.
\sinnamon{}, as with the previous approach, randomly projects coordinates from the original space
to the sketch space. But the sketch space is a union of two subspaces: One that records the upper-bound
on coordinate values and another that registers the lower-bound instead. It was shown that
reconstructing a sparse vector from the sketch approximates inner product with any arbitrary query
with high accuracy.

Bruch et al.~\cite{bruch2023sinnamon} couple the sketches with an inverted index,
and empirically evaluate a coordinate-at-a-time algorithm for sparse MIPS.
They show considerable compression rate in terms of the size of the index
as well as latencies that are sometimes an order of magnitude better than WAND
on embedding vectors produced by \splade{}~\cite{formal2021splade,formal2022splade}.

\subsection{Dense MIPS}

Let us note that there exists an extremely vast body of works on approximate nearest neighbor
(ANN) search that is in and of itself an interesting area of research. Strictly speaking,
however, MIPS is a fundamentally different (and, in fact, a much harder) 
problem because inner product is not a proper metric; in fact, maximum cosine similarity
search and ANN with Euclidean distance are special cases of MIPS.
In spite of this, many MIPS solutions for dense vectors adapt ANN solutions to
inner product, often without any theoretical justification.

Consider, for example, the family of MIPS solutions that is based on proximity graphs
such as IP-NSW~\cite{ip-nsw18} and its many
derivatives~\cite{tan2021norm_adjusted_ipnsw,liu2019understanding,zhou2019mobius}.
These classes of algorithms construct a graph where each data point is a node
in the graph and two nodes are connected if they are deemed ``similar.''
Typically, similarity is based on Euclidean distance. But the authors of~\cite{ip-nsw18}
show that when one uses inner product (albeit improperly) to construct the graph,
the resulting structure is nonetheless capable of finding the maximizers of
inner product rather quickly and accurately.

Graph-based methods may work well but they come with two serious issues.
First, while we can reason about their performance in the Euclidean space,
we can say very little about why they do or do not work for inner product,
and under what conditions they may fail. It is difficult, for example,
to settle on a configuration of hyperparameters without conducting
extensive experiments and evaluation on a validation dataset.
The second and even more limiting challenge is the poor scalability
and slow index construction of graph methods.

Another family of MIPS algorithms can best be described as
different realizations of Locality Sensitive Hashing
(LSH)~\cite{indyk1998mips,shrivastava2014lsh_mips,
neyshabur2015lsh-mips,huang2015query_aware_lsh,song2021promips,wu2019local,ma2021sparse_binary_code_mips,
yan2018norm_ranging_lsh}. The idea is to project data points such that
``similar'' points are placed into the same ``bucket.'' Doing so enables
sublinear search because, during retrieval, we limit the search to the
buckets that collide with the query.

Many LSH methods for MIPS transform the problem to Euclidean
or angular similarity search first, in order to then recycle existing hash functions.
One of the main challenges with this way of approaching MIPS is that inner product
behaves oddly in high dimensions, in a way that is different from, say, Euclidean distance:
the maximum inner product between vectors is typically much smaller than the average vector norm.
Making LSH-based MIPS accurate requires an increasingly larger number of projections,
which leads to an unreasonable growth in index size~\cite{tiwari2023faster}.

Another method that is borrowed from the ANN literature
is search using an inverted file (IVF).
This method takes advantage of the geometrical structure of vectors
to break a large collection into smaller partitions. Points within each partition
are expected to result in a similar inner product with an arbitrary query
point---though there are no theoretical guarantees that that phenomenon actually
materializes. Despite that, clustering-based IVF is a simple and widely-adopted
technique~\cite{pq2011,Johnson2021faiss}, and has been shown to perform well
for MIPS~\cite{auvolat2015clustering}. Its simplicity and well-understood behavior
are the reasons we study this particular technique in this work.

Finally, in our review of the dense MIPS literature, we exclusively described
space partitioning algorithms that reduce the search space through some form of
partitioning or hashing, or by organizing vectors in a graph structure and traversing the edges
towards the nearest neighbors of a given query. It should be noted, however,
that the other and often critical aspect of MIPS is the actual computation of
inner product. There are many works that address that particular
challenge often via quantization (see~\cite{scann} and references therein)
but that are beyond the scope of this article.

\section{Notation and Experimental Setup}
\label{section:setup}

We begin by laying out our notation and terminology. Furthermore, throughout this work,
we often interleave theoretical and empirical analysis. To provide sufficient context
for our arguments, this section additionally gives details on our empirical setup
and evaluation measures.

\subsection{Notation}
Suppose we have a collection $\mathcal{X} \subset \mathbb{R}^{m+N}$
of possibly \emph{hybrid} vectors. That means, if $x \in \mathcal{X}$,
then $x$ is a vector that is comprised of an $m$-dimensional \emph{dense},
an $N$-dimensional \emph{sparse} array of coordinates,
where dense and sparse are as defined in Section~\ref{section:introduction}.
We abuse terminology and call the dense part of $x$ its ``dense vector''
and denote it by $x^d \in \mathbb{R}^m$. Similarly, we call the sparse part,
$x^s \in \mathbb{R}^N$, its ``sparse vector.'' We can write $x = x^d \oplus x^s$,
where $\oplus$ denotes concatenation.

The delineation above will prove helpful later when we discuss the status quo and our proposal
within one mathematical framework. Particularly, we can say that a sparse retrieval algorithm
operates on the sparse collection $\mathcal{X}^s = \{ x^s \;|\; x = x^d \oplus x^s \in \mathcal{X} \}$,
and similarly dense retrieval algorithms operate on $\mathcal{X}^d$, defined symmetrically.
Hybrid vectors collapse to dense vectors when $N=0$
(or when $x^s = \mathbf{0}$ for all $x \in \mathcal{X}$),
and reduce to sparse vectors when $m=0$ (or $x^d = \mathbf{0}\; \forall x \in \mathcal{X}$).

In our notation, MIPS aims to solve the following problem:
\begin{equation}
    \mathcal{S} = \argmax^{(k)}_{x \in \mathcal{X}} \; \langle q \; , \; x \rangle
    \label{equation:mips}
\end{equation}
to find, from $\mathcal{X}$, the set $\mathcal{S}$ of top $k$ vectors whose inner product
with the query vector $q = q^d \oplus q^s \in \mathbb{R}^{m+N}$ is maximal. Sparse and dense
MIPS are then special cases of the formulation above, when query and document vectors are
restricted to their sparse or dense subspaces respectively.

We write $nz(u)$ for the set of non-zero coordinates in a sparse vector,
$nz(u) = \{ i \;|\; u_i \neq 0 \}$, and denote the average number of non-zero
coordinates with $\psi = \mathbbm{E}[|nz(X)|]$ for a random vector $X$.
We denote coordinate $i$ of a vector $u$ using subscripts: $u_i$.
To refer to the $j$-th vector in a collection of vectors, we use superscripts: $u^{(j)}$.
We write $\langle u, v \rangle$ to express the inner product of two vectors $u$ and $v$.
We denote the set of consecutive natural numbers $\{ 1, 2, \ldots, m \}$
by $[m]$ for brevity.
Finally, we reserve capital letters to denote random variables (e.g., $X$)
and calligraphic letters for sets (e.g., $\mathcal{X}$).

\subsection{Experimental Configuration}

\begin{table*}[t]
\caption{Datasets of interest along with select statistics.
The rightmost two columns report the average number of non-zero
entries in documents and, in parentheses, queries for sparse vector
representations of the datasets.}
\label{table:dataset-stats}
\begin{center}
\begin{sc}
\begin{tabular}{c|cc|cc}
\toprule
Dataset & Document Count & Query Count & \splade{} & \esplade{}\\
\midrule
\textsc{MS Marco} Passage& $8.8$M & $6{,}980$ & 127 (49) & 185 (5.9) \\
NQ & $2.68$M & $3{,}452$ & 153 (51) & 212 (8) \\
\textsc{Quora} & $523$K & $10{,}000$ & 68 (65) & 68 (8.9) \\
\textsc{HotpotQA} & $5.23$M & $7{,}405$ & 131 (59) & 125 (13) \\
\textsc{Fever} & $5.42$M & $6{,}666$ & 145 (67) & 140 (8.6) \\
\textsc{DBPedia} & $4.63$M & $400$ & 134 (49) & 131 (5.9) \\
\bottomrule
\end{tabular}
\end{sc}
\end{center}
\end{table*}

\subsubsection{Datasets}
We perform our empirical analysis on a number of publicly available
datasets, summarized in Table~\ref{table:dataset-stats}.
The largest dataset used in this work is the \textsc{MS Marco}\footnote{Available at \url{https://microsoft.github.io/msmarco/}} Passage Retrieval v1 dataset~\cite{nguyen2016msmarco},
a retrieval and ranking collection from Microsoft.
It consists of about $8.8$ million short passages which, along with queries in natural language,
originate from Bing. The queries are split into train, dev, and eval non-overlapping subsets.
We use the small dev query set (consisting of $6{,}980$ queries) in our analysis.

We also experiment with $5$ datasets from the BeIR~\cite{thakur2021beir} collection\footnote{Available at \url{https://github.com/beir-cellar/beir}}: Natural Questions (NQ, question answering), \textsc{Quora} (duplicate detection), \textsc{HotpotQA} (question answering), \textsc{Fever} (fact extraction),
and \textsc{DBPedia} (entity search). For a more detailed description of each dataset,
we refer the reader to~\cite{thakur2021beir}.

\subsubsection{Sparse Vectors}
We convert the datasets above into sparse vectors by
using \splade{}~\cite{formal2022splade} and \esplade{}~\cite{lassance2022sigir}.
\textbf{\splade{}}\footnote{Pre-trained checkpoint from HuggingFace available at \url{https://huggingface.co/naver/splade-cocondenser-ensembledistil}}~\cite{formal2022splade}
is a deep learning model that produces sparse representations for text.
The vectors have roughly $30{,}000$ dimensions, where each dimension corresponds
to a term in the BERT~\cite{devlin2019bert} WordPiece~\cite{wordpiece} vocabulary.
Non-zero entries in a vector reflect learnt term importance weights.

\splade{} representations allow us to test the behavior of our algorithm
on query vectors with a large number of non-zero entries. However, we 
also create another set of vectors using a more efficient variant of \splade{},
called \textbf{\esplade{}}\footnote{Pre-trained checkpoints for document and
query encoders were obtained from \url{https://huggingface.co/naver/efficient-splade-V-large-doc} and \url{https://huggingface.co/naver/efficient-splade-V-large-query}, respectively}~\cite{lassance2022sigir}.
This model produces queries that have far fewer non-zero entries than the original
\splade{} model, but documents that may have a larger number of non-zero entries.

These two models give us a range of sparsity rates to work with
and examine our algorithms on. As a way to compare and contrast
the more pertinent properties of the learnt sparse representations,
Table~\ref{table:dataset-stats} shows the differences in the
sparsity rate of the two embedding models for all datasets considered
in this work.

\subsubsection{Evaluation}
Our main metric of interest is the \textbf{accuracy}\footnote{What we call ``accuracy''
in this work is also known as ``recall'' in the ANN literature. However, ``recall'' is an overloaded
term in the IR literature as it also refers to the portion of \emph{relevant} documents
returned for a query. We use ``accuracy'' instead to avoid that confusion.}
of approximate algorithms,
measured as follows: For every test query, we obtain the exact solution to MIPS
by exhaustively searching over the entire dataset. We then obtain approximate set of
top-$k$ documents using a system of interest. Accuracy is then measured as
the ratio of exact documents that are present in the approximate set.
This metric helps us study the impact of the different sources of error.

We also report \textbf{throughput} as queries per second (QPS) in a subset of our experiments
where efficiency takes center stage. When computing QPS, we include the
time elapsed from the moment query vectors are presented
to the algorithm to the moment the algorithm returns the requested top $k$ document vectors
for all queries---we emphasize that the algorithms used in this work do not
operate in batch mode.
We note that, because this work is a study of retrieval of vectors,
we do not factor into throughput the time it takes to embed a given piece of text.

\subsubsection{Hardware and Code}
We conduct experiments on a commercially available platform with
an Intel Xeon Platinum 8481C Processor (Sapphire Rapids) with a clock rate of $1.9$GHz,
$20$ virtual CPUs ($2$ vCPUs per physical core), and $44$GB of main memory.
This setup represents a typical server in a production environment---in fact,
we rented this machine from the Google Cloud Platform.

We further note that, we implemented all the methods discussed in this
work in the Rust programming language. We rely on the Rust compiler
for any platform-specific optimization and do not otherwise optimize the code
for the Intel platform (such as by developing SIMD code).

\section{Analysis of Random Projections for Sparse Vectors}
\label{section:sketching}

As noted earlier, the historical bifurcation of the retrieval machinery can, in no small part,
be attributed to the differences between sparse and dense vectors---in addition to
the application domain. For example, sparse vectors
are plagued with a much more serious case of the \emph{curse of dimensionality}.
In extremely high-dimensional spaces where one may have thousands to millions of dimensions,
the geometrical properties and probabilistic certainty that power clustering start to break down.
So does our intuition of the space.

The high dimensionality of sparse vectors poses another challenge: greater computation required
to perform basic operations. While optimized implementations (see, e.g.,~\cite{KIM2020113288}
and references therein) of spherical KMeans exist for sparse vectors, for example,
their efficiency nonetheless grows with the number of dimensions.
Standard KMeans is even more challenging: Cluster centroids are likely
to be high-dimensional \emph{dense} vectors, leading to orders of magnitude
more computation to perform cluster assignments in each iteration of the algorithm.

These difficulties---computational complexity and geometrical oddities---pose a
fundamental challenge to clustering over sparse vectors.
That leads naturally to dimensionality reduction, and in particular
\emph{sketching}~\cite{woodruff2014sketching}:
Summarizing a high-dimensional vector into a lower-dimensional space such
that certain properties, such as the distance between points or inner
products, are preserved with some quantifiable error.

The reason sketching is appealing is that
the mathematics behind it offer guarantees in an \emph{oblivious} manner:
with no further assumptions on the source and nature of the vectors
themselves or their distribution. Additionally, sketching a vector
is often fast since it is a requisite for their application in streaming algorithms.
Finally, the resulting \emph{sketch} in a (dense and) low-dimensional space
facilitates faster subsequent computation in exchange for a controllable error.

In this work, we explore two such sketching functions ($\phi(\cdot)$ in the notation
of Algorithm~\ref{algorithm:indexing}):
One classical result that has powered much of the
research on sketching is the linear Johnson-Lindenstrauss (JL)
transform~\cite{JLLemma1984ExtensionsOL}, which produces dense sketches of its input
and enables computing an unbiased estimate of inner product (or Euclidean distance). Another,
is the non-linear \sinnamon{} function~\cite{bruch2023sinnamon} that produces
sparse sketches of its input that enable deriving upper-bounds on inner product.

In the remainder of this section, we review these two algorithms in depth
and compare and contrast their performance. Importantly, we consider the
approximation error in isolation: How does sketching affect MIPS if our
MIPS algorithm itself were exact? In other words, if we searched exhaustively
for the top $k$ maximizers of inner product with a query, what accuracy
may be expect if that search were performed on sketches of vectors versus the
original vectors?

\subsection{The Johnson-Lindenstrauss Transform}

\subsubsection{Review}
Let us repeat the result due to Johnson and Lindenstrauss~\cite{JLLemma1984ExtensionsOL}
for convenience:

\begin{lemma}[Johnson-Lindenstrauss]
For $0 < \epsilon < 1$ and any set $\mathcal{V}$ of $|\mathcal{V}|$ points
in $\mathbb{R}^N$, and an integer $n = \Omega(\epsilon^{-2} \ln |\mathcal{V}|)$,
there exists a Lipschitz mapping $f: \mathbb{R}^N \rightarrow \mathbb{R}^n$ such that
\begin{equation*}
    (1 - \epsilon) \lVert u - v \rVert_2^2 \leq \lVert f(u) - f(v) \rVert_2^2
    \leq (1 + \epsilon) \lVert u - v \rVert_2^2,
\end{equation*}
for all $u, v \in \mathcal{V}$.
\end{lemma}

This result has been extensively studied and further developed since its introduction.
Using simple proofs, for example, it can be shown that the mapping $f$ may be
a linear transformation by an $n \times N$ random matrix $\Phi$ drawn
from a certain class of distributions. Such a matrix $\Phi$ is
said to form a JL transform~\cite{woodruff2014sketching}.


There are many constructions of $\Phi$ that form a JL transform.
It is trivial to show that when the entries of $\Phi$ are independently drawn from
$\mathcal{N}(0, \frac{1}{n})$, then $\Phi$ is a JL transform with parameters
$(\epsilon, \delta, \theta)$ if $n = \Omega(\epsilon^{-2} \ln (\theta / \delta))$.
$\Phi=\frac{1}{\sqrt{n}} R$, where $R_{n \times N}$ is a matrix whose entries are
independent Rademacher random variables, is another simple-to-prove example of a JL transform.
The literature offers a large number of other, more efficient constructions
such as the Fast JL Transform~\cite{ailon2006fjlt}, as well as specific
theoretical results for sparse vectors (e.g.,~\cite{baraniuk2006cs}).
We refer the interested reader to~\cite{woodruff2014sketching} for an excellent
survey of these results.

\subsubsection{Theoretical Analysis}
In this work, we are interested in the transformation in the context of inner product
rather than the $\ell 2$ norm and Euclidean distance.
Let us take $\phi(u) = R u$, with $R \in \{-1/\sqrt{n}, 1/\sqrt{n}\}^{n \times N}$,
as one candidate sketching function in Algorithm~\ref{algorithm:indexing}
and state the following results for our particular construction:

\begin{theorem}
\label{theorem:jl-variance-fixed-vectors}
Fix two vectors $u$ and $v \in \mathbb{R}^N$.
Define $Z_\textsc{Sketch} = \langle \phi(u), \phi(v) \rangle$ as the random
variable representing the inner product of sketches of size $n$,
prepared using the projection $\phi(u) = R u$, with $R \in \{-1/\sqrt{n}, 1/\sqrt{n}\}^{n \times N}$ being a random Rademacher matrix.
$Z_\textsc{Sketch}$ is an unbiased estimator of $\langle u, v \rangle$.
Its distribution tends to a Gaussian with variance:
\begin{equation}
\frac{1}{n} \big( \lVert u \rVert_2^2 \lVert v \rVert_2^2 + \langle u, v \rangle^2 - 2 \sum_i u_i^2 v_i^2  \big).
\end{equation}
\end{theorem}

We give our proof of the claim above in Appendix~\ref{appendix:jl-variance-fixed-vectors}.
We next make the following claim for a fixed query vector $q$ and a \emph{random} document
vector, thereby taking it a step closer to the MIPS setup. We present a proof in
Appendix~\ref{appendix:jl-variance-fixed-query}.

\begin{theorem}
\label{theorem:jl-variance-fixed-query}
Fix a query vector $q \in \mathbb{R}^N$ and let $X$ be
a random vector drawn according to the following probabilistic model.
Coordinate $i$, $X_i$, is non-zero with probability $p_i > 0$ and,
if it is non-zero, draws its value from a distribution with mean $\mu$
and variance $\sigma^2$. $Z_\textsc{Sketch} = \langle \phi(q), \phi(X) \rangle$,
with $\phi(u) = R u$ and $R \in \{-1/\sqrt{n}, 1/\sqrt{n}\}^{n \times N}$,
has expected value $\mu \sum_i p_i q_i$ and variance:
\begin{equation}
    \frac{1}{n} \big[
    (\mu^2 + \sigma^2)\big( \lVert q \rVert_2^2 \sum_i p_i - \sum_i p_i q_i^2 \big) +
    \mu^2 \big( (\sum_i q_i p_i)^2 - \sum_i (q_i p_i)^2 \big)
    \big].
\end{equation}
\end{theorem}

Consider the special case where $p_i = \psi / N$ for some constant $\psi$ for all dimensions $i$.
Further assume, without loss of generality, that the (fixed) query vector has
unit norm: $\lVert q \rVert_2 = 1$. It can be observed that the
variance of $Z_\textsc{Sketch}$ decomposes into a term that is
$(\mu^2 + \sigma^2) (1 - 1/N) \psi/n$,
and a second term that is a function of $1/N^2$.
The mean is a linear function of the non-zero coordinates in the query:
$(\mu \sum_i q_i) \psi/N $.
As $N$ grows, the mean of $Z_\textsc{Sketch}$ tends to $0$ at
a rate proportional to the sparsity rate ($\psi/N$), while its variance tends to
$(\mu^2 + \sigma^2) \psi/n$.

The analysis above suggests that the ability of $\phi(\cdot)$, as defined in this
section, to preserve the inner product of a query vector with a randomly drawn
document vector deteriorates as a function of the number of non-zero coordinates.
For example, when the number of non-zero coordinates becomes larger,
$\langle \phi(q), \phi(X) \rangle$ for a
fixed query $q$ and a random vector $X$ becomes less reliable because
the variance of the approximation increases.
Nonetheless, as we see later in this work, the degree of noise is
often manageable in practice as evidenced by the accuracy of
Algorithm~\ref{algorithm:retrieval}.

\subsection{The \sinnamon{} Transform}

\subsubsection{Review}
Like JL transform, \sinnamon{}~\cite{bruch2023sinnamon} aims to
reduce the dimensionality of (sparse) vectors. Unlike JL transform,
it does so through a non-linear mapping.

\sinnamon{} uses half the sketch to record upper-bounds
on the values of non-zero coordinates in a vector, and the other
half to register lower-bounds. For notational convenience, let us
assume that the sketch size is $n=2m$.
Given a vector $u \in \mathbb{R}^N$ and $h$ independent random mappings
$\pi_o: [N] \rightarrow [m]$ ($1 \leq o \leq h$), \sinnamon{} constructs
the upper-bound sketch $\overline{u} \in \mathbb{R}^m$ where its
$k$-th coordinate is assigned the following value:
\begin{equation}
    \overline{u}_k \leftarrow \max_{\{ i \in nz(u) \;|\; \exists \;o\; \mathit{s.t.}\; \pi_o(i) = k \}} u_i.
\end{equation}
The lower-bound sketch, $\underline{u}$, is filled in a symmetric manner, in the sense that
the algorithmic procedure is the same but the operator changes from $\max(\cdot)$ to $\min(\cdot)$.

Computing the inner product between a query vector $q \in \mathbb{R}^N$ and
a vector $u$ given its sketch ($\phi(u) = \underline{u} \oplus \overline{u}$) uses the
following procedure:
Positive query values are multiplied by the least upper-bound from $\overline{u}$,
and negative query values by the greatest lower-bound from $\underline{u}$:
\begin{equation}
    \label{equation:sinnamon:general-ip}
    \sum_i \mathbbm{1}_{i \in nz(u)}
        q_i \big(
        \mathbbm{1}_{q_i > 0} \min_{k \in \{ \pi_o(i) \; 1 \leq o \leq h\}} \overline{u}_k +
        \mathbbm{1}_{q_i < 0} \max_{k \in \{ \pi_o(i) \; 1 \leq o \leq h\}} \underline{u}_k
        \big).
\end{equation}

The indicator $\mathbbm{1}_{i \in nz(u)}$, which is kept in conjunction with the sketch,
guarantees that the partial inner product between a query coordinate $q_i$
and the sketch of a document vector (i.e., individual summands in Equation~(\ref{equation:sinnamon:general-ip}))
is $0$ if $i \notin nz(u)$. That pairing of the sketch with the
indicator function improves the bound on error dramatically while
maintaining a large compression rate. For formal results on the
probability of the inner product error, we refer the reader to the
original work \cite{bruch2023sinnamon}.

\subsubsection{Theoretical Analysis}
In this work, we use a simplified instance of \sinnamon{}, which we call \weaksinnamon{},
by (a) setting the number of random mappings to $1$, which we denote by $\pi$;
and (b) removing $\mathbbm{1}_{i \in nz(u)}$ from the inner product computation.
These two reductions have important side effects that ultimately enable us
to apply existing clustering algorithms and compute inner product between vectors.

Let us focus on the upper-bound sketch to illustrate these differences;
similar arguments can be made for the lower-bound sketch.
First, notice that the upper-bound sketch of a document vector simplifies to $\overline{u}$ where:
\begin{equation}
    \overline{u}_k \leftarrow \max_{\{ i \in nz(u) \;|\; \pi(i) = k \}} u_i,
\end{equation}
and that the upper-bound sketch of a query vector, $\overline{q}$, becomes:
\begin{equation}
    \overline{q}_k \leftarrow \sum_{\{ i \in nz(q) \;|\; \pi(i) = k \;\land\; q_i > 0 \}} q_i.
\end{equation}
We denote the former by $\phi_d(\cdot)$ (for document) and the latter by $\phi_q(\cdot)$
(for query).

Second, the inner product computation between the sketches of query and document vectors
reduces to:
\begin{equation}
\label{equation:sinnamon:reduced-ip}
    \langle \phi_q(q), \phi_d(u) \rangle = \langle \overline{q}, \overline{u} \rangle + \langle \underline{q}, \underline{u} \rangle
    = \sum_{i: \; q_i > 0} q_i \overline{u}_{\pi(i)} + \sum_{i: \; q_i < 0} q_i \underline{u}_{\pi(i)} .
\end{equation}

We now extend the analysis in~\cite{bruch2023sinnamon} to the setup above.
We begin by stating the following claim that is trivially true:

\begin{theorem}
    \label{theorem:sinnamon:ip-upperbound}
    For a query vector $q$ and document vector $u$,
    $\langle q, u \rangle \leq \langle \phi_q(q), \phi_d(u) \rangle$.
\end{theorem}

Importantly, the inner product between query and document sketches is not an unbiased
estimator of the inner product between the original vectors. Let us now model the
probability of the approximation error.

Consider the upper-bound sketch first. Using a similar argument to Theorem~5.4 of~\cite{bruch2023sinnamon},
we state the following result and provide a proof in Appendix~\ref{appendix:sinnamon:upper-bound-sketch-error}:

\begin{theorem}
\label{theorem:sinnamon:upper-bound-sketch-error}
Let $X$ be a random vector drawn according to the following probabilistic model.
Coordinate $i$, $X_i$, is non-zero with probability $p_i > 0$ and,
if it is non-zero, draws its value from a distribution with PDF $\phi$ and CDF $\Phi$.
Then:
\begin{equation}
\label{equation:sinnamon:upper-bound-error}
    \mathbb{P}[\overline{X}_{\pi(i)} - X_i \leq \delta] \approx
        (1 - p_i) \big( e^{-\frac{1}{m} (1 - \Phi(\delta)) \sum_{j \neq i} p_j} \big) +
        p_i \int e^{-\frac{1}{m} (1 - \Phi(\alpha + \delta)) \sum_{j \neq i} p_j} \phi(\alpha) d \alpha
\end{equation}
\end{theorem}
A symmetric argument can be made for the error of the lower-bound sketch.
Crucially, given the result above, which formalizes the CDF of the sketching
approximation error, we can obtain the expected value and variance of the
random variables $\overline{X}_{\pi(i)} - X_i$ and $\underline{X}_{\pi(i)} - X_i$
for all dimensions $i$. From there, and following similar arguments as the proof of Theorem~5.8
of~\cite{bruch2023sinnamon}, it is easy to show that the approximation error
takes on a Gaussian distribution with mean:
\begin{equation*}
 \sum_{i: \; q_i > 0} q_i \mathbb{E}[\overline{X}_{\pi(i)} - X_i] + \sum_{i: \; q_i < 0} q_i \mathbb{E}[\underline{X}_{\pi(i)} - X_i]   
\end{equation*}
and variance that is:
\begin{equation*}
    \sum_{i: \; q_i > 0} q_i^2 \mathit{Var}[\overline{X}_{\pi(i)} - X_i] + \sum_{i: \; q_i < 0} q_i^2 \mathit{Var}[\underline{X}_{\pi(i)} - X_i].
\end{equation*}

Let us illustrate the implications of Theorem~\ref{theorem:sinnamon:upper-bound-sketch-error}
by considering the special case where $p_i = \psi/N$ for all dimensions $i$.
As the sparsity rate increases and $N$ grows,
the second term in Equation~(\ref{equation:sinnamon:upper-bound-error}) tends to $0$ at
a rate proportional to $\psi/N$, while the first term dominates,
tending approximately to $\exp\big(-(1 - \Phi(\delta)) \psi/m\big)$.
By making $\psi/m$ smaller, we can control the approximation error and have it concentrate on smaller magnitudes.
That subsequently translates to a more accurate inner product between a fixed query and
a randomly drawn document vector.

As a final remark on \weaksinnamon{}, we note that when $n$ is larger than the
number of non-zero coordinates in a document vector, the resulting sketch itself
is \emph{sparse}. Furthermore, sketching using \weaksinnamon{} only requires $\mathcal{O}(\psi)$ operations,
with $\psi$ denoting the number of non-zero coordinates, while the JL transform
has a sketching complexity of $\mathcal{O}(n\psi)$.
As we explain later, these properties will play a key role in the efficiency
of sparse MIPS.

\subsection{Empirical Comparison}
Our results from the preceding sections shed light on how JL and \weaksinnamon{}
transformations are expected to behave when applied to sparse vectors.
Our main conclusion is that the sparsity rate heavily affects the approximation error.
In this section, we design experiments that help us observe the expected behavior
in practice and compare the two dimensionality reduction algorithms on real data.

Given a sparse dataset and a set of queries, we first obtain the exact top-$1$ document
for each query by performing an exhaustive search over the entire collection.
We then create a second dataset wherein each vector is a sketch of a vector
in the original dataset. We now perform exact search over the sketch dataset
to obtain top-$k^\prime$ ($k^\prime \geq 1$) documents, and report the accuracy
of the approximate retrieval.

\begin{figure}[t]
\begin{center}
\centerline{
\subfloat[\textsc{Quora}]{
\includegraphics[width=0.32\linewidth]{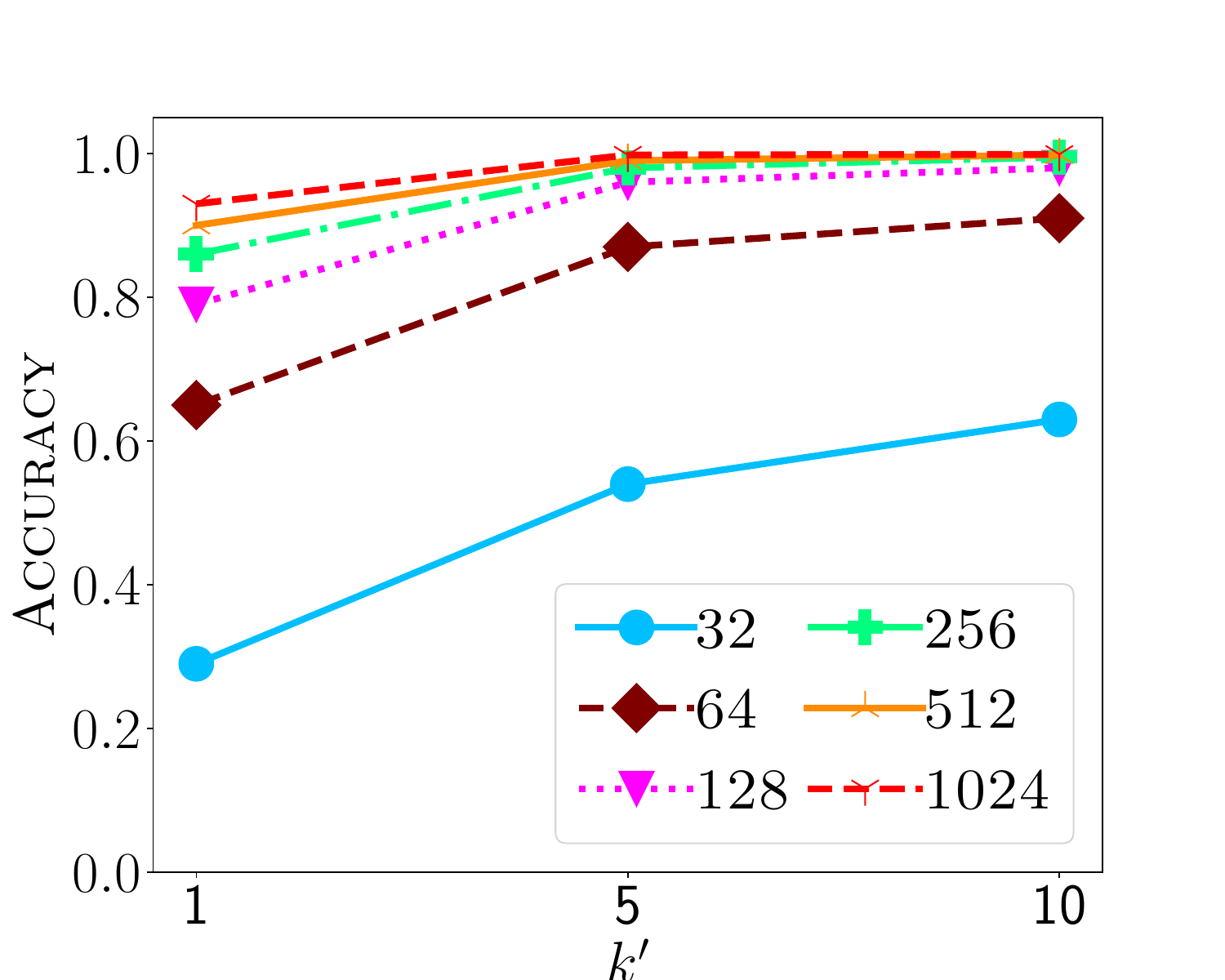}
\includegraphics[width=0.32\linewidth]{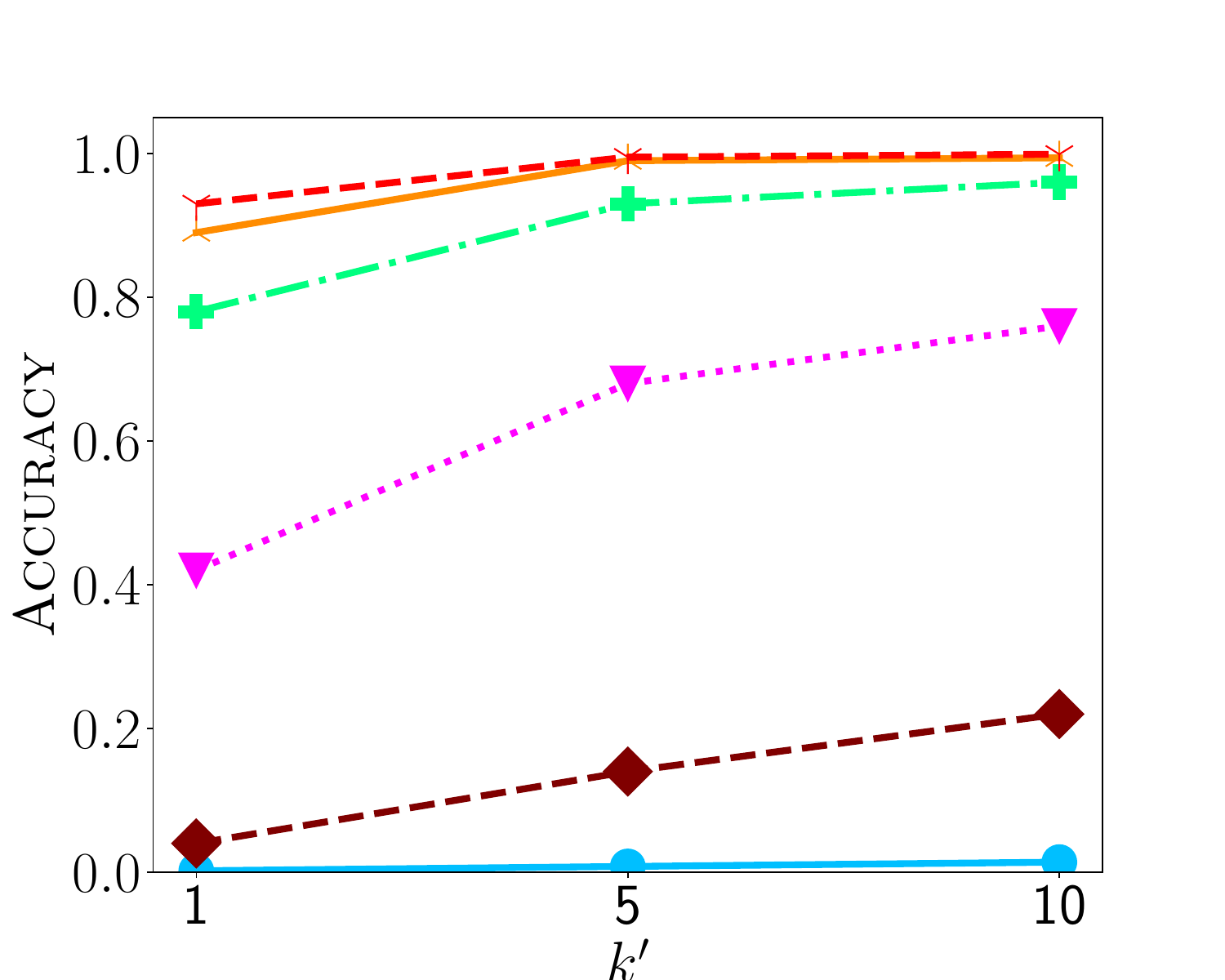}
\includegraphics[width=0.32\linewidth]{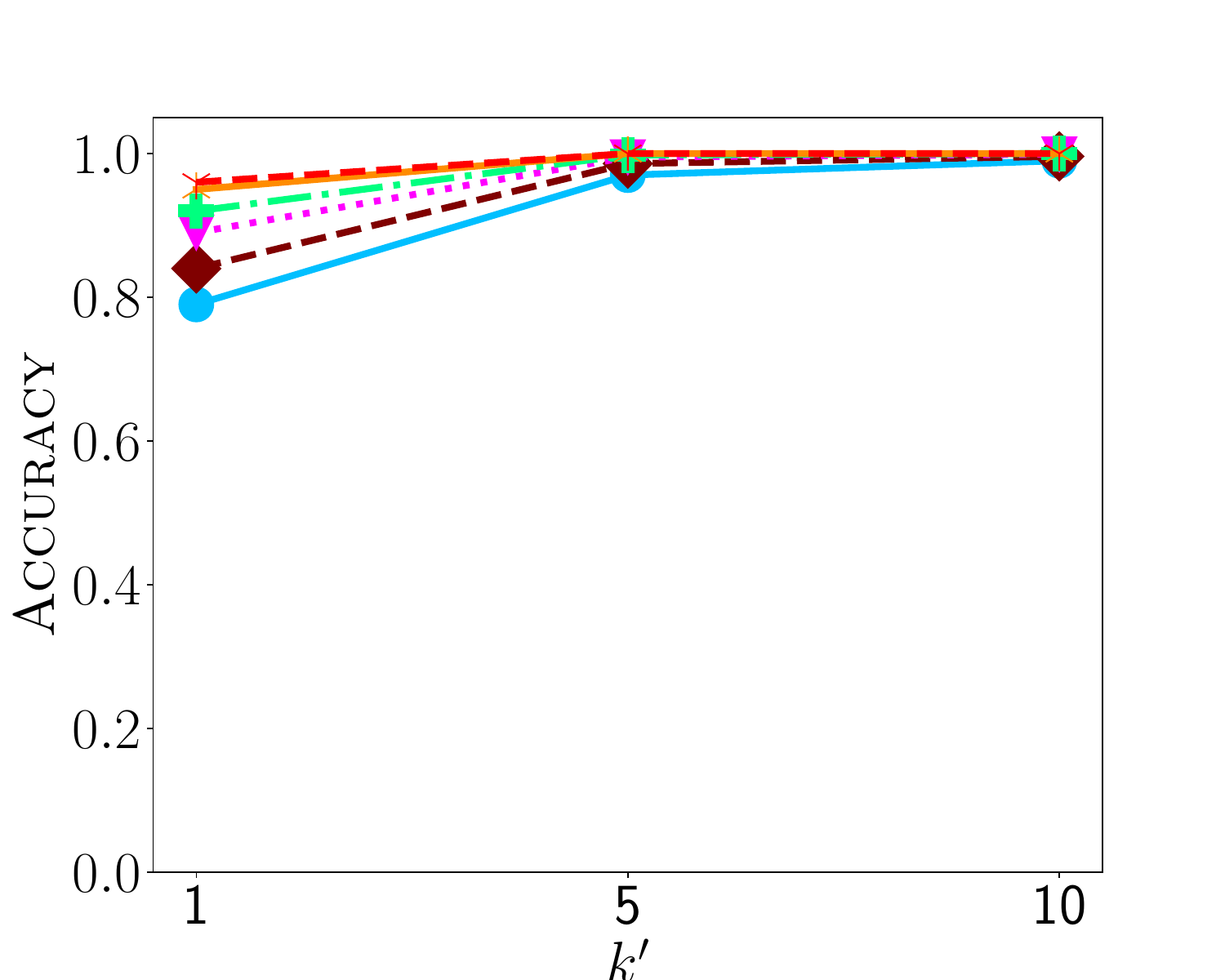}
}}
\vspace{-0.5cm}
\centering{
\subfloat[\textsc{NQ}]{
\includegraphics[width=0.32\linewidth]{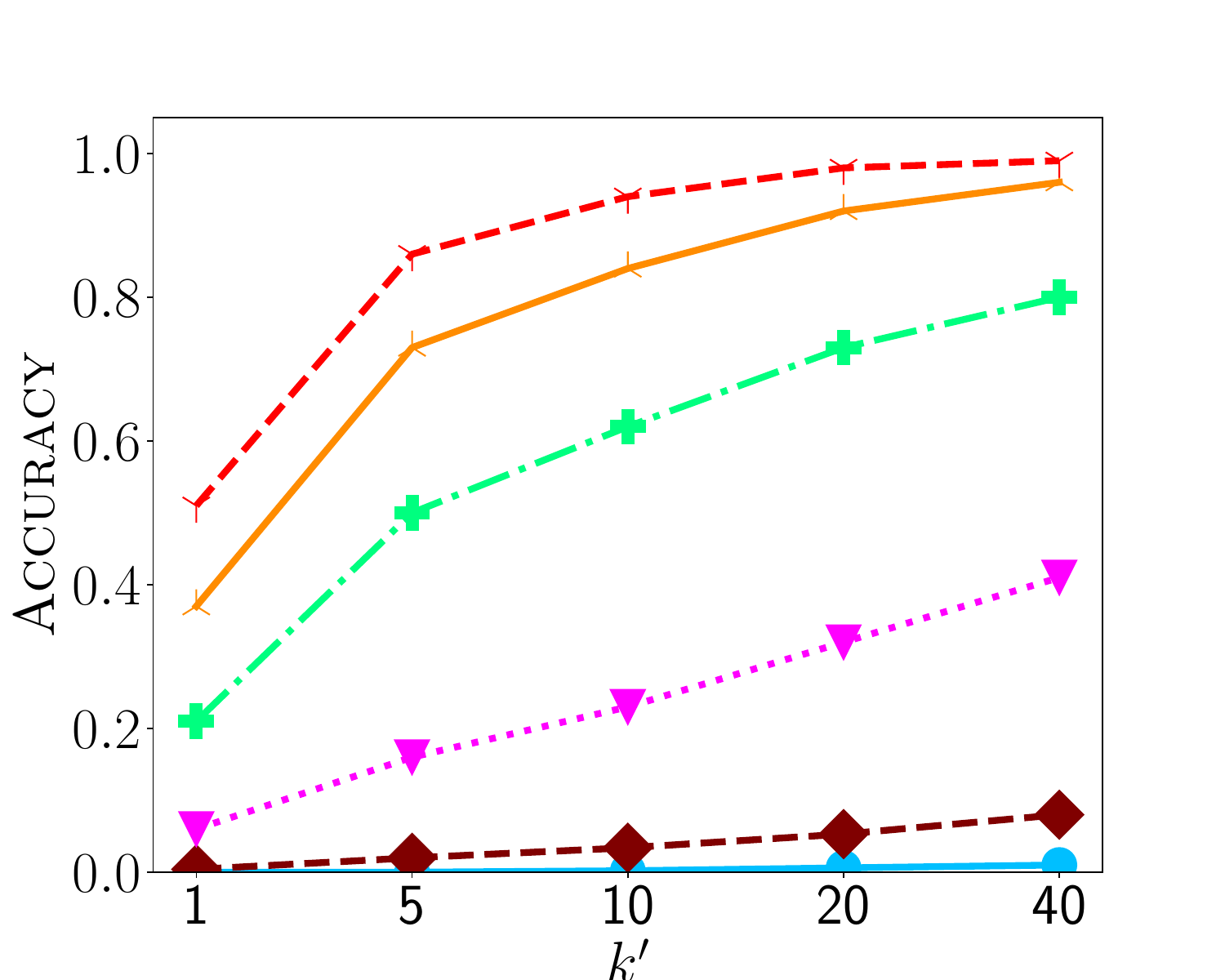}
\includegraphics[width=0.32\linewidth]{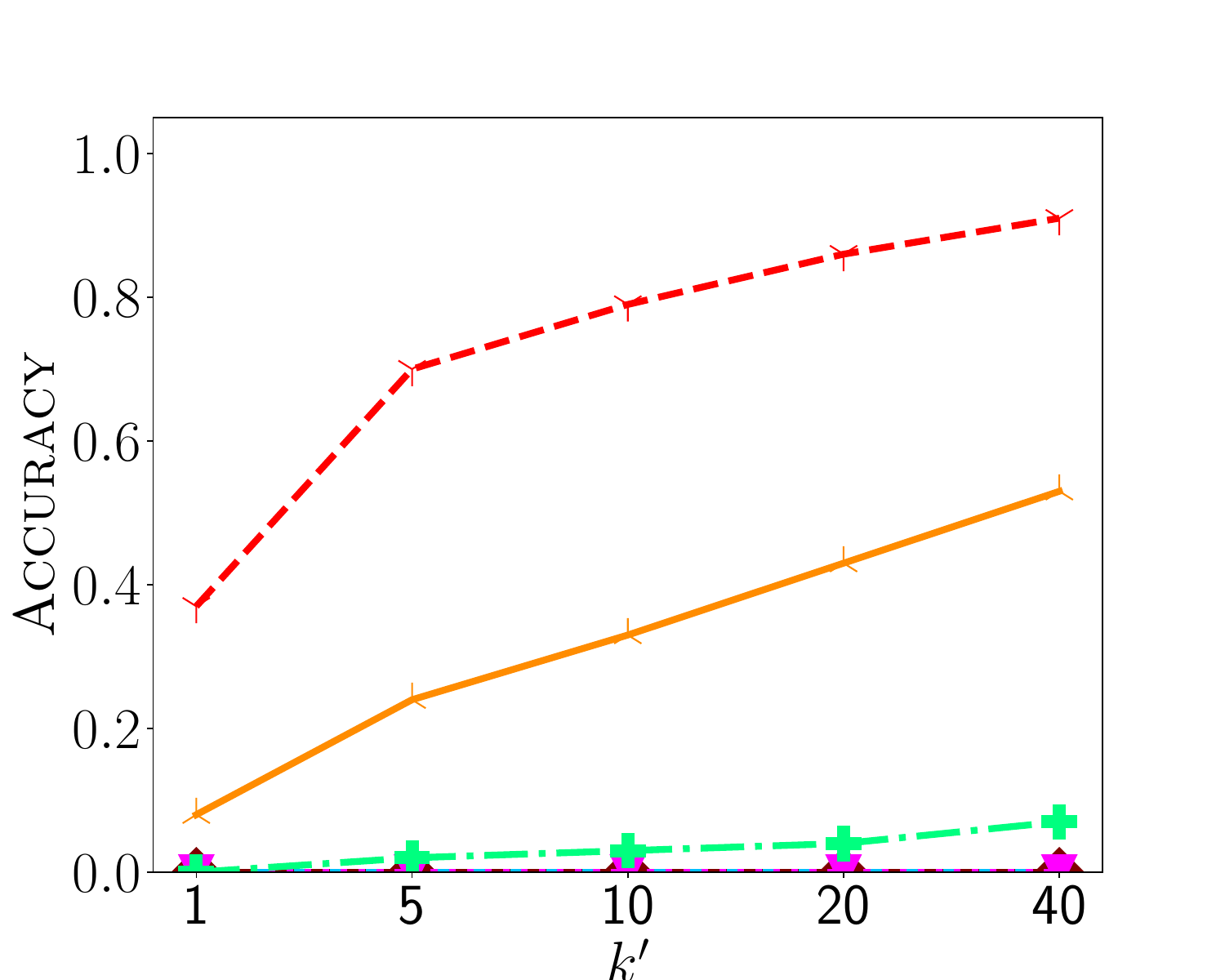}
\includegraphics[width=0.32\linewidth]{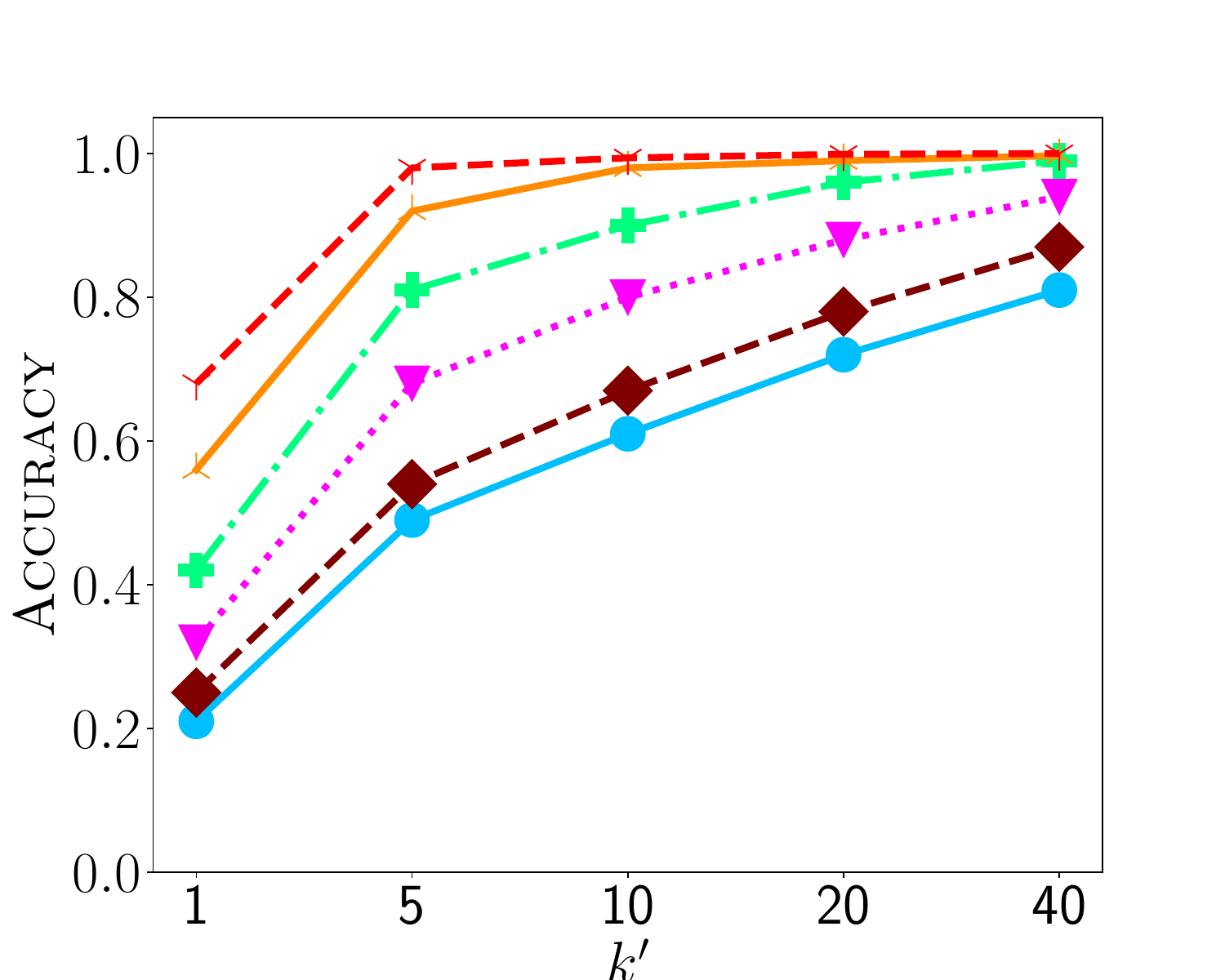}
}}
\vspace{0.2cm}
\caption{Top-$1$ accuracy of retrieval for test queries over sketches produced by JL transform (left column),
\weaksinnamon{} (middle column), and, as a point of reference, the original \sinnamon{} algorithm (right column).
We retrieve the top-$k^\prime$ documents by performing an exhaustive search over the sketch collection
and re-ranking the candidates by exact inner product to obtain the top-$1$ document and compute accuracy.
Each line in the figures represents a different sketch size $n$. We note that \weaksinnamon{}
and \sinnamon{} only use half the sketch
to record upper-bounds but leave the lower-bound sketch unused because \splade{} vectors
are non-negative. That implies that their effective sketch size is half that of the JL transform's.
}
\label{figure:sketching-quality}
\end{center}
\end{figure}

There are two parameters in the setup above that are of interest to us.
First is the sketch size, $n$. By fixing the dataset (thus its sparsity rate)
but increasing the sketch size, we wish to empirically quantify the effect of
using larger sketches on the ability of each algorithm to preserve inner product.
Note that, because the vectors are non-negative, \weaksinnamon{} only uses half the
sketch capacity to form the upper-bound sketch---reducing its effective
sketch size to $n/2$.

The second factor is $k^\prime$ which controls how ``hard'' a retrieval
algorithm must work to compensate for the approximation error. Changing
$k^\prime$ helps us understand if the error introduced by a particular
sketch size can be attenuated by simply retrieving more candidates and
later re-ranking them according to their exact score.

The results of our experiments are presented in Figure~\ref{figure:sketching-quality}
for select datasets embedded with the \splade{} model.
We chose these datasets because they have very different sizes and sparsity rates,
as shown in Table~\ref{table:dataset-stats}, with \textsc{Quora} having
the largest sparsity rate and fewest documents, and \textsc{NQ} the smallest sparsity rate
and a medium collection size.

\begin{figure}[t]
\begin{center}
\centerline{
\subfloat[\splade{}]{
\includegraphics[width=0.32\linewidth]{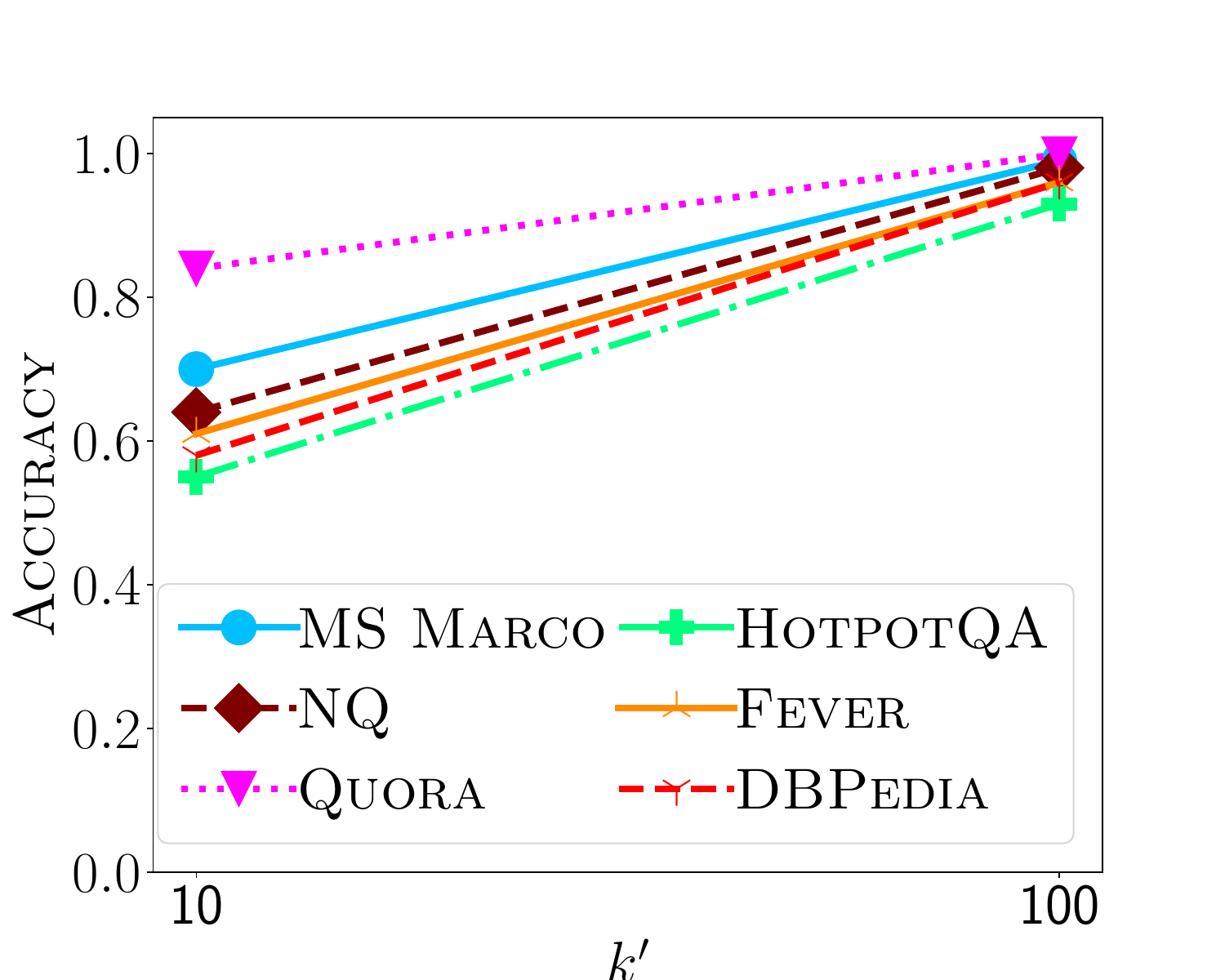}
\includegraphics[width=0.32\linewidth]{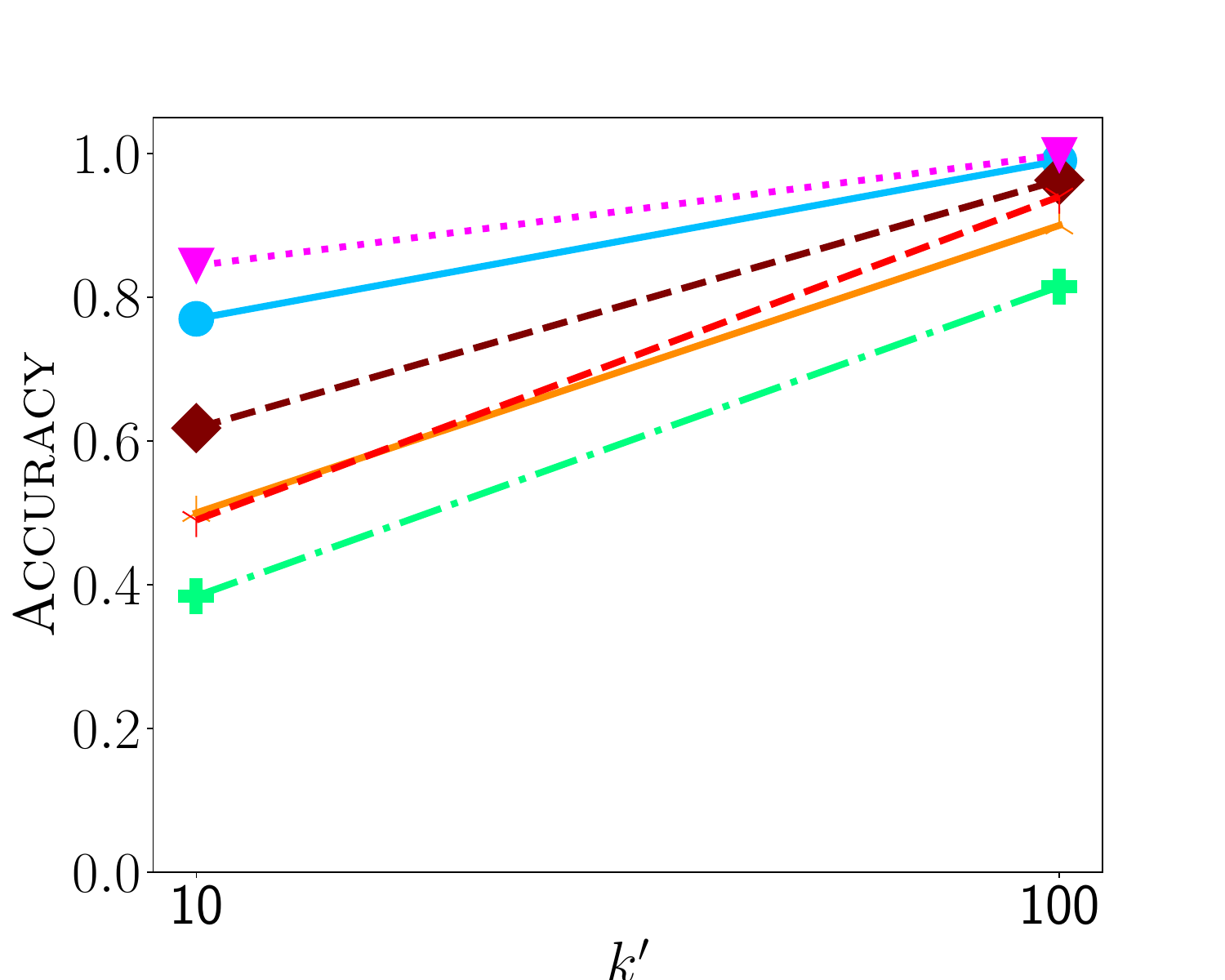}
\includegraphics[width=0.32\linewidth]{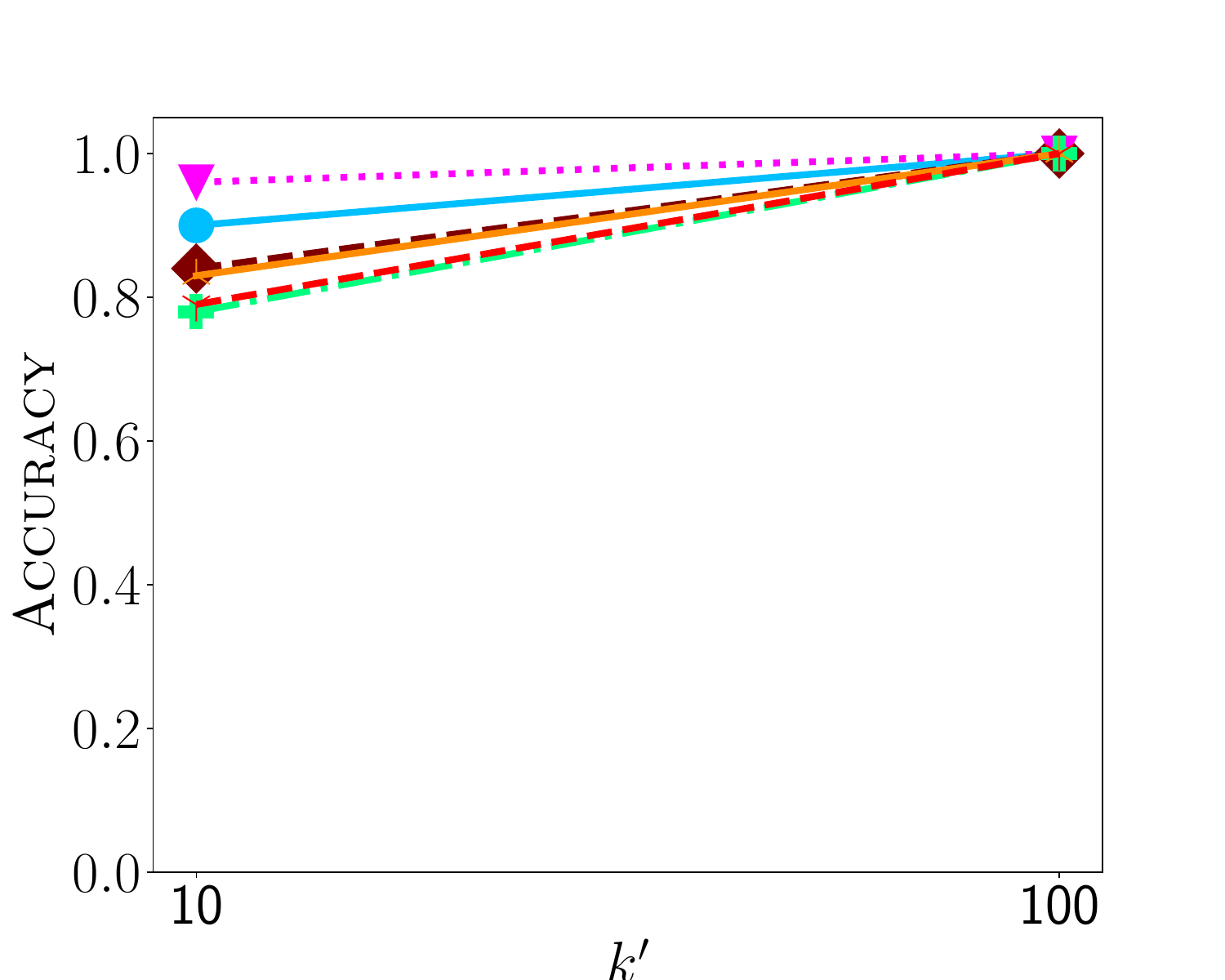}
}}
\vspace{-0.5cm}
\centering{
\subfloat[\esplade{}]{
\includegraphics[width=0.32\linewidth]{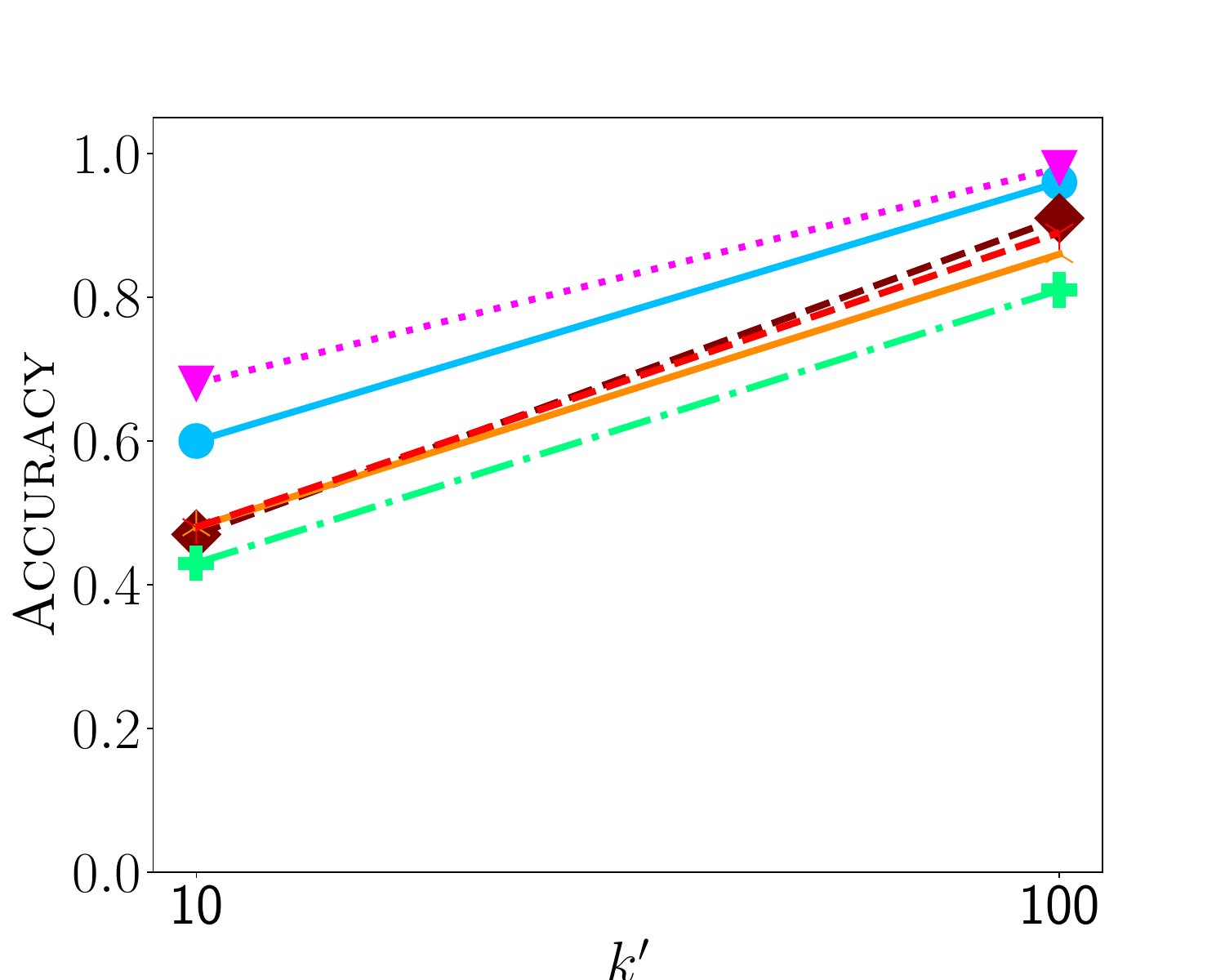}
\includegraphics[width=0.32\linewidth]{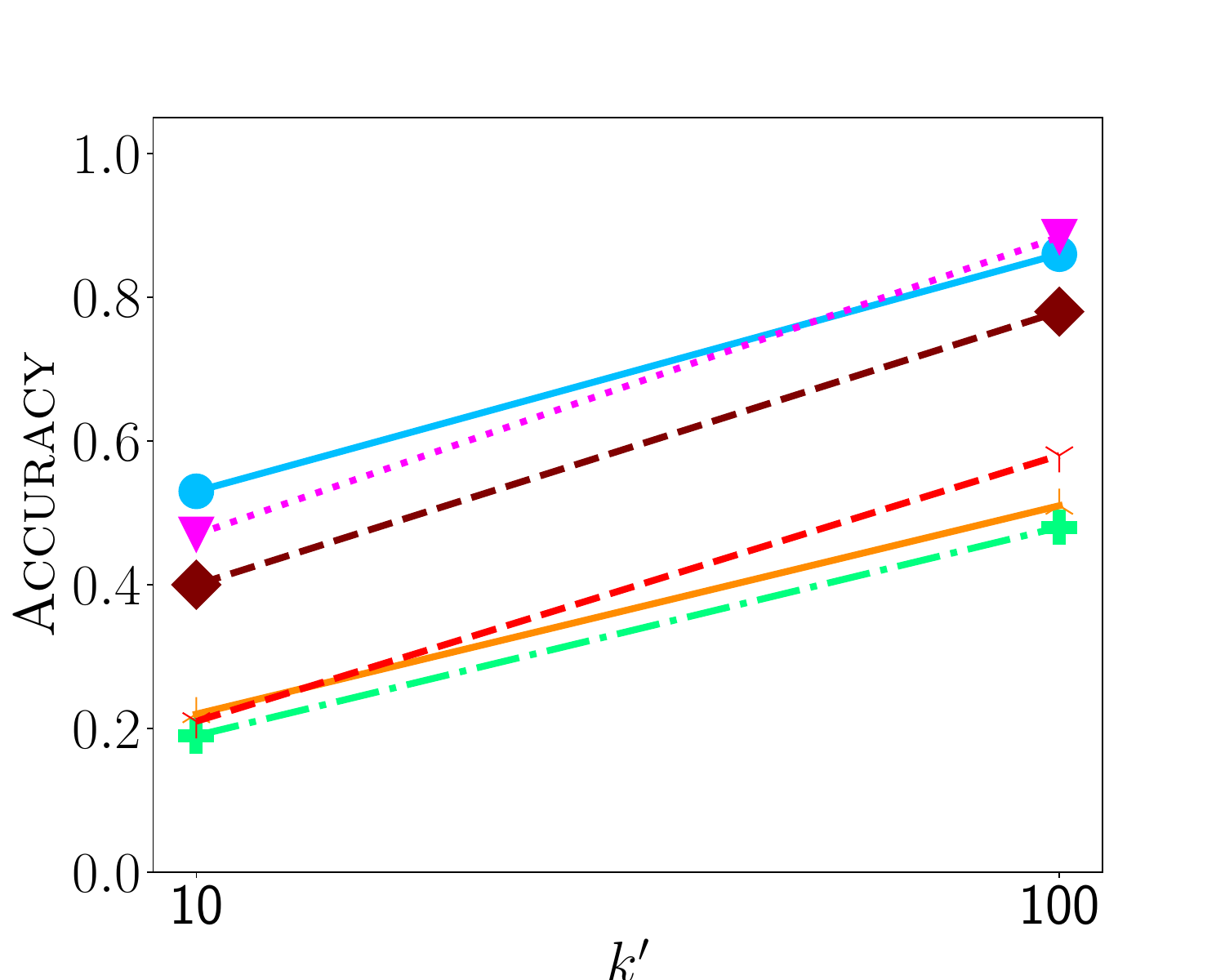}
\includegraphics[width=0.32\linewidth]{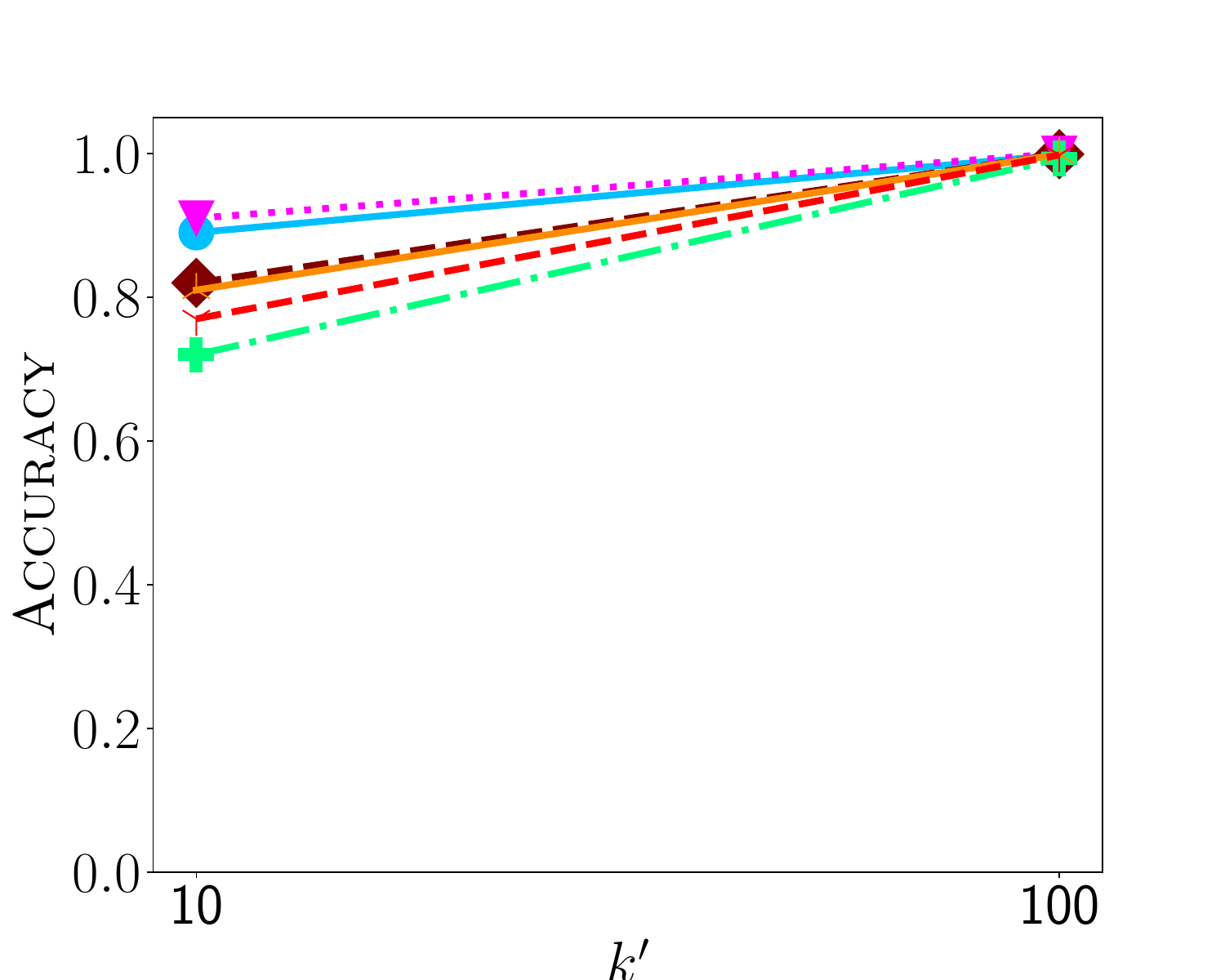}
}}
\vspace{0.2cm}
\caption{Top-$10$ accuracy of retrieval for test queries over sketches of size $n=1024$ produced by JL transform (left column),
\weaksinnamon{} (middle column), and, for reference, the original \sinnamon{} algorithm (right column).
As in Figure~\ref{figure:sketching-quality}, we retrieve the top-$k^\prime$ documents by performing an exhaustive
search over the sketch collection and re-ranking the candidates by exact inner
product to obtain the top-$10$ documents and compute accuracy.
Similarly, each line in the figures represents a different sketch size $n$. In these experiments,
however, we adjust the effective sketch size of \weaksinnamon{} and \sinnamon{} to match that of the JL transform's.}
\label{figure:sketching-quality:top10}
\end{center}
\end{figure}

Naturally, our observations are consistent with what the theoretical results predict.
The sketch quality improves as its size increases. That shows the effect of the
parameter $n$ on the approximation variance of the JL transform and the concentration
of error in \weaksinnamon{} sketches.

Another unsurprising finding is that \weaksinnamon{}'s sensitivity to the $\psi/n$
factor becomes evident in \textsc{NQ}: When the ratio between the number of non-zero
coordinates and the sketch size ($\psi/n$) is large, the variance of the approximation error becomes larger.
The reason is twofold: more non-zero coordinates are likely to collide as vectors
become more dense; and, additionally, sketches themselves become more dense,
thereby increasing the likelihood of error for \emph{inactive} coordinates.
To contextualize \weaksinnamon{} and the effects of our modifications to the
original algorithm on the approximation error, we also plot in Figure~\ref{figure:sketching-quality}
the performance of \sinnamon{}.

While increasing the sketch size is one way to lower the probability of error,
casting a wider net (i.e., $k^\prime > k$) followed by re-ranking appears to
also improve retrieval quality.

Now that we have a better understanding of the effect of the parameters
on the quality of the sketching algorithms, let us choose one configuration
and repeat the experiments above on all our datasets. One noteworthy adjustment
is that we set \weaksinnamon{}'s effective sketch size to match that of the JL transform's:
As we noted, because \weaksinnamon{} leaves the lower-bound sketch unused for non-negative vectors,
we re-allocate it for the upper-bound sketch, in effect giving \weaksinnamon{}'s upper-bound sketch
$n$ dimensions to work with. Another change is that we use a more challenging configuration and
perform top-$10$ retrieval. Finally, we also include \esplade{} for completeness.

Figure~\ref{figure:sketching-quality:top10} shows the results of these experiments.
The general trends observed in these figures are consistent with the findings of
Figure~\ref{figure:sketching-quality}: Obtaining a larger pool of candidates
from sketches and re-ranking them according to their exact inner product is a
reliable way of countering the approximation error; and, \weaksinnamon{} generally
underperforms the JL transform in preserving inner product between vectors.
Additionally, as vectors become more dense, the sketching quality degrades,
leading to a higher approximation error.

Another interesting but expected phenomenon is that sketching performs
comparatively poorly on \esplade{}. That is because, query vectors generated
by the \esplade{} model are more sparse than those made by \splade{}. When a
query has few non-zero coordinates, the expected inner product becomes small
while the variance of JL transform sketches concentrates around a constant,
as predicted by Theorem~\ref{theorem:jl-variance-fixed-query}.
As for \weaksinnamon{}, when queries have a large number of non-zero coordinates,
the shape of the distribution of error becomes less sensitive to the approximation
error of individual coordinates; with fewer non-zero coordinates in the query vector,
the opposite happens.

\begin{figure}[t]
\begin{center}
\centerline{
\subfloat[\splade{}]{
\includegraphics[width=0.49\linewidth]{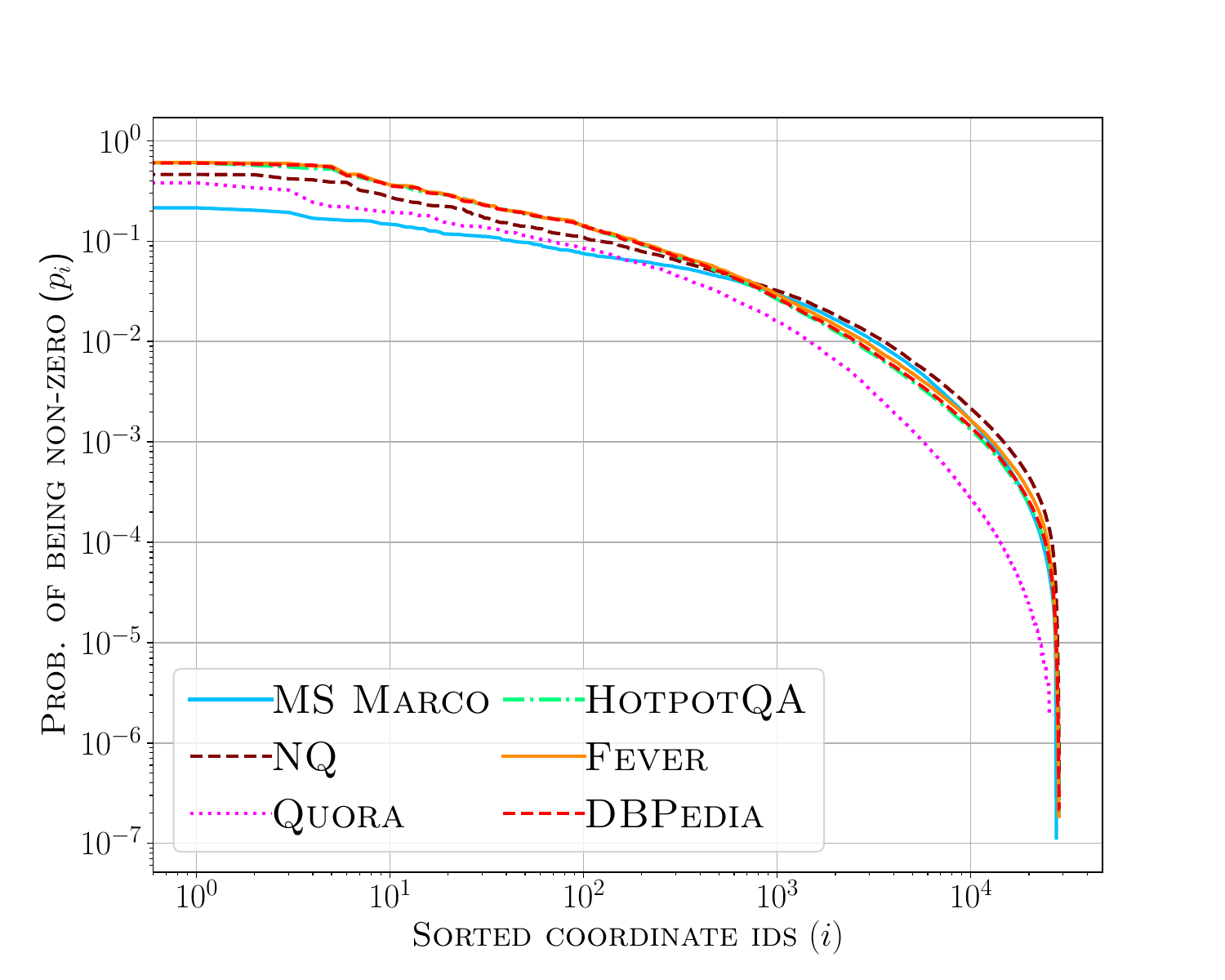}
}
\subfloat[\esplade{}]{
\includegraphics[width=0.49\linewidth]{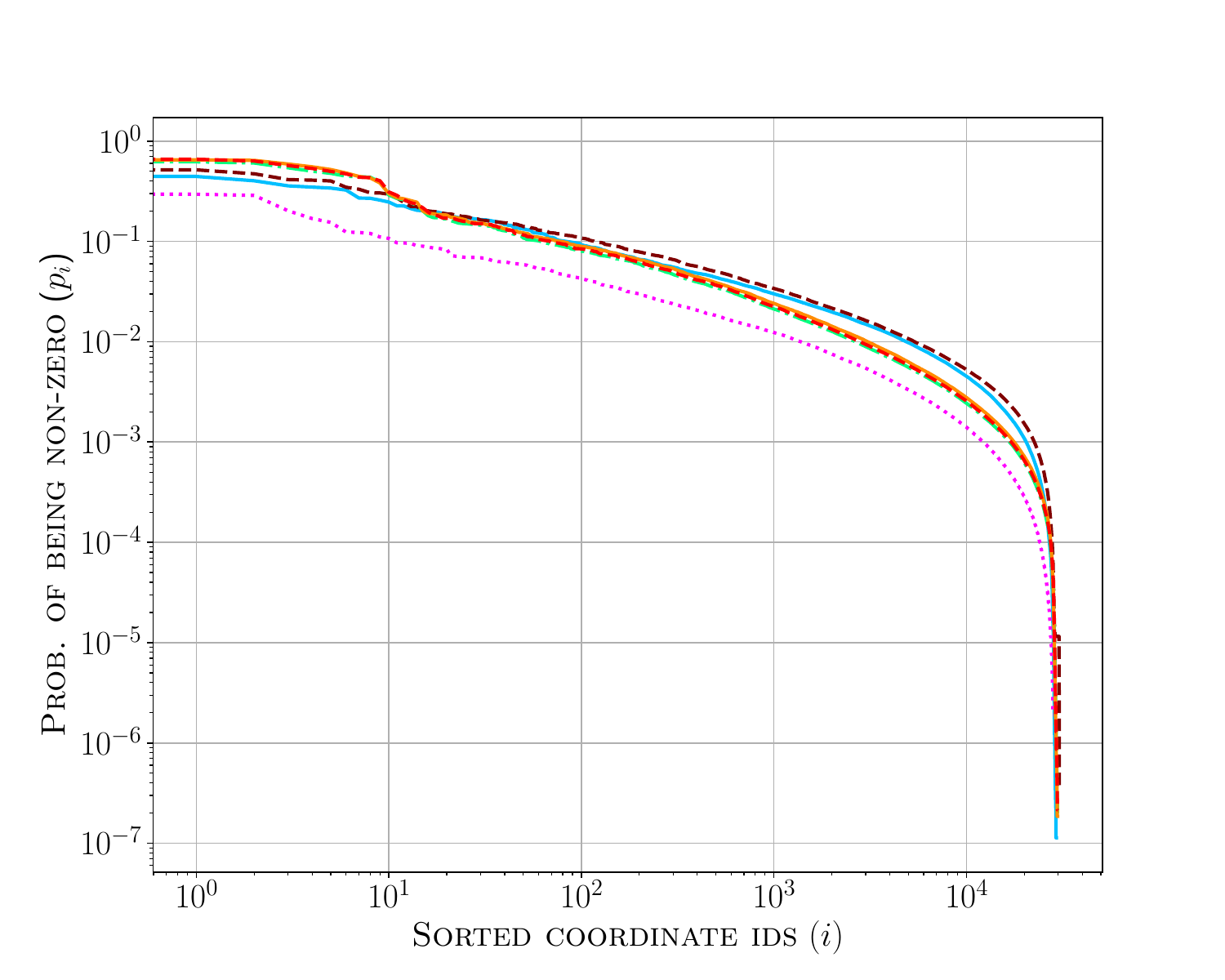}
}}
\vspace{0.2cm}
\caption{Probability of each coordinate being non-zero ($p_i$ for coordinate $i$)
for \splade{} and \esplade{} vectors of several datasets. To aid visualization,
we sort the coordinates by $p_i$'s in descending order. A Zipfian distribution
would manifest as a line in the log-log plot. Notice that, this distribution
is closer to uniform for \textsc{MS Marco} than others.}
\label{figure:sketching-quality:dist-non-zeros}
\end{center}
\end{figure}

As a final observation, we notice that retrieval accuracy is generally
higher for \textsc{Quora}, \textsc{MS Marco}, and \textsc{NQ} datasets.
That is easy to explain for \textsc{Quora} as it is a more sparse dataset
with a much smaller $\psi/n$. On the other hand, the observed trend is
rather intriguing for a larger and more dense dataset such as \textsc{MS Marco}.
On closer inspection, however, it appears that the stronger performance
can be attributed to the probabilities of coordinates being non-zero (i.e., $p_i$'s).
In Figure~\ref{figure:sketching-quality:dist-non-zeros},
we plot the distribution of $p_i$'s but, to make
the illustration cleaner, sort the coordinates by their $p_i$ in descending order.
Interestingly, the distribution of $p_i$'s is closer to uniform for \textsc{MS Marco}
and \textsc{NQ}, while it is more heavily skewed for \textsc{Fever}, \textsc{DBPedia},
and \textsc{HotpotQA}.

\section{Evaluation of Clustering over Sketches of Sparse Vectors}
\label{section:clustering}

In the preceding section, we were squarely concerned with the ability of
the two sketching algorithms in approximately preserving inner product between
a query vector and an arbitrary document vector. That analysis is relevant
if one were to directly operate on sketches as opposed to
the original vectors when, say, building a graph-based nearest neighbor search
index such as HNSW~\cite{malkov2016hnsw} or IP-NSW~\cite{ip-nsw18}.
In this work, our primary use for sketches is to form partitions
in the context of Algorithms~\ref{algorithm:indexing} and~\ref{algorithm:retrieval}:
Whether $\mathcal{R}$ searches over sketches or the original vectors
is left as a choice.

In that framework, Section~\ref{section:sketching} has already studied the first line of the two algorithms:
sketching the sparse vectors. In this section, we turn to the clustering procedure
and empirically evaluate two alternatives: Standard and spherical KMeans.
Note that, the clustering choice is the last piece required to complete
the two algorithms and apply IVF-style search to sparse vectors.

Standard KMeans is an iterative protocol that partitions the input
data into a predefined number of clusters, $K$.
It first samples $K$ arbitrary points, called ``centroids,'' from the data distribution
at random---though there are other
initialization protocols available, such as KMeans++~\cite{kmeansplusplus}.
It then repeats until convergence two steps:
It assigns each data point to the nearest centroid by their Euclidean distance to form partitions in the first step;
and, in the second step, recomputes the centroids to be the mean of the
mass of all data points assigned to each partition.
While this Expectation-Maximization procedure may fall into local optima,
it generally produces partitions that approximate Voronoi regions in a dataset.

Spherical KMeans works similarly, with the notable exception that at the end of each iteration,
it normalizes the centroids so that they are projected onto the unit sphere.
This form of clustering has been used in the past for a topical analysis
of text documents~\cite{sphericalKMeans} among other applications.

Both of these clustering algorithms are popular choices in the IVF-based
approximate nearest neighbor search as evidenced by their integration into
commonly used software packages such as FAISS~\cite{Johnson2021faiss}.
As such, we plug the two methods into Algorithms~\ref{algorithm:indexing}~and~\ref{algorithm:retrieval}
and apply them to our datasets. Our objective is to understand
the differences between the two clustering choices in terms of their role
in the overall retrieval quality as well as their sensitivity to the choice
of sketching algorithm.

\subsection{Empirical Comparison}

\begin{figure}[t]
\begin{center}
\centerline{
\subfloat[\textsc{MS Marco}]{
\includegraphics[width=0.32\linewidth]{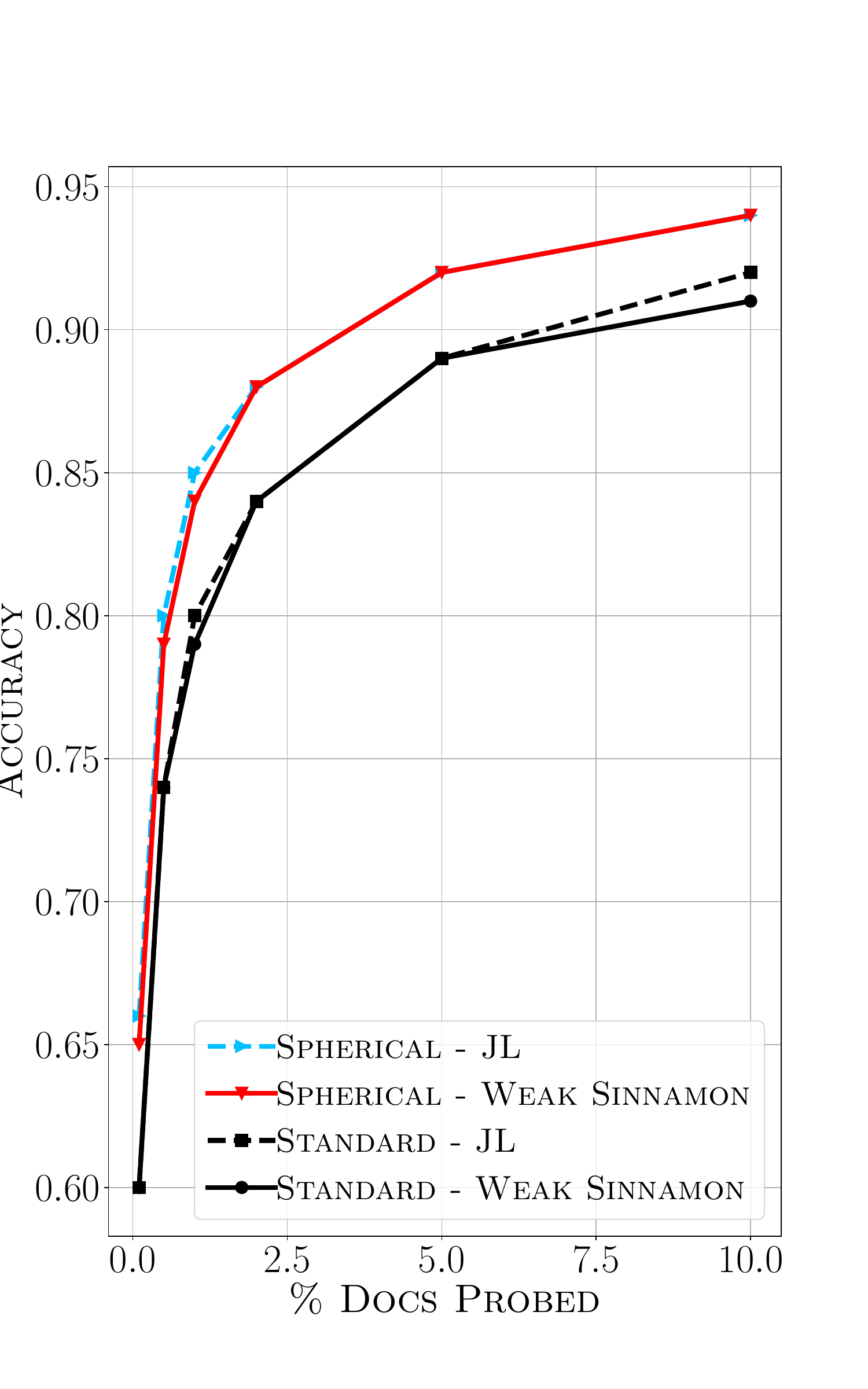}
}
\subfloat[\textsc{NQ}]{
\includegraphics[width=0.32\linewidth]{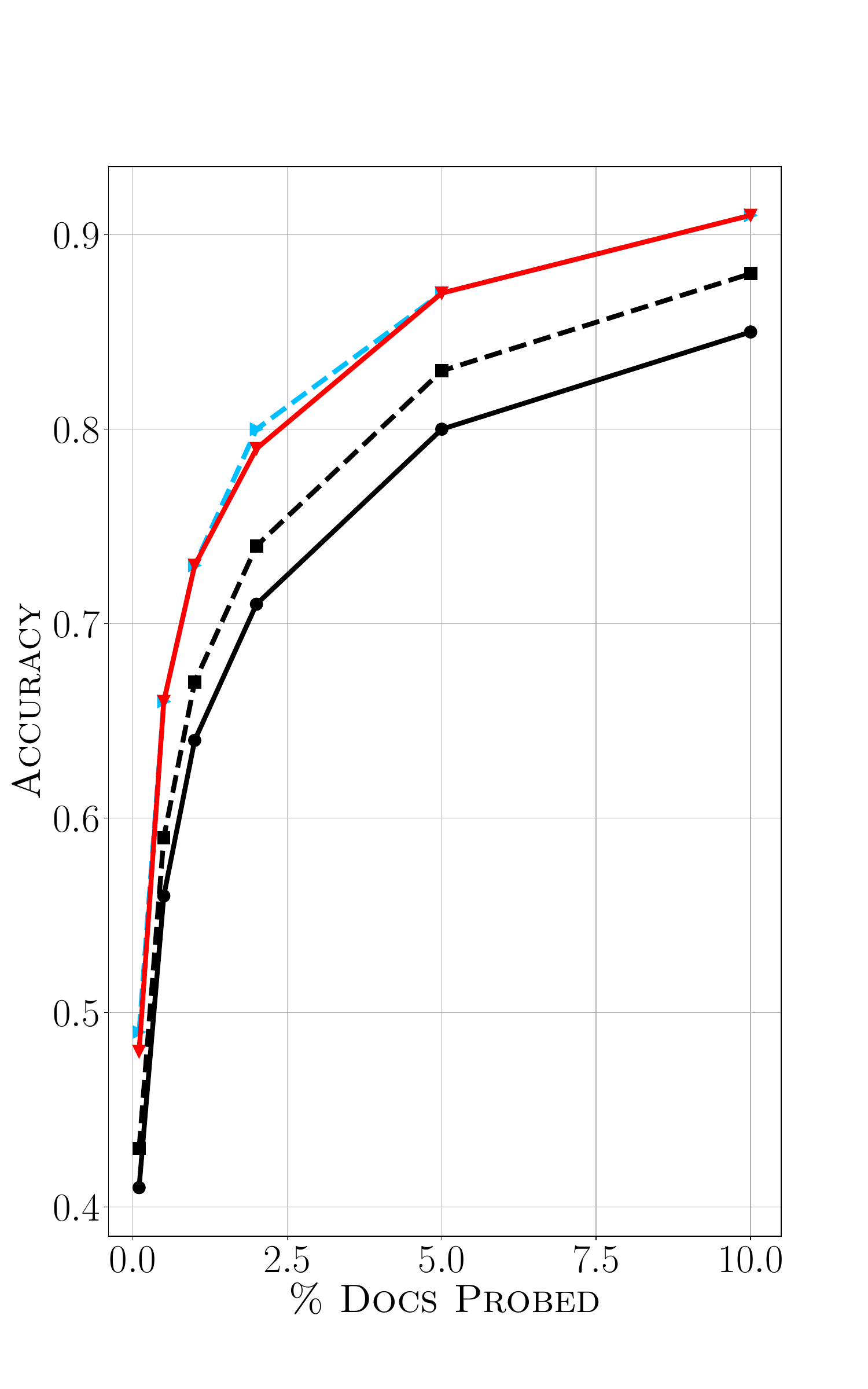}
}
\subfloat[\textsc{Quora}]{
\includegraphics[width=0.32\linewidth]{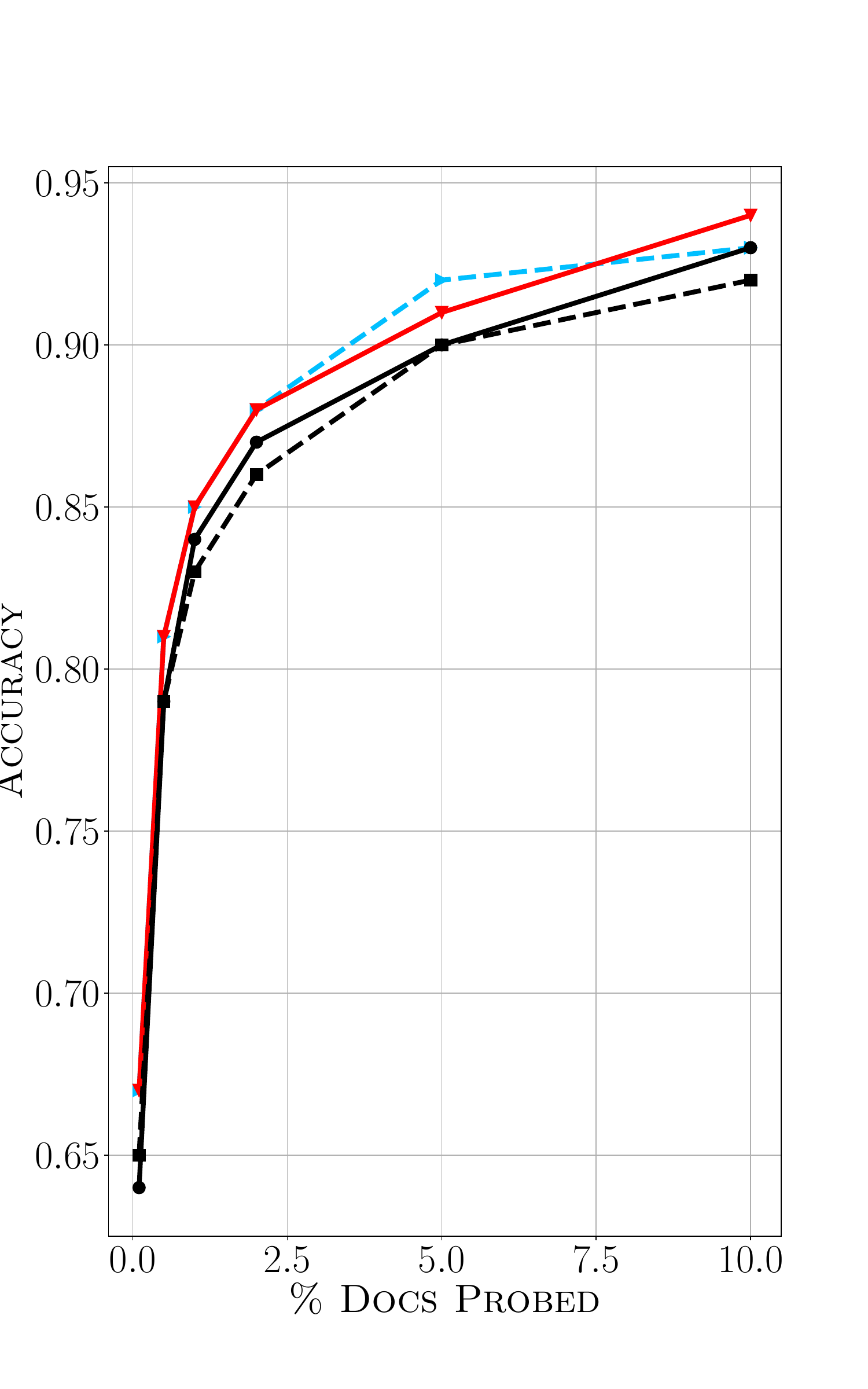}
}}
\centerline{
\subfloat[\textsc{HotpotQA}]{
\includegraphics[width=0.32\linewidth]{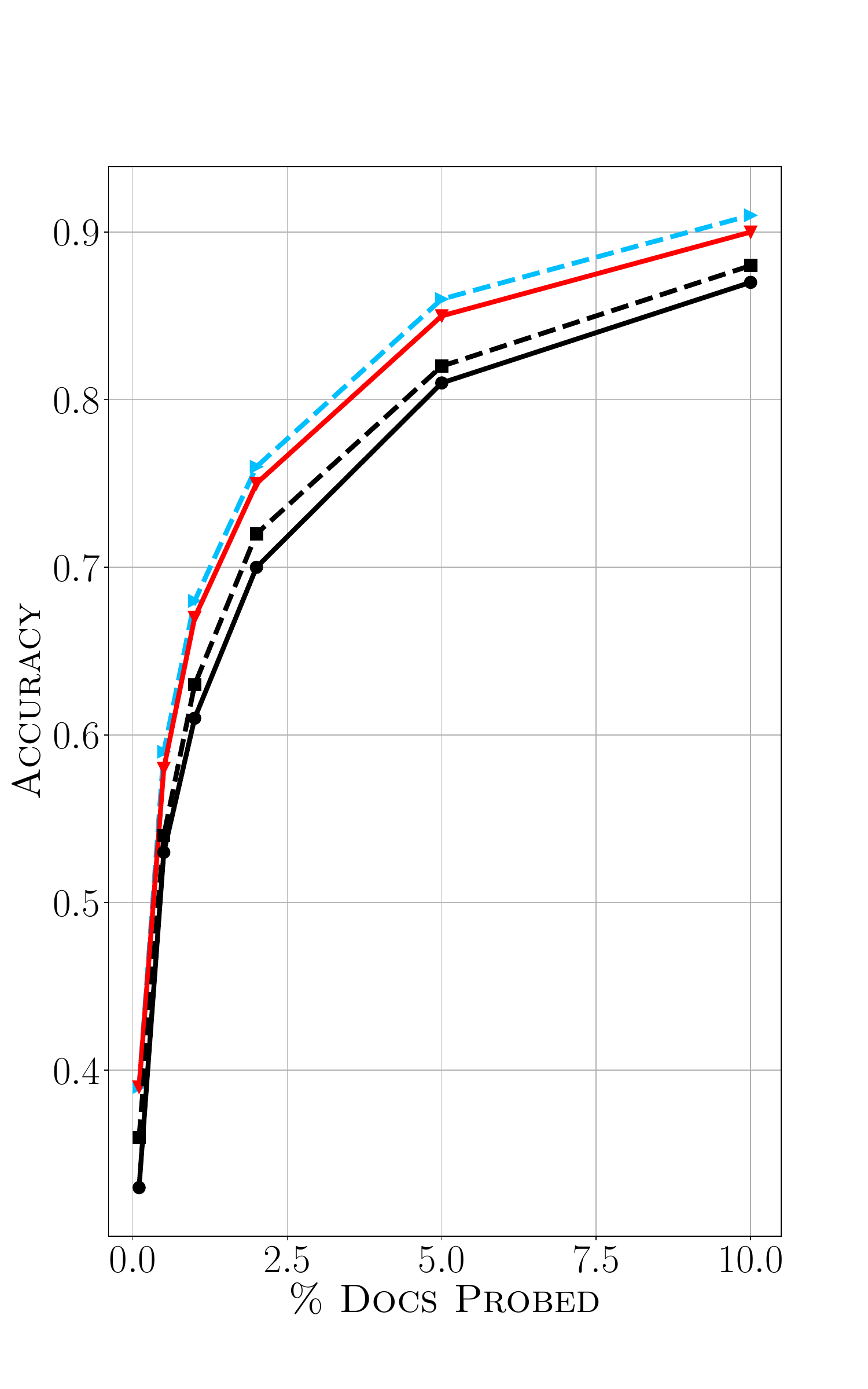}
}
\subfloat[\textsc{Fever}]{
\includegraphics[width=0.32\linewidth]{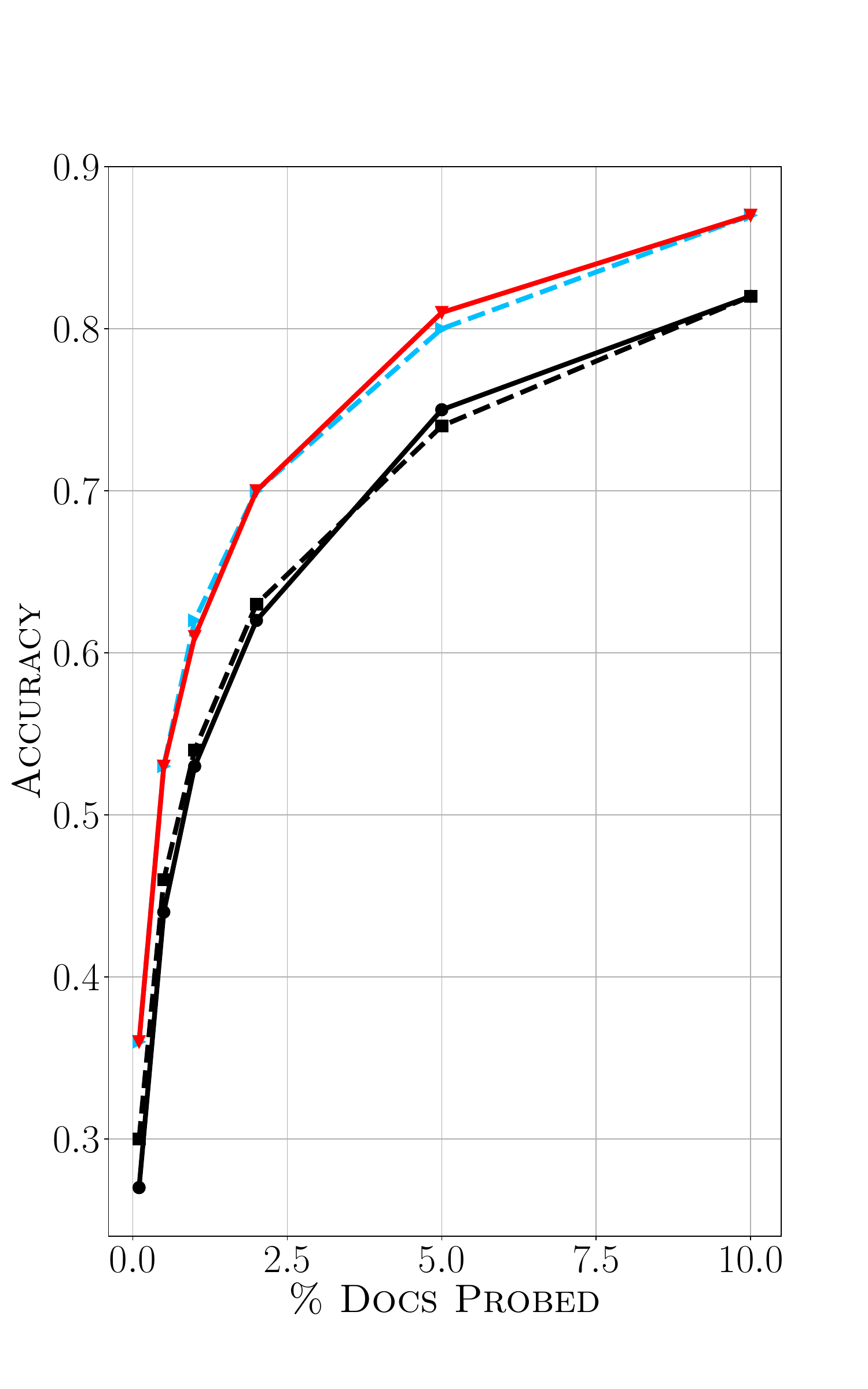}
}
\subfloat[\textsc{DBPedia}]{
\includegraphics[width=0.32\linewidth]{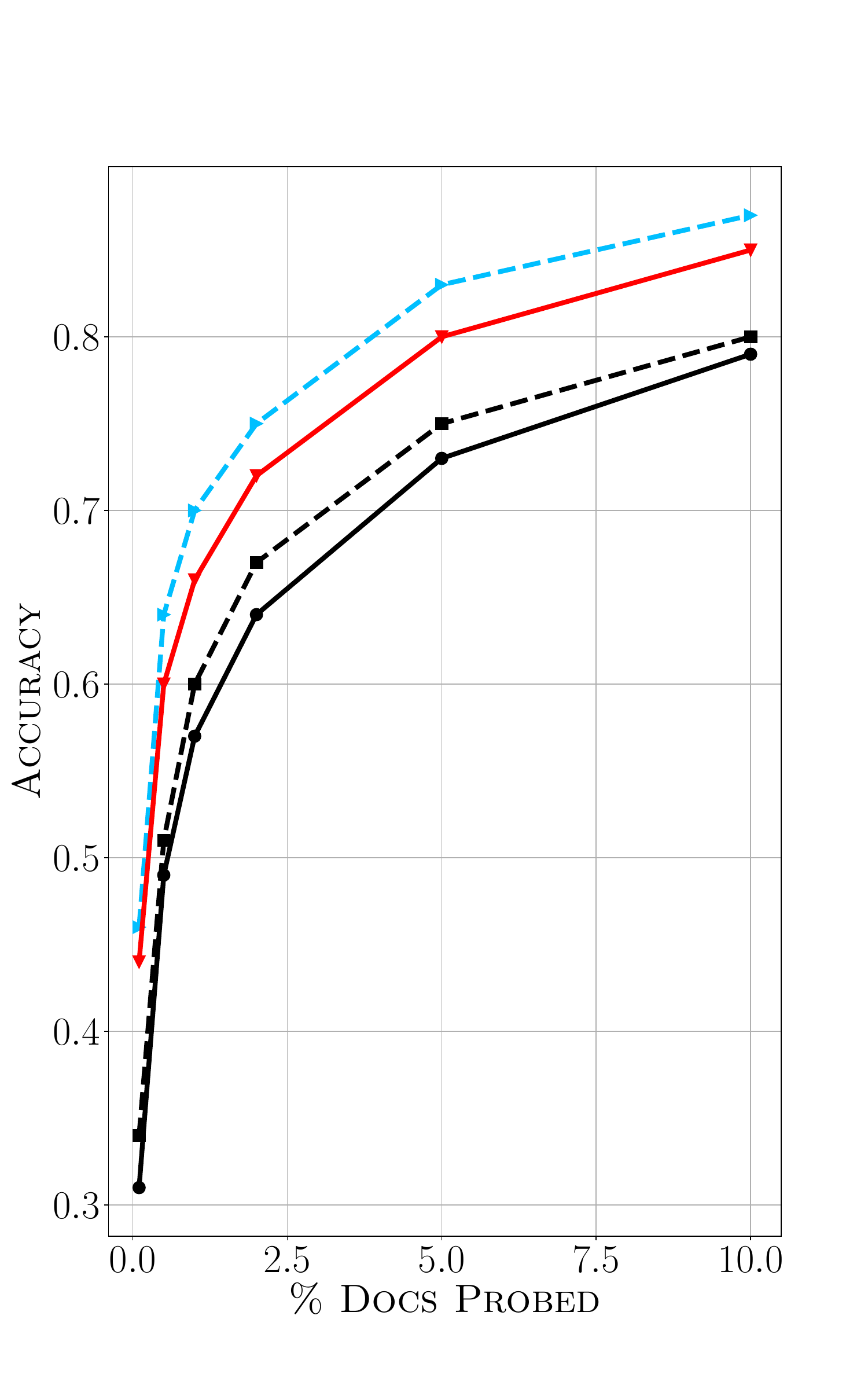}
}}
\vspace{0.2cm}
\caption{Top-$10$ accuracy of Algorithm~\ref{algorithm:retrieval} for \splade{}
vectors versus the number of documents examined ($\ell$)---
expressed as percentage of the size of the collection---for different clustering algorithms
(standard and spherical KMeans)
and different sketching mechanisms (JL transform and \weaksinnamon{}, with sketching size of $1024$).
Note that the vertical axis is not consistent across figures.}
\label{figure:clustering-quality:splade}
\end{center}
\end{figure}

\begin{figure}[t]
\begin{center}
\centerline{
\subfloat[\textsc{MS Marco}]{
\includegraphics[width=0.32\linewidth]{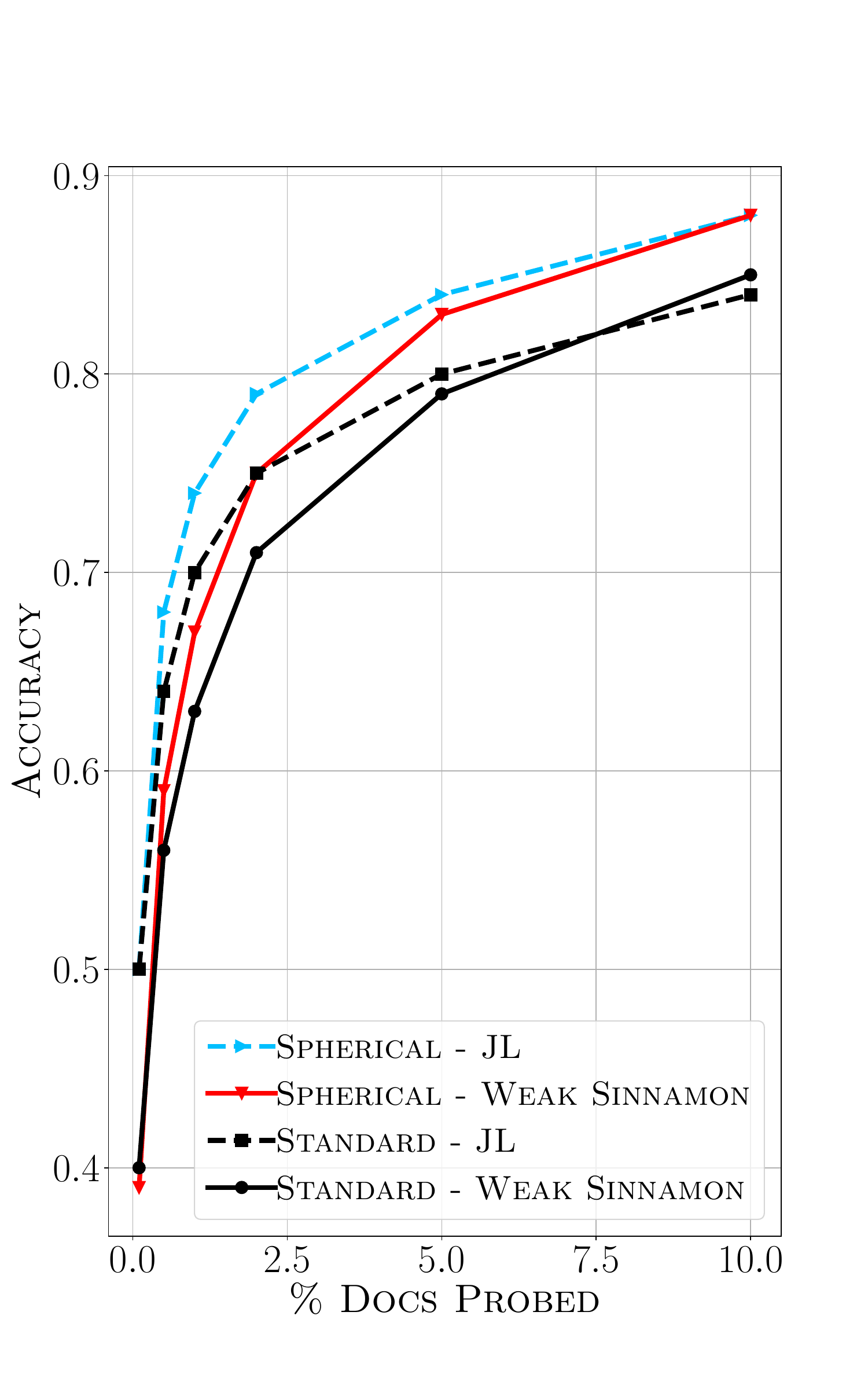}
}
\subfloat[\textsc{NQ}]{
\includegraphics[width=0.32\linewidth]{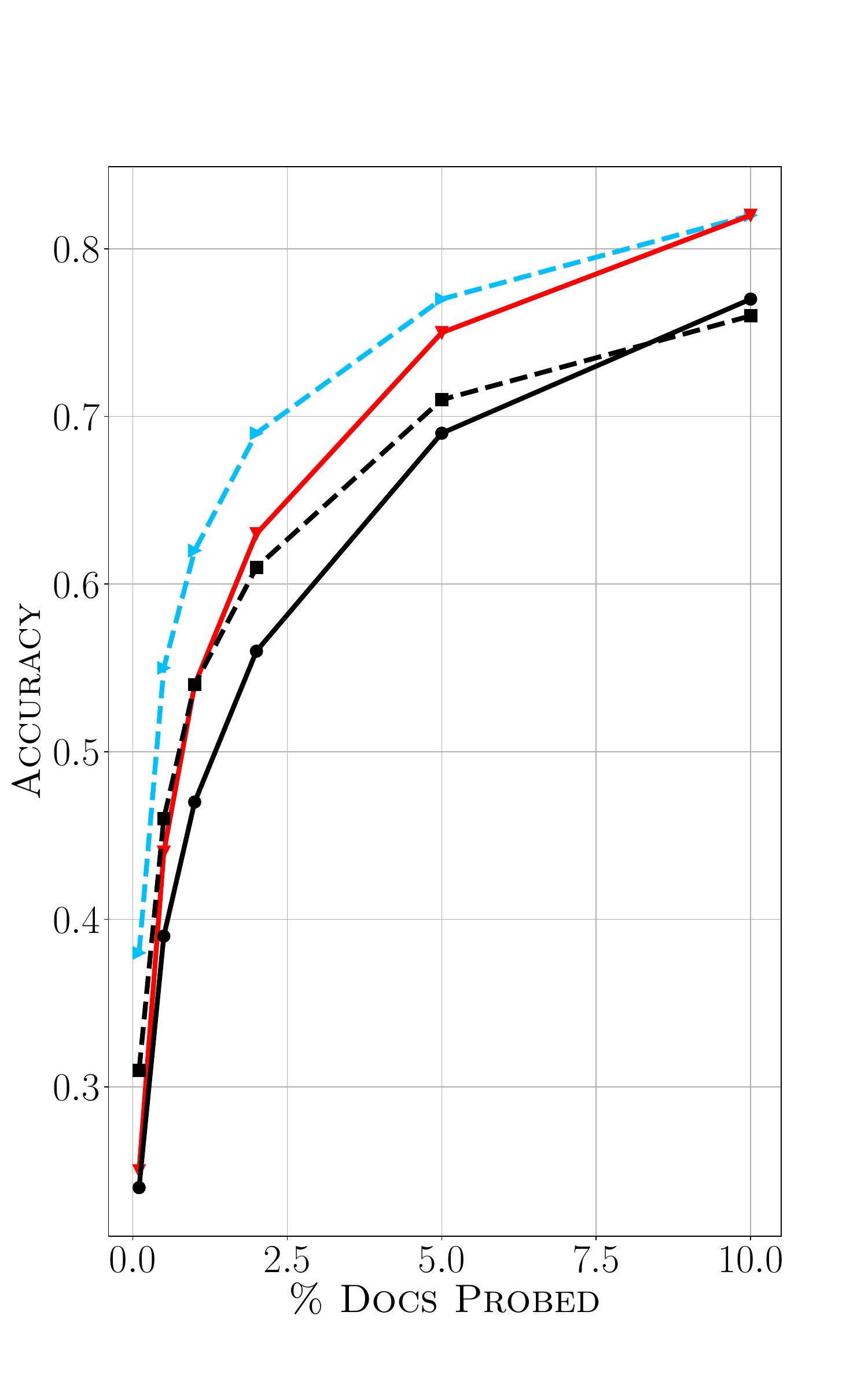}
}
\subfloat[\textsc{Quora}]{
\includegraphics[width=0.32\linewidth]{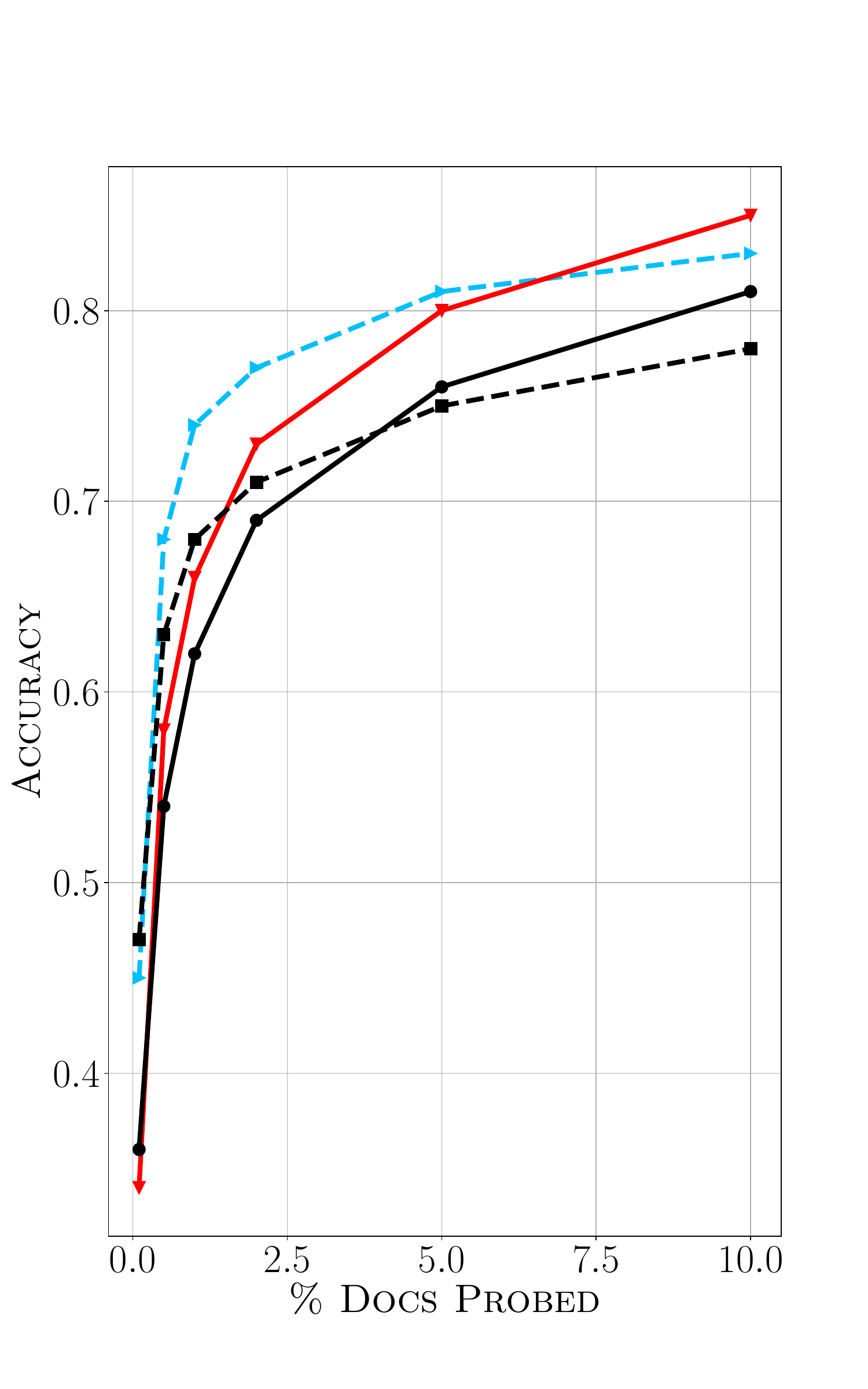}
}}
\centerline{
\subfloat[\textsc{HotpotQA}]{
\includegraphics[width=0.32\linewidth]{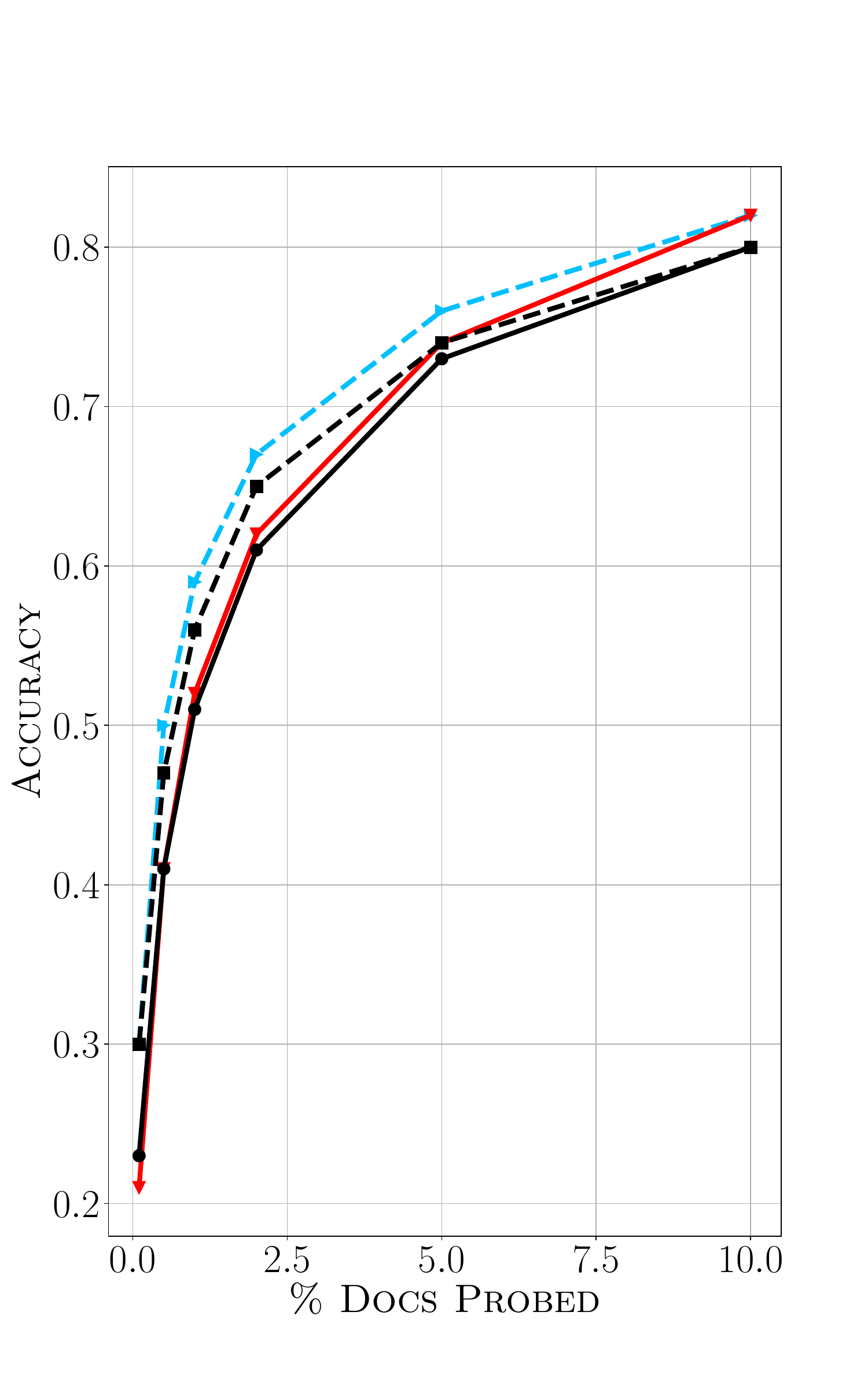}
}
\subfloat[\textsc{Fever}]{
\includegraphics[width=0.32\linewidth]{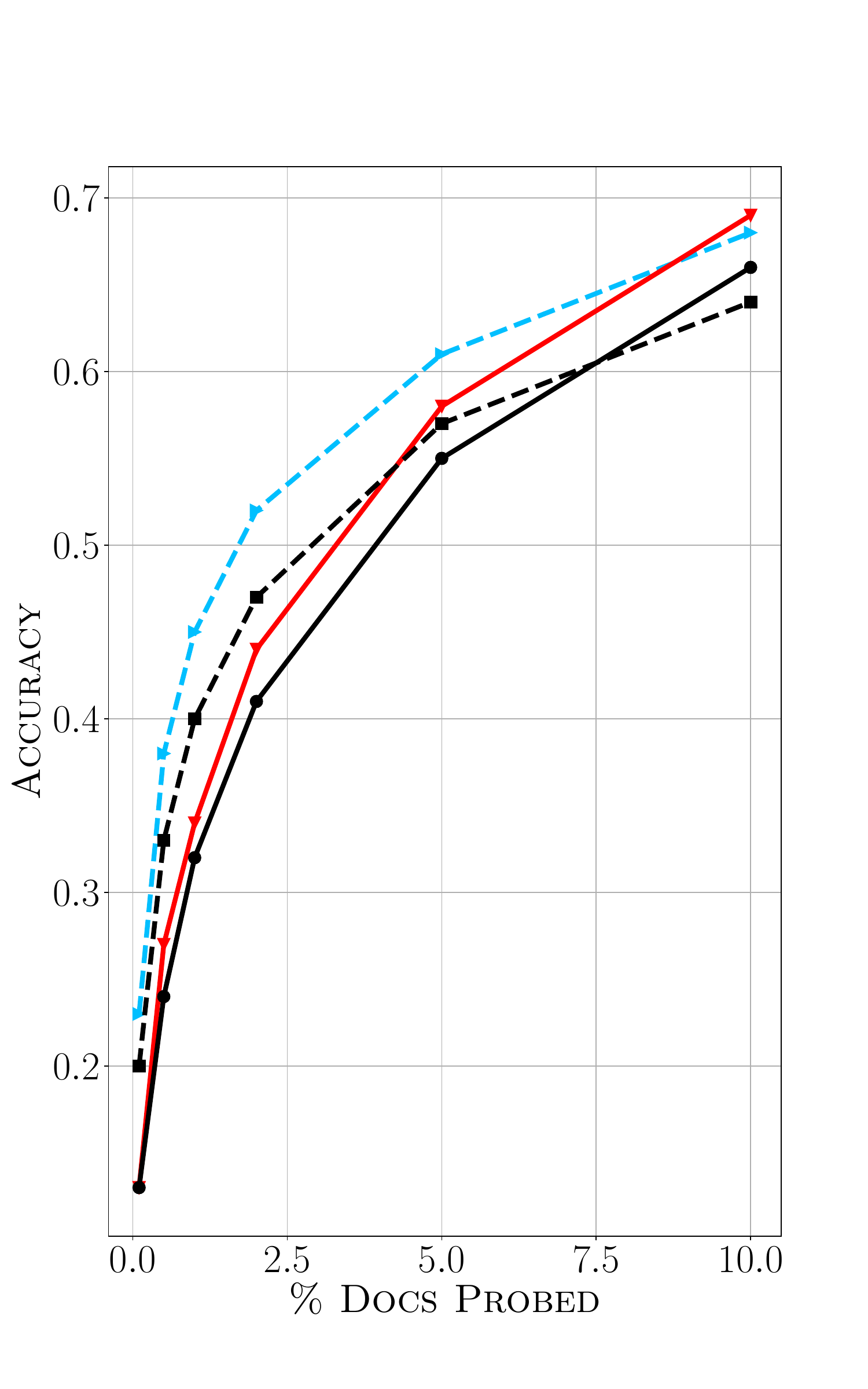}
}
\subfloat[\textsc{DBPedia}]{
\includegraphics[width=0.32\linewidth]{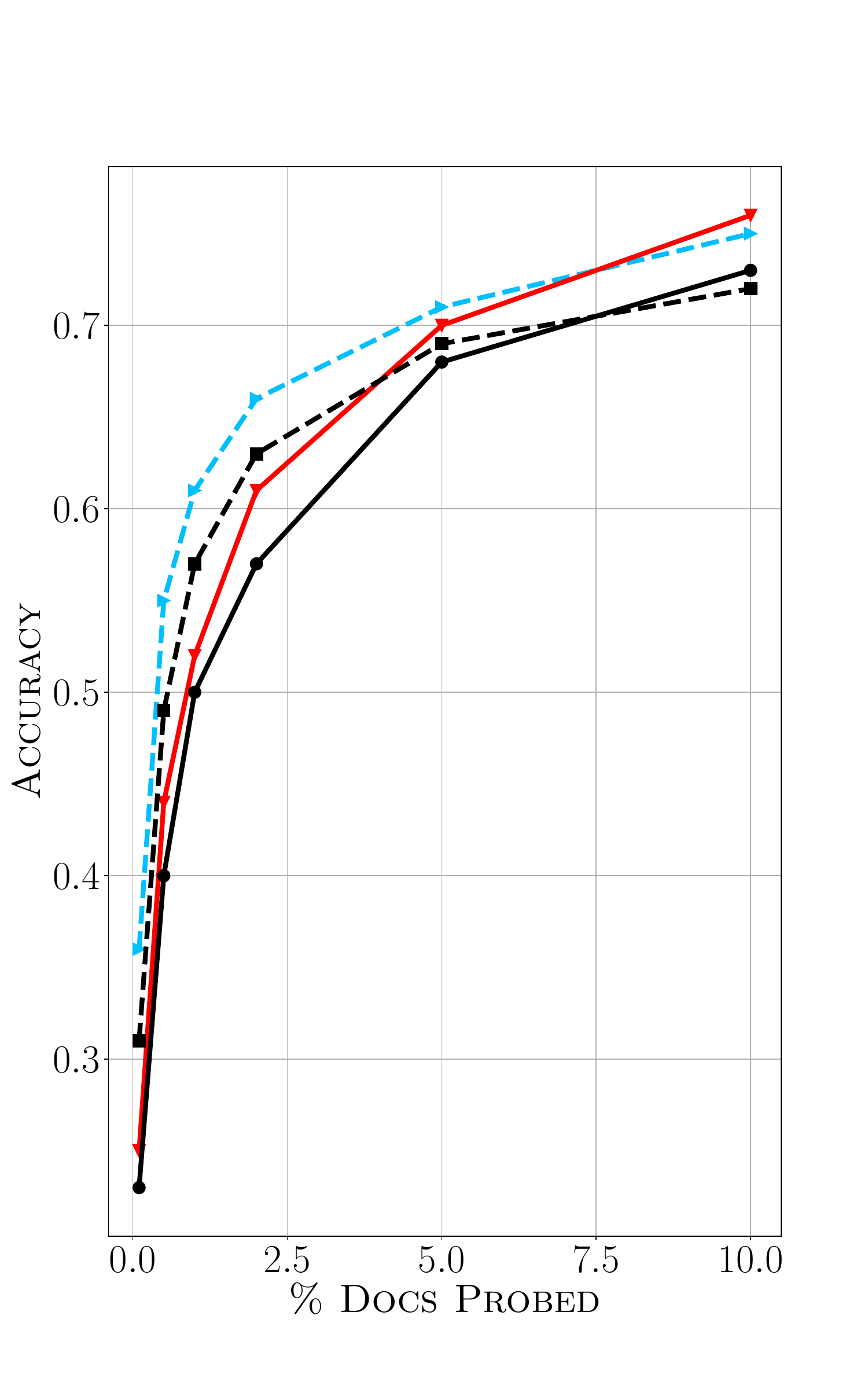}
}}
\vspace{0.2cm}
\caption{Top-$10$ accuracy of Algorithm~\ref{algorithm:retrieval} for \esplade{}
vs. the number of documents examined ($\ell$).}
\label{figure:clustering-quality:esplade}
\end{center}
\end{figure}

We begin by emphasizing that, in this particular section,
we do not pay attention to speed and only report accuracy as a function of
the total number of documents examined, $\ell$, in Algorithm~\ref{algorithm:retrieval}.
Additionally, we use an \emph{exact}, exhaustive search algorithm as $\mathcal{R}$ over
the original vectors to find the final top-$k$ candidates once the
$\ell$-subset of a dataset has been identified.

Before we state our findings, a note on our choice of
``the number of documents examined'' ($\ell$) versus the more familiar
notion of ``the number of clusters searched'' (known commonly as \textsc{nProbe}): The standard
KMeans algorithm is highly sensitive to vector norms. That is natural as the algorithm cares solely
about the Euclidean distance between points within a partition. When it operates on a collection
of vectors with varying norms, then, it is intuitive that it tends to isolate high-normed points
in their own, small partitions, while lumping together the low-normed vectors into massive
clusters. As a result of this phenomenon, partitions produced by standard KMeans
are often imbalanced. Probing a fixed number of partitions at search time
puts KMeans at an unfair disadvantage compared to its spherical variant.
By choosing to work with $\ell$ rather than fixating on the number of top clusters
we remove that variable from the equation.

Figure~\ref{figure:clustering-quality:splade} summarizes our results for the \splade{}-generated
vectors. We plot one figure per dataset, where each figure depicts the relationship between
top-$10$ accuracy and $\ell$ (expressed as percentage of the total number of documents).
When applying Algorithm~\ref{algorithm:indexing} to the datasets, we set the sketch size to $1024$
as per findings of Section~\ref{section:sketching}. Additionally, we fix the number of partitions $P$
to $4\sqrt{\lvert \mathcal{X} \rvert}$ where $\lvert \mathcal{X} \rvert$ is the number of documents
in a dataset $\mathcal{X}$. Plots for \esplade{} are shown separately in Figure~\ref{figure:clustering-quality:esplade}.

One of the most striking observations is that spherical KMeans appears to
be a better choice universally on the vector datasets we examine in this work.
By partitioning the data with spherical KMeans in Algorithm~\ref{algorithm:indexing}
and examining at most $10\%$ of the collection, we often reach a top-$10$
accuracy well above $0.8$ and often $0.9$. This is in contrast to the performance
of standard KMeans which often lags behind.

We are also surprised by how little the choice of JL transform versus \weaksinnamon{}
appears to matter, in the high-accuracy regime, for the purposes of partitioning with
spherical KMeans and retrieval over the resulting partitions. When the clustering
method is the standard KMeans, on the other hand, the difference between the two
sketching algorithms is sometimes more noticeable. Additionally, and perhaps unsurprisingly,
the difference between the two sketching methods is more pronounced
in experiments on the \esplade{} vector datasets.

\section{Clustering as Dynamic Pruning for the Inverted Index}
\label{section:dynamic-pruning}
Throughout the previous sections, we simply assumed that once
Algorithm~\ref{algorithm:retrieval} has identified the top partitions
and accumulated the $\ell$-subset of documents to examine,
the task of actually finding the top-$k$ vectors from that
restricted subset would be delegated to a secondary MIPS algorithm, $\mathcal{R}$,
which we have thus far ignored. We now wish to revisit $\mathcal{R}$.

There are many ways one could design and implement $\mathcal{R}$
and apply it to the set of partitions $\mathcal{P}_\mathcal{I}$
on Line~\ref{algorithm:retrieval:restricted-mips} of Algorithm~\ref{algorithm:retrieval}.
For example, $\mathcal{R}$ may be an exhaustive search---an option
we used previously because we argued we were assessing retrieval quality alone
and did not concern ourselves with efficiency.
As another example, if partitions are stored on separate physical (or logical) retrieval nodes in
a distributed system, each node could use
an inverted index-based algorithm to find the top-$k$ candidates from
their partition of the index. This section proposes a novel alternative for $\mathcal{R}$
that is based on the insight that clustering documents for IVF-based search
and dynamic pruning algorithms in the inverted index-based top-$k$ retrieval literature
are intimately connected.

\subsection{Partitioning Inverted Lists}
Consider an \emph{optimal} partitioning $\mathcal{P}^\ast$ of a collection $\mathcal{X}$ of
sparse vectors into $P$ clusters with a set of representative points $\mathcal{C}^\ast$.
In the context of MIPS, optimality implies that for any given sparse query $q$,
we have that the solution to
$\mathcal{C}_i = \argmax_{c \in \mathcal{C}^\ast} \langle q, c_i\rangle$
represents the partition $\mathcal{P}_i$ in which we can find the maximizer of
$\argmax_{x \in \mathcal{X}} \langle q, x \rangle$. That implies that, when performing
MIPS for a given query, we \emph{dynamically prune} the set of documents in
$\mathcal{X} \setminus \mathcal{P}_i$; the procedure
is dynamic because $\mathcal{P}_i$ depends on the query vector.

Consider now an inverted index that represents $\mathcal{X}$.
Typically, its inverted lists are sorted either by document identifiers
or by the ``impact'' of each document on the final inner product
score~\cite{tonellotto2018survey}. The former is consequential for
compression~\cite{pibiri2020compression} and document-at-a-time
dynamic pruning algorithms~\cite{tonellotto2018survey}, while the latter
provides an opportunity for early-termination of score
computation---we reiterate that, all of these techniques work only on
non-negative vectors or that their extension to negative vectors in non-trivial.
But, as we explain, $\mathcal{P}^\ast$ induces another organization
of inverted lists that will enable fast, approximate retrieval in
the context of Algorithm~\ref{algorithm:retrieval} for general sparse vectors.

\begin{algorithm}[!t]
\SetAlgoLined
{\bf Input: }{Collection of sparse vectors, $\mathcal{X} \subset \mathbb{R}^{N}$;
Clusters $\mathcal{P}$ obtained from Algorithm~\ref{algorithm:indexing}.}\\
\KwResult{Inverted index, $\mathcal{I}$; Skip list, $\mathcal{S}$.}

\begin{algorithmic}[1]

\STATE $\mathcal{I} \leftarrow \emptyset$ \Comment*[r]{Initialize the inverted index}
\STATE $\mathcal{S} \leftarrow \emptyset$ \Comment*[r]{Initialize the skip list}

\FOR{$\mathcal{P}_i \in \mathcal{P}$}
    \STATE $\mathbf{SortAscending}(\mathcal{P}_i)$ \Comment*[r]{Sort partition by document identifier}
    \FOR{$j \in \mathcal{P}_i$}
        \FOR{$t \in \mathit{nz}(x^{(j)})$}
            \STATE $\mathcal{S}[t].\textsc{Append}(i, \lvert \mathcal{I}[t] \rvert)$ \textbf{if}
            it is the first time a document from $\mathcal{P}_i$ is recorded in $\mathcal{I}[t]$
            \STATE $\mathcal{I}[t].\textsc{Append}(j, x^{(j)}_t)$ \Comment*[r]{Append document identifier and value to list}
        \ENDFOR
    \ENDFOR
\ENDFOR

\RETURN $\mathcal{I}$, $\mathcal{S}$
\end{algorithmic}
\caption{Constructing a partitioned inverted index}
\label{algorithm:inverted-index}
\end{algorithm}

\begin{algorithm}[!t]
\SetAlgoLined
{\bf Input: }{Inverted index, $\mathcal{I}$; Skip list, $\mathcal{S}$ obtained
from Algorithm~\ref{algorithm:inverted-index};
Sparse query vector, $q$; Set of partitions to probe, $\mathcal{P}_\mathcal{I}$
from Algorithm~\ref{algorithm:retrieval}}.\\
\KwResult{Top $k$ vectors.}

\begin{algorithmic}[1]

\STATE $\mathit{scores} \leftarrow \emptyset$ \Comment*[r]{A mapping from documents to scores}

\FOR{$t \in \mathit{nz}(q)$}
    \STATE $\mathit{SLPosition} \leftarrow 0$ \Comment*[r]{Pointer into the skip list $\mathcal{S}[t]$}
    \FOR{$\mathcal{P}_i \in \mathcal{P}_\mathcal{I}$}
        \STATE Advance $\mathit{SLPosition}$ until partition of $\mathcal{S}[t][\mathit{SLPosition}]$ matches $\mathcal{P}_i$
        \STATE $\mathit{begin} \leftarrow \mathcal{S}[t][\mathit{SLPosition}].\textsc{Offset}$
        \STATE $\mathit{end} \leftarrow \mathcal{S}[t][\mathit{SLPosition} + 1].\textsc{Offset}$
        \FOR{$(\mathit{docid}, \mathit{value}) \in \mathcal{I}[t][\mathit{begin} \ldots \mathit{end} ]$}
            \STATE $\mathit{scores}[\mathit{docid}] \leftarrow \mathit{scores}[\mathit{docid}] + q_t \times \mathit{value}$
        \ENDFOR
    \ENDFOR
\ENDFOR

\RETURN Top $k$ documents given $\mathit{scores}$
\end{algorithmic}
\caption{Query processing over partitioned inverted lists}
\label{algorithm:query-processing}
\end{algorithm}

Our construction, detailed in Algorithm~\ref{algorithm:inverted-index},
is straightforward. At a high level, when forming an inverted list for a coordinate $t$,
we simply iterate through partitions and add vectors
from that partition whose coordinate $t$ is non-zero to the inverted list.
As we do so, for each inverted list, we record the \emph{offsets} within the list
of each partition in a separate \emph{skip} list. Together the two structures
enable us to traverse the inverted lists by only evaluating documents
in a given set of partitions.

An alternative way of viewing the joint inverted and skip lists is to
think of each inverted list as a set of variable-length segments or blocks,
where documents within each block are grouped according to a clustering
algorithm.

Before we demonstrate the retrieval logic, we must remark
on the space complexity of the resulting structure. There are two
factors to comment on. First, sorting the inverted lists by partition identifier
rather than document identifier may lead to suboptimality for compression
algorithms. That is because, the new arrangement of documents may distort
the $d$-gaps (i.e., the difference between two consecutive document identifiers
in an inverted list); compression algorithms perform better when $d$-gaps
are smaller and when there is a \emph{run} of the same $d$-gap in the list.
But we can address that concern trivially through document identifier
reassignment: After partitioning is done by Algorithm~\ref{algorithm:indexing},
we assign new identifiers to documents such that documents within a partition
have consecutive identifiers.

The second factor is the additional data stored in $\mathcal{S}$.
In the worst case, each inverted list will have documents from every partition.
That entails that each $\mathcal{S}[t]$ records $P$ additional pairs of integers
consisting of partition identifier and the offset within the inverted list where
that partition begins. As such, in the worst case, the inverted index is inflated
by the size of storing $2NP$ integers. However, given that $P$ is orders of magnitude
smaller than the total number of non-zero coordinates in the collection,
and as such $2NP \ll \psi \lvert \mathcal{X} \rvert$, the increase to the total size
of the inverted index is mild at worst.
Moreover, skip lists can be further compressed using an integer or integer-list codec.

\subsection{Query Processing over Partitioned Inverted Lists}
When Algorithm~\ref{algorithm:retrieval} gives us a set of partitions
$\mathcal{P}_\mathcal{I}$ to probe, we use a simple coordinate-at-a-time
scheme to compute the scores of documents in $\bigcup \mathcal{P}_\mathcal{I}$
and return the top-$k$ vectors.

When processing coordinate $t$ and accumulating partial inner product scores,
we have two operations to perform. First, we must take the intersection of the skip list
and the list of whitelisted partitions: $\mathcal{P}_\mathcal{I} \cap \mathcal{S}[t].\textsc{PartitionId}$ (where the operator $\textsc{PartitionId}$ returns the
partition identifier of every element in the skip list).
Only then do we traverse the inverted list $\mathcal{I}[t]$ by looking at the
offsets of partitions in the intersection set. One possible instance of
this procedure is described in Algorithm~\ref{algorithm:query-processing}.

\subsection{Empirical Evaluation}
There are four key properties that we wish to evaluate.
Naturally, we care about the efficiency of Algorithms~\ref{algorithm:inverted-index}
and~\ref{algorithm:query-processing} when we use them as $\mathcal{R}$ in Algorithm~\ref{algorithm:retrieval}.
But, seeing as the partitioning performed by Algorithm~\ref{algorithm:indexing}
is not guaranteed to be the optimal partitioning $\mathcal{P}^\ast$,
we understand there is a risk of losing retrieval accuracy by probing a fraction
of partitions, as demonstrated
in Section~\ref{section:clustering}. As such, the second important property
is the effectiveness of the methods presented here.
We thus report throughput versus accuracy as one trade-off space of interest.

We also presented Algorithms~\ref{algorithm:inverted-index}
and~\ref{algorithm:query-processing} as a new dynamic pruning method
for the inverted index. To show that for different levels of accuracy,
we indeed prune the inverted lists, we additionally report the size of
the pruned space as we process queries.

A third factor is the size of the inverted index and the inflation
due to (a) the additional data structure that holds skip pointers
and (b) the partition centroids produced by Algorithm~\ref{algorithm:indexing}.
We also evaluate this aspect, but we do not apply compression anywhere
in our evaluation: We consider compression to be orthogonal to this work and only
report the overhead.

Finally, we implemented Algorithms~\ref{algorithm:indexing} through~\ref{algorithm:query-processing} by enabling parallelism within and across queries.
We believe, therefore, it is important to measure the effect of
the number of CPU cores on throughput. As such, we present throughput
measurements by changing the number of cores we make available to the algorithms.

\subsubsection{Baseline Retrieval Algorithm}
As argued earlier, we are interested in general sparse vectors,
such as those produced by \splade{}, which exhibit distributional
properties that differ from traditional sparse vectors based on lexical models
of relevance. It has been noted by others~\cite{bruch2023sinnamon,mackenzie2021wacky}
that an exhaustive disjunctive query processing over the inverted
index---a method Bruch et al. referred to as \linscan{}---outpeforms
all dynamic pruning-based optimization methods and represents a strong baseline.
We therefore use \linscan{} as our baseline system.

\linscan{} is a safe algorithm as it evaluates every qualified document
(i.e., documents that contain at least one non-zero coordinate of the query vector).
But as Bruch et al. show in~\cite{bruch2023sinnamon}, there is a simple
strategy to turn \linscan{} into an approximate algorithm: By giving the
algorithm a time budget, we can ask it to process as many
coordinates as possible until the budget has been exhausted. At that point,
\linscan{} returns the approximate top-$k$ set according to the accumulated
partial inner product scores. We use this variant to obtain approximate
top-$k$ sets for comparison with our own approximate algorithms.

\begin{figure}[t]
\begin{center}
\centerline{
\subfloat[\textsc{MS Marco}]{
\includegraphics[width=0.32\linewidth]{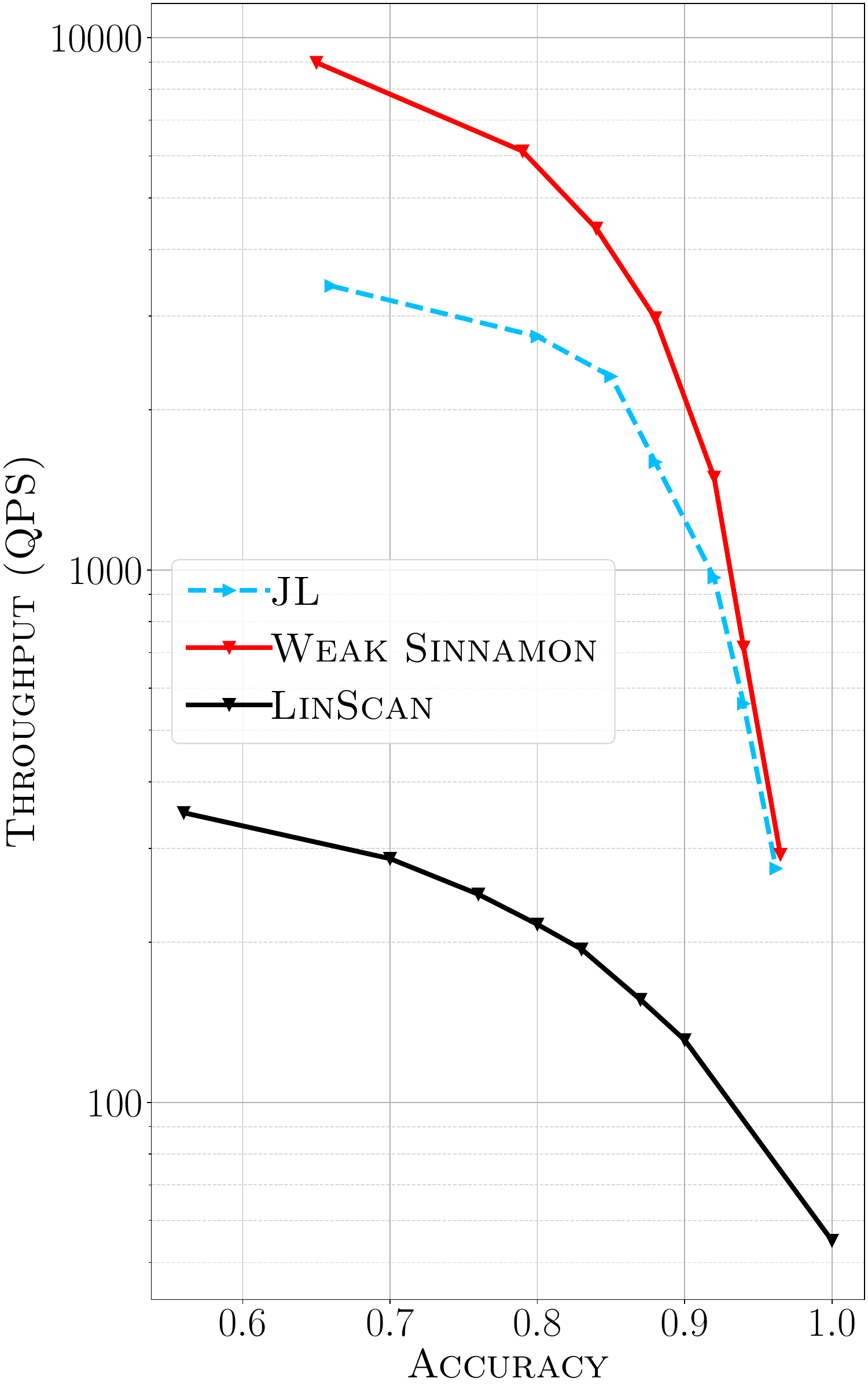}
}
\subfloat[\textsc{NQ}]{
\includegraphics[width=0.32\linewidth]{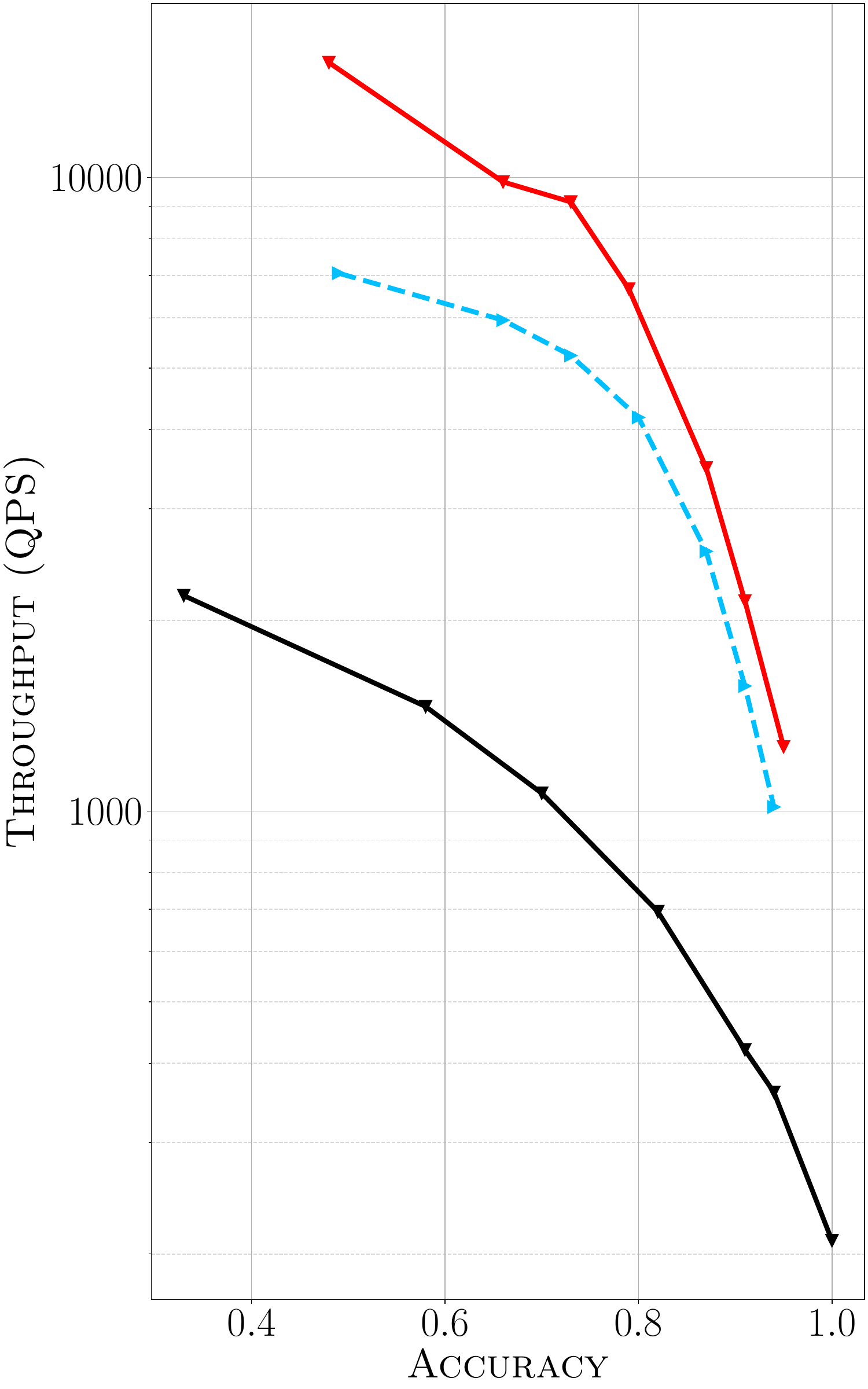}
}
\subfloat[\textsc{Quora}]{
\includegraphics[width=0.32\linewidth]{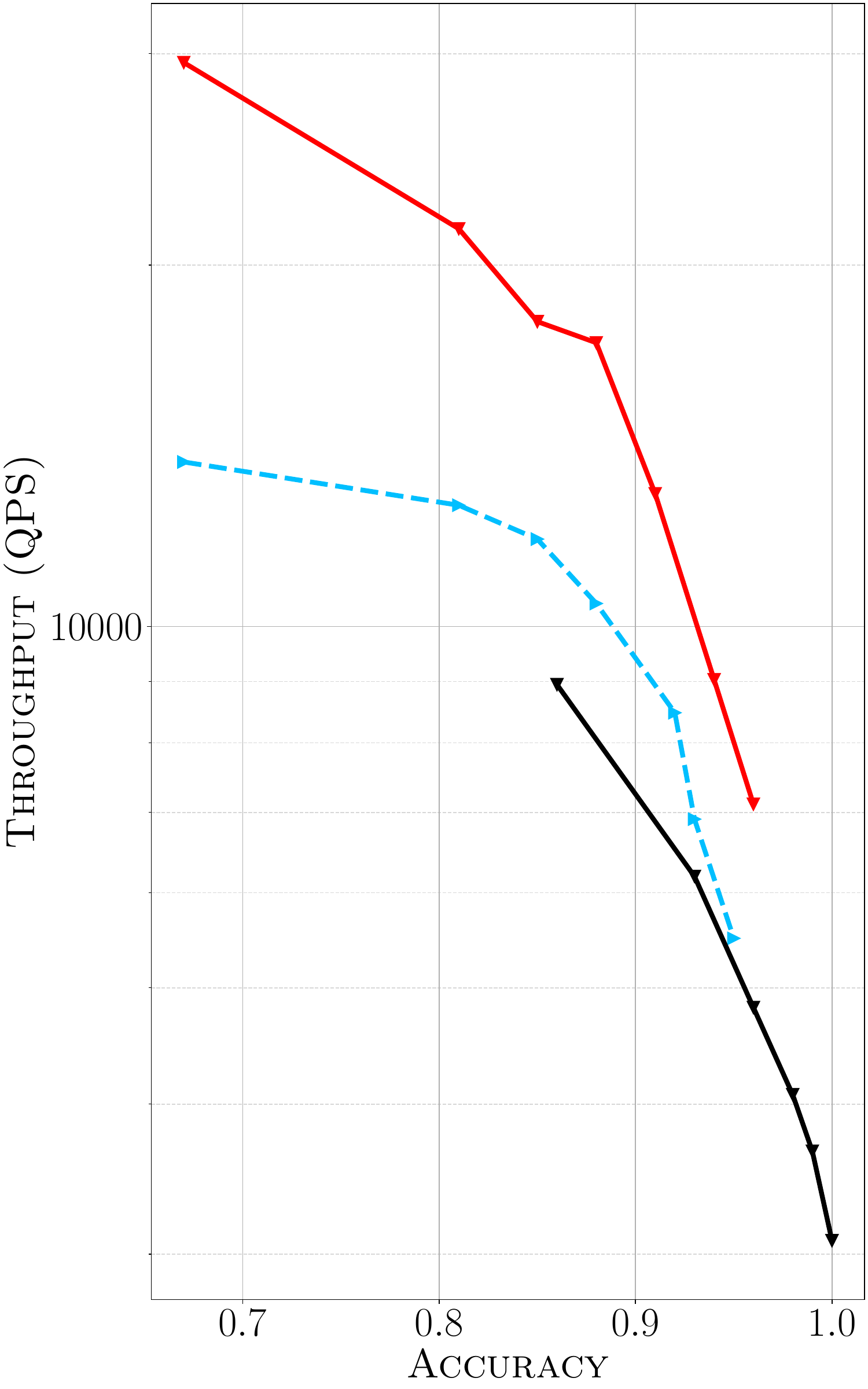}
}}
\centerline{
\subfloat[\textsc{HotpotQA}]{
\includegraphics[width=0.32\linewidth]{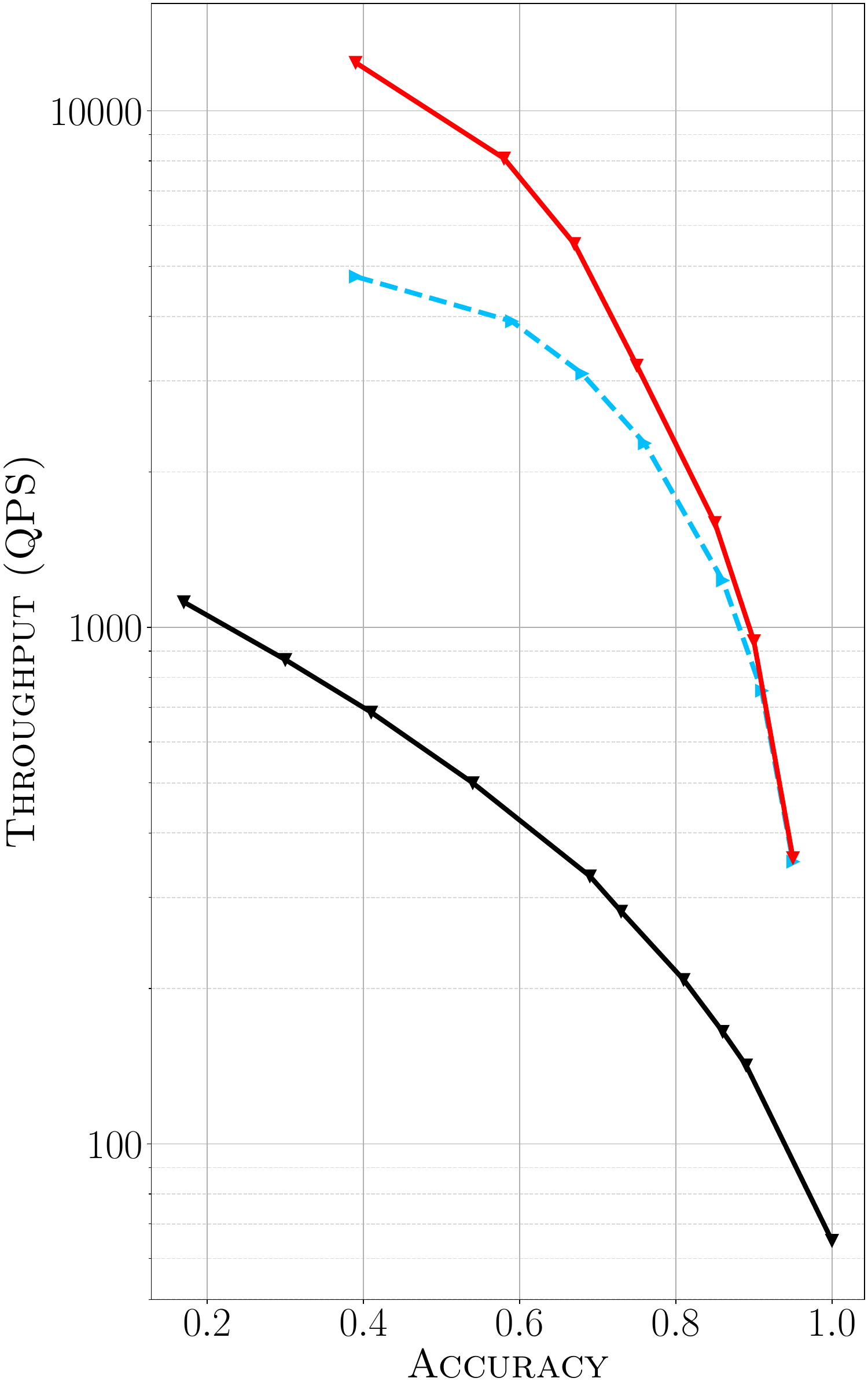}
}
\subfloat[\textsc{Fever}]{
\includegraphics[width=0.32\linewidth]{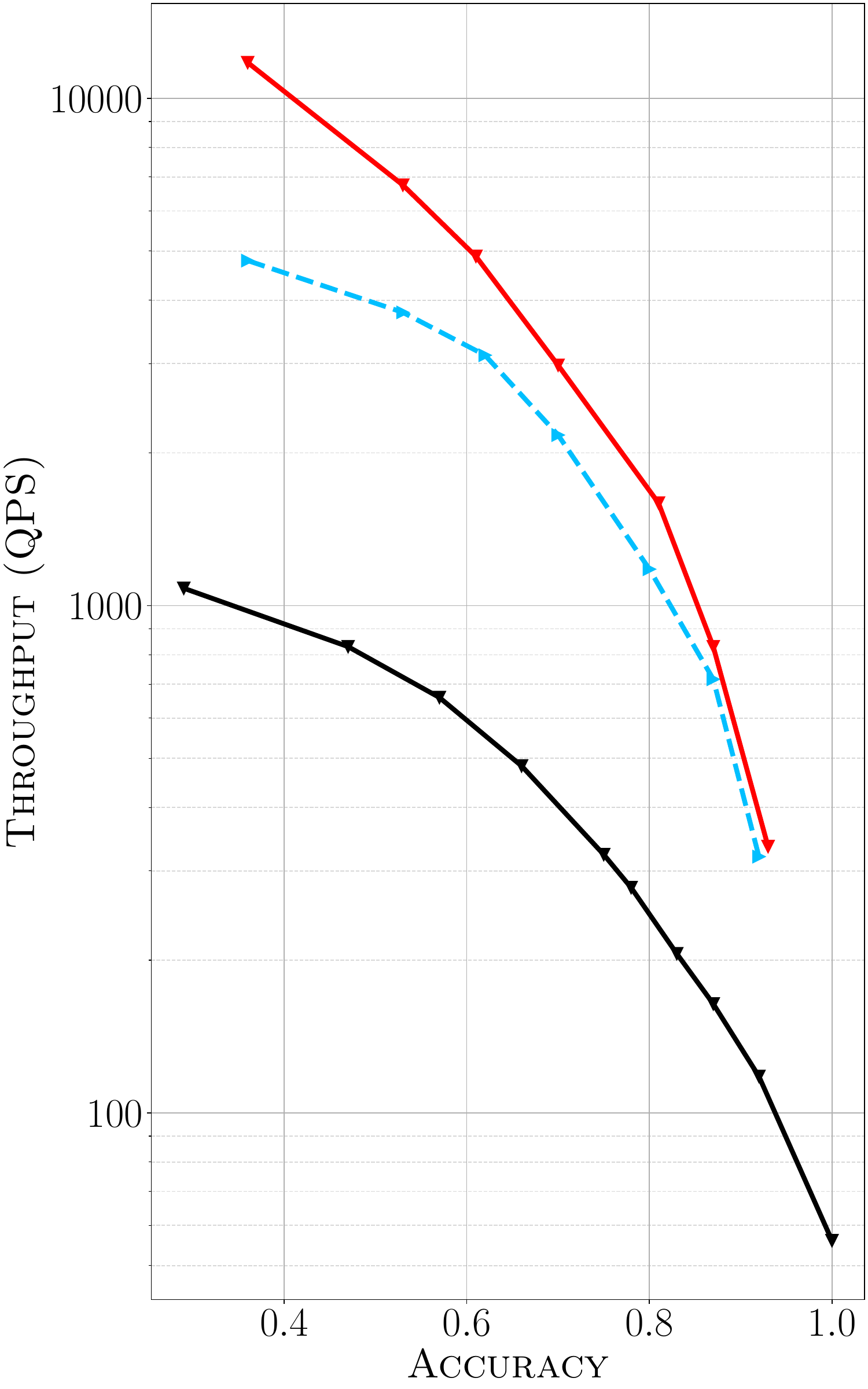}
}
\subfloat[\textsc{DBPedia}]{
\includegraphics[width=0.32\linewidth]{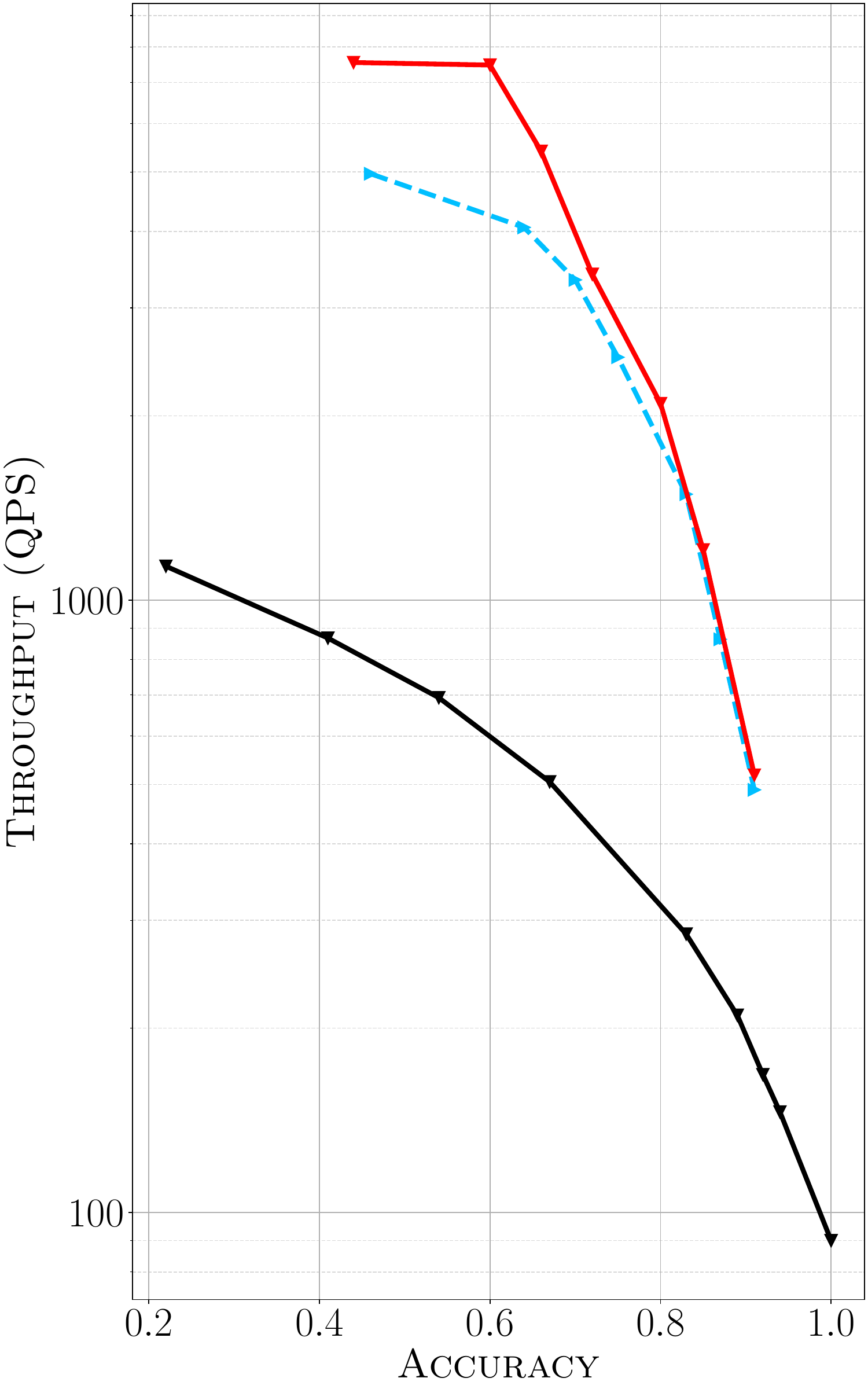}
}}
\vspace{0.2cm}
\caption{Throughput (as queries per second) versus top-$10$ retrieval accuracy
on \splade{}-encoded datasets.
We limit the experiments to an instance of Algorithm~\ref{algorithm:indexing} that
uses spherical KMeans. Included here is an approximate
variant of an exhaustive disjunctive query processor (\linscan{}).
We use $20$ CPU cores and repeat each experiment $10$
times for a more reliable throughput measurement.
Axes are not consistent across figures.}
\label{figure:qps-accuracy:splade}
\end{center}
\end{figure}

\begin{figure}[t]
\begin{center}
\centerline{
\subfloat[\textsc{MS Marco}]{
\includegraphics[width=0.32\linewidth]{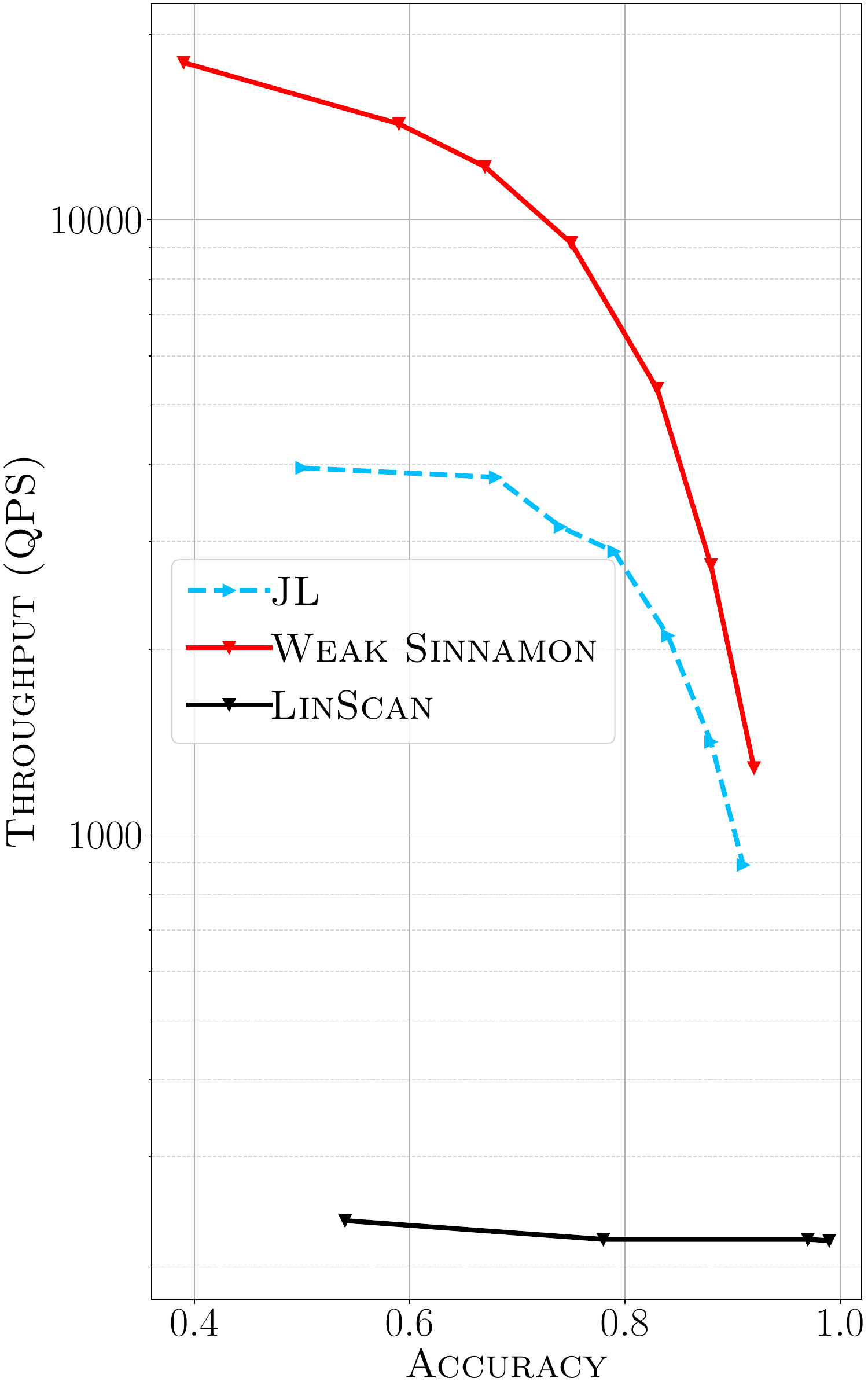}
}
\subfloat[\textsc{NQ}]{
\includegraphics[width=0.32\linewidth]{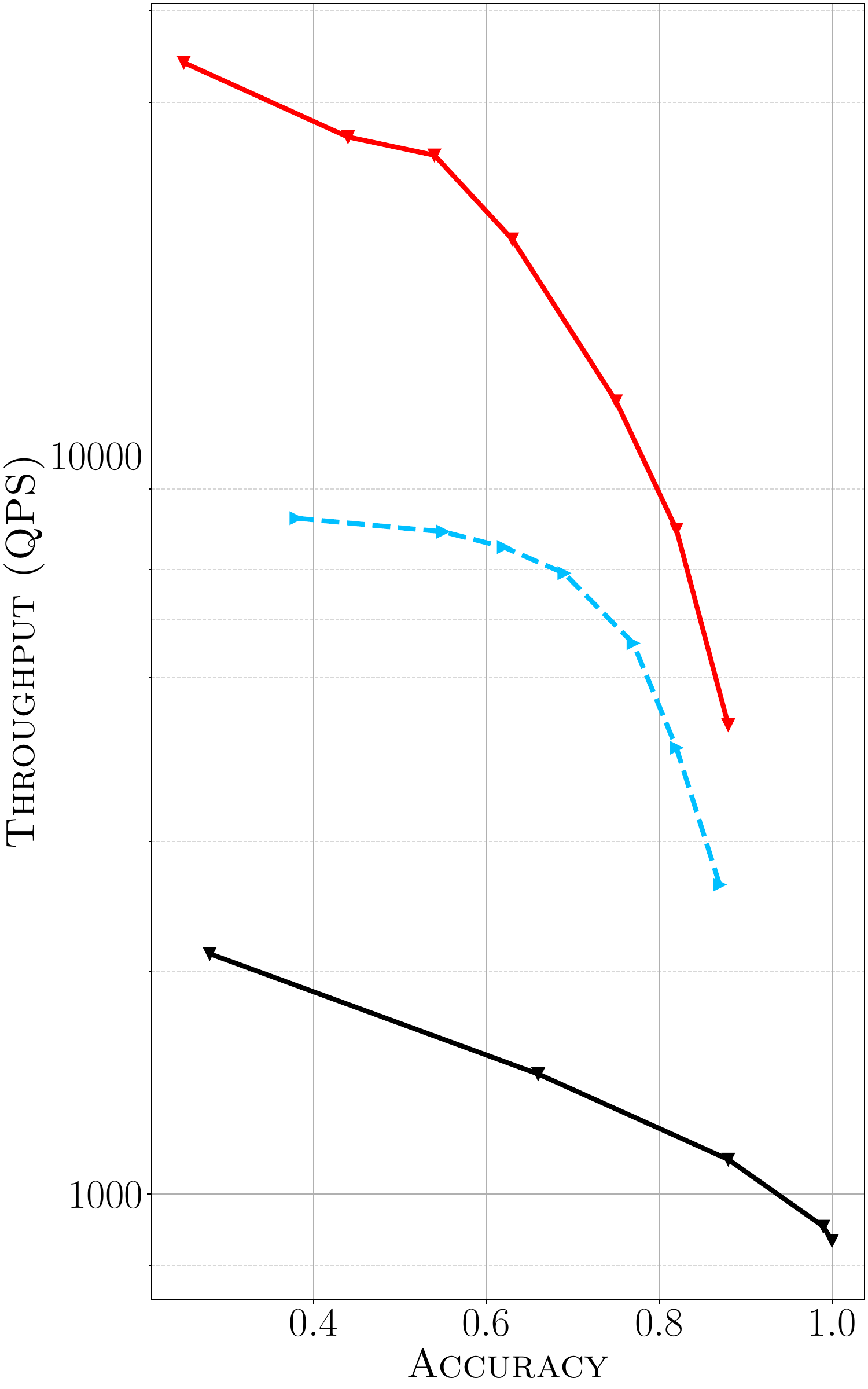}
}
\subfloat[\textsc{Quora}]{
\includegraphics[width=0.32\linewidth]{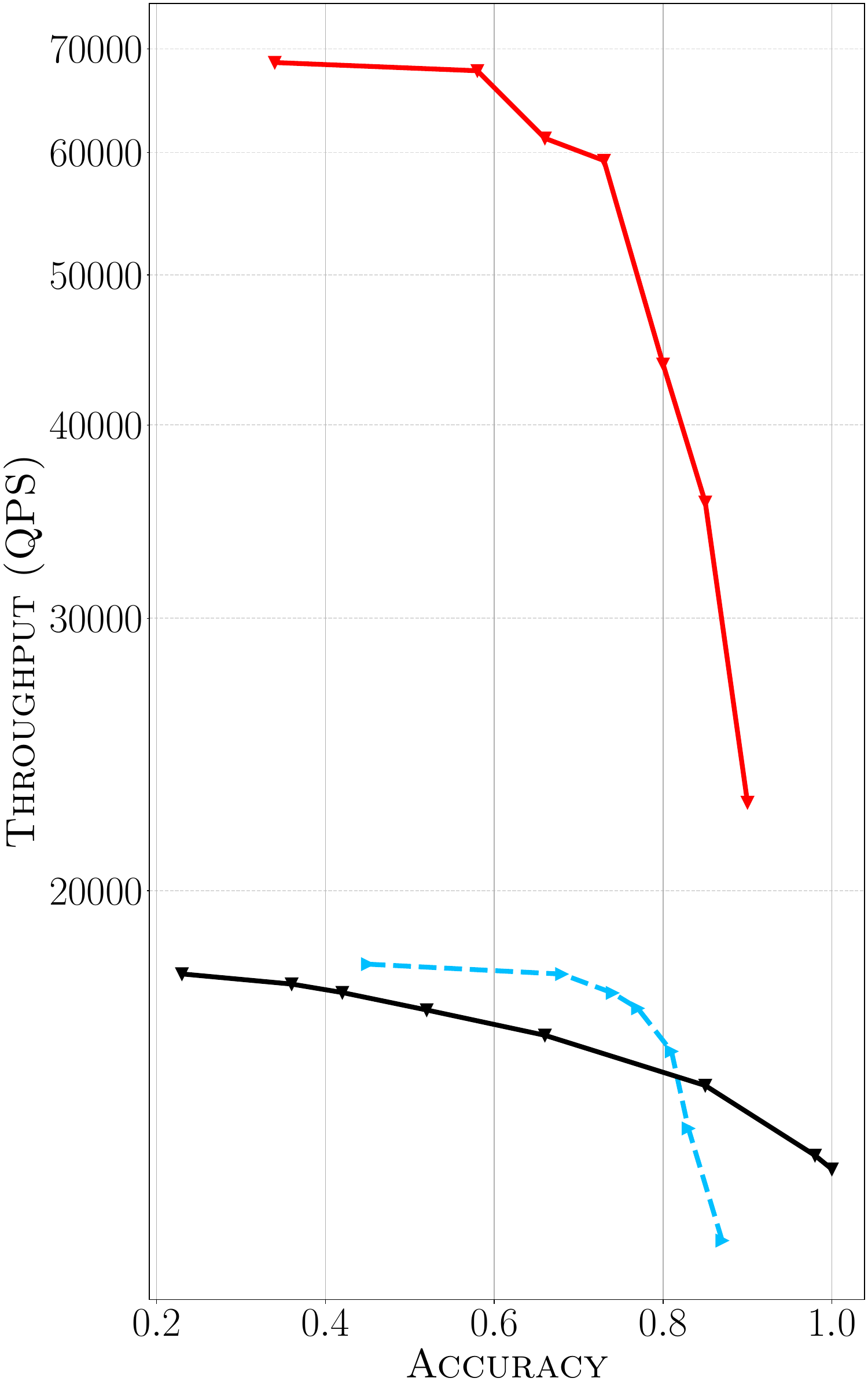}
}}
\centerline{
\subfloat[\textsc{HotpotQA}]{
\includegraphics[width=0.32\linewidth]{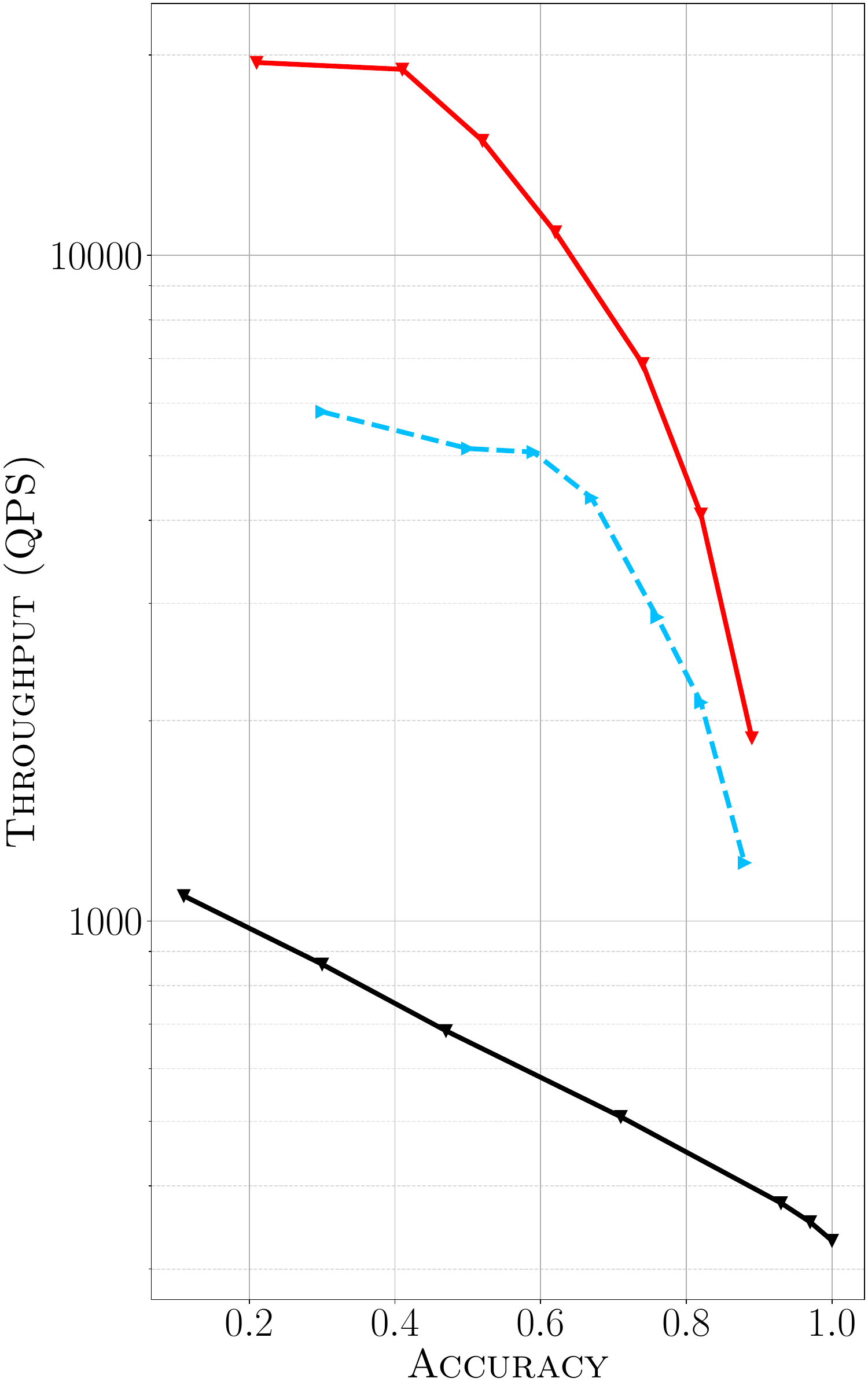}
}
\subfloat[\textsc{Fever}]{
\includegraphics[width=0.32\linewidth]{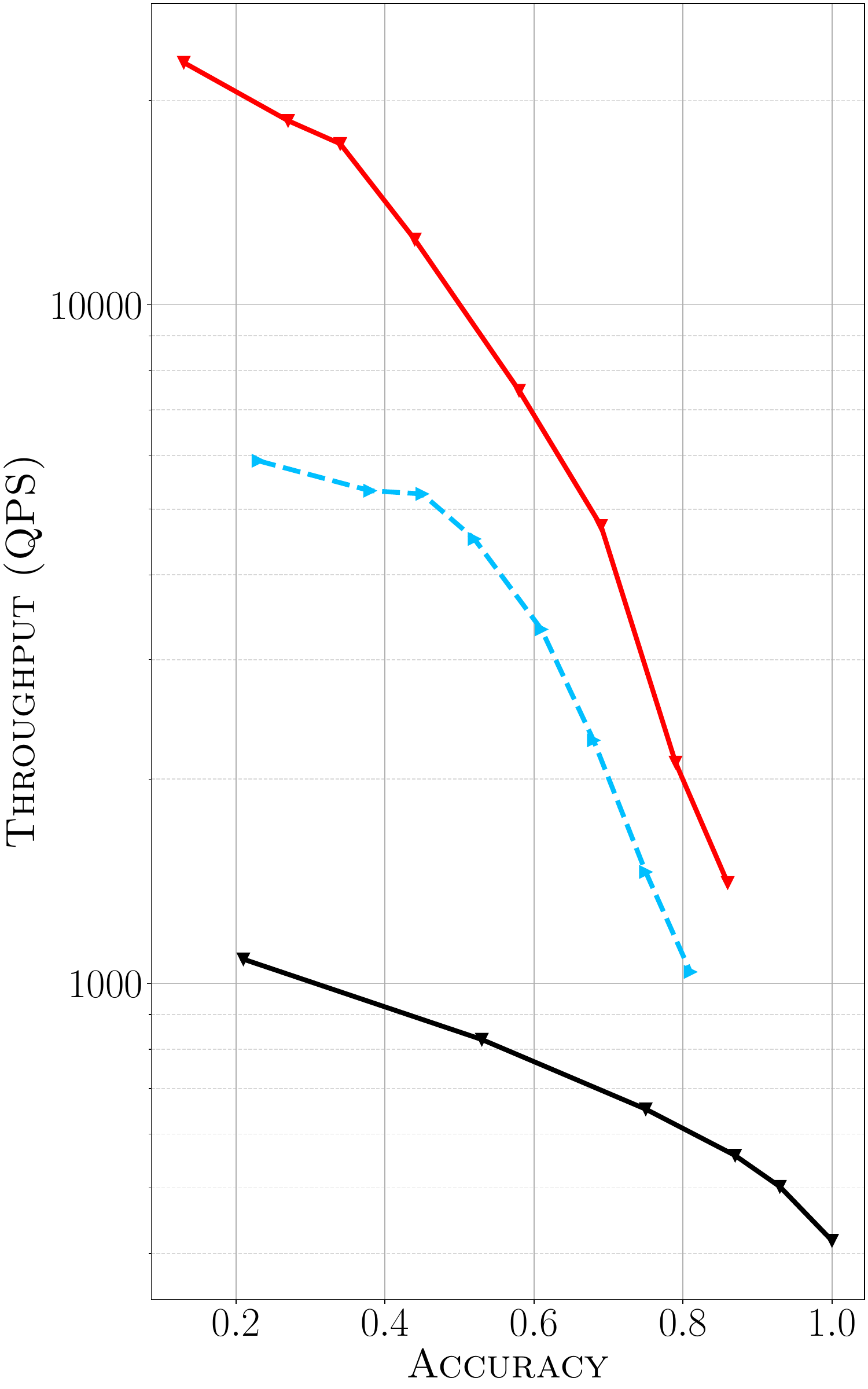}
}
\subfloat[\textsc{DBPedia}]{
\includegraphics[width=0.32\linewidth]{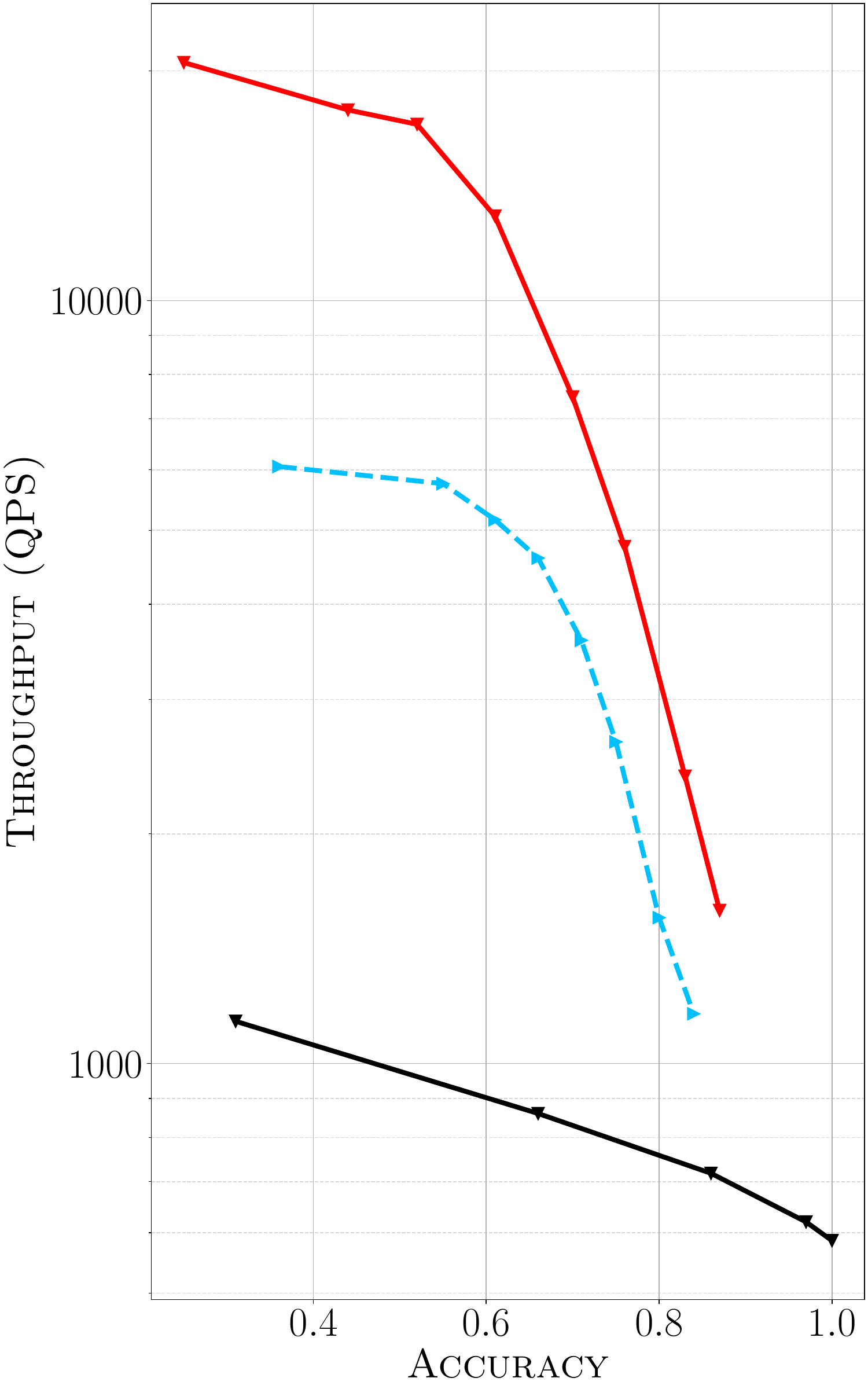}
}}
\vspace{0.2cm}
\caption{Throughput vs. top-$10$ retrieval accuracy
on \esplade{}-encoded datasets. Setup is as in Figure~\ref{figure:qps-accuracy:splade}.}
\label{figure:qps-accuracy:esplade}
\end{center}
\end{figure}

\subsubsection{Throughput versus Accuracy}
The first topic of evaluation is the trade-off between throughput and accuracy.
We can trade one factor off for the other by adjusting the parameter $\ell$
in Algorithm~\ref{algorithm:retrieval}: A smaller $\ell$ will result in
probing fewer partitions, which in turn leads to faster retrieval but
lower quality. Letting $\ell$ approach the size of the collection, on the other hand,
results in the algorithm probing every partition, leading to a slower but
higher-quality retrieval.

We tune this knob as we perform top-$10$ retrieval over our datasets.
We use \splade{} and \esplade{} vectors as input to the algorithms,
sketch them using the JL and \weaksinnamon{} transforms, but partition
the data only using spherical KMeans. The results of our experiments
are shown in Figures~\ref{figure:qps-accuracy:splade}
and~\ref{figure:qps-accuracy:esplade}.

In order to digest the trends, we must recall that the throughput
of our retrieval method is affected by two factors: the time it takes
to perform inner product of a query vector with cluster centroids, and
the time it takes to execute algorithm $\mathcal{R}$ on the subset
of partitions identified from the previous step. In the low-recall regime,
we expect the first factor to make up the bulk of the processing time,
while in the high-recall regime the cost of executing $\mathcal{R}$ starts
to dominate the overall processing time.

That phenomenon is evident in the figures for both \splade{} and \esplade{}
experiments. That also explains why when sketching is done with \weaksinnamon{},
throughput is much better than the JL transform: \weaksinnamon{} creates
sparse query sketches which lead to faster inner product computation with
partition centroids.

What is also clear from our experiments is that our approximate
method always compares favorably to the approximate baseline.
In fact, for the same desired accuracy, our method often reaches
a throughput that is orders of magnitude larger than that of the baseline's.
For instance, on \textsc{MS Marco} encoded with \splade{}, an instance
of our algorithm that operates on \weaksinnamon{} sketches processes
queries at an extrapolated rate of approximately $2{,}000$ queries per second
and delivers $90\%$ accuracy, while the baseline method yields a throughput of
roughly $150$ queries per second. At lower recalls, the gap is substantially wider.

As we require a higher accuracy, all methods become slower.
Ultimately, of course, if we set $\ell$ too high,
our algorithms become slower than the exact baseline. That is because,
our approximate algorithms have to pay the price of computing inner product
with centroids \emph{and} must execute the additional step of intersecting
$\mathcal{P}_\mathcal{I}$ with the skip lists. We do not show this empirically,
however.

\begin{figure}[t]
\begin{center}
\centerline{
\subfloat[\splade{}]{
\includegraphics[width=0.48\linewidth]{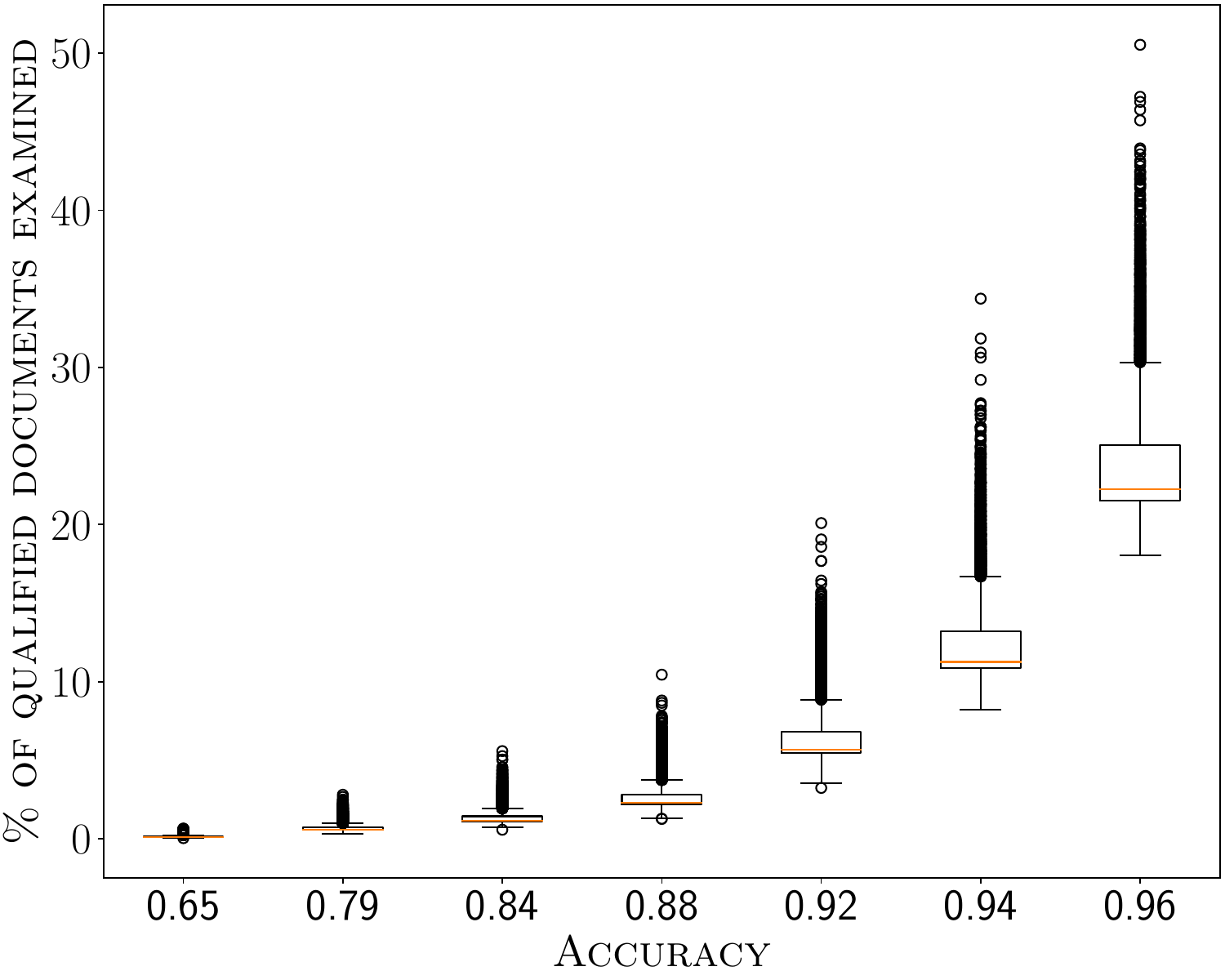}
}
\subfloat[\esplade{}]{
\includegraphics[width=0.48\linewidth]{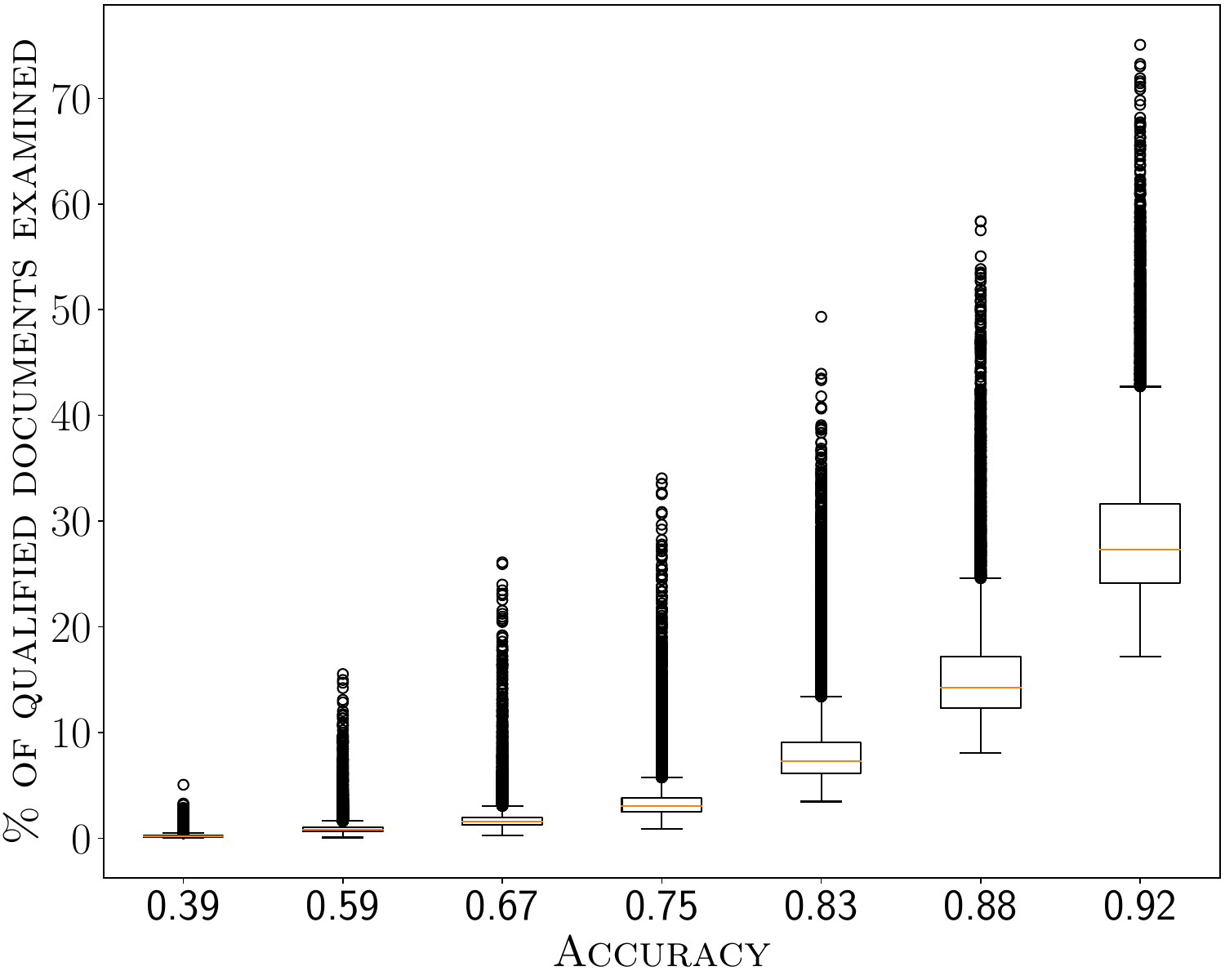}
}}
\vspace{0.2cm}
\caption{Percentage of qualified documents (i.e., documents that contain at least one
non-zero coordinate of the query) pruned versus top-$10$ accuracy for the \textsc{MS Marco}
dataset. In this setup, Algorithm~\ref{algorithm:indexing} uses \weaksinnamon{} along
with spherical KMeans for partitioning. Note the irregular spacing of the horizontal axes.}
\label{figure:dynamic-pruning}
\end{center}
\end{figure}

\subsubsection{Effect of Dynamic Pruning}
As we already explained, when we adjust the parameter
$\ell$ in Algorithm~\ref{algorithm:retrieval},
we control the number of documents the sub-algorithm $\mathcal{R}$
is allowed to evaluate. While we studied the impact of $\ell$
on efficiency as measured by throughput, here we wish to understand
its effect in terms of the amount of pruning it induces.
While throughput measurements depend on our specific implementation of
Algorithm~\ref{algorithm:query-processing}, measuring the portion
of documents pruned is implementation-agnostic and, as such,
serves as a more definitive measure of efficiency.

To that end, we count, for each query, the actual number of documents
evaluated by Algorithm~\ref{algorithm:query-processing} as we
gradually increase $\ell$.
We plot this quantity in Figure~\ref{figure:dynamic-pruning} for
\textsc{MS Marco} from a configuration of our algorithms that uses
\weaksinnamon{} and spherical KMeans. To improve visualization,
we show not raw counts, but the percentage of qualified documents---defined,
once again, as the number of documents that contain at least one non-zero coordinate
of the query---that Algorithm~\ref{algorithm:query-processing} evaluates.
That is indicative of how much of the inverted lists the algorithm
manages to skip.

As one observes, in the low-recall region, the algorithm probes only a fraction
of the inverted lists. On \splade{} dataset, the algorithm reaches a top-$10$ accuracy
of $0.94$ by merely evaluating, on average, about $10\%$ of the total number of documents in
the inverted lists. On \esplade{}, as expected, the algorithm is relatively less effective.

These results are encouraging. It shows the potential that a clustering-based
organization of the inverted index has for dynamic pruning in approximate MIPS.
Importantly, this method does not require the vectors to follow certain
distributions or be non-negative.

\subsubsection{Index Size Overhead}
As we mentioned earlier, our algorithms add overhead to the index structure
required for query processing. If our reference point is the \linscan{} algorithm
with a basic (uncompressed) inverted index, our methods introduce two
additional structures: (a) the skip list, $\mathcal{S}$, in
Algorithm~\ref{algorithm:inverted-index}; and, (b) the array of
$4\sqrt{\lvert \mathcal{X} \rvert}$ centroids produced
by Algorithm~\ref{algorithm:indexing}. We next measure this overhead.

We report our findings in Table~\ref{table:index-sizes} for \splade{}
and \esplade{} vector datasets, measured in GB of space after serialization to disk.
We reiterate that, we do not apply compression to the index. That is because
there is an array of compression techniques that can be applied to the different
parts of the data structure (such as quantization, approximation, and $d$-gap compression).
Choosing any of those would arbitrarily conflate the inflation due to the overhead
and the compression rate.

We observe that the overhead of our method on larger datasets is relatively mild.
The increase in size ranges from $6\%$ to $10\%$ (\textsc{Quora} excluded) for the
\splade{}-encoded datasets and a slightly wider and large range for \esplade{}-encoded
datasets.

\begin{table}[t]
\caption{Index sizes in GB. The index in \linscan{} is made up of an inverted index
with document identifiers and floating point values (uncompressed). The index in our
method stores $4\sqrt{\lvert \mathcal{X} \rvert}$ centroids from the application of
spherical KMeans to \weaksinnamon{} for dataset $\mathcal{X}$, an inverted index
with the same size as \linscan{}, and the skip list structure $\mathcal{S}$.}
\label{table:index-sizes}
\begin{center}
\begin{sc}
\begin{footnotesize}

\begin{tabular}{cc|cccccc}
& Method & \textsc{MS Marco} & \textsc{NQ} & \textsc{Quora} & \textsc{HotpotQA} & \textsc{Fever} & \textsc{DBPedia} \\
\midrule
\parbox[t]{10mm}{\multirow{2}{*}{\footnotesize \splade{}}}
& \linscan{} & $8.4$ & $3.1$ & $0.27$ & $5.1$ & $5.9$ & $4.7$ \\
& Ours & $9.0 (+7\%)$ & $3.43 (+10\%)$ & $0.32 (+18\%)$ & $5.5 (+8\%)$ & $6.3 (+7\%)$ & $5.0 (+6\%)$ \\
\midrule
\parbox[t]{10mm}{\multirow{2}{*}{\footnotesize \textsc{E. Splade}}}
& \linscan{} & $12$ & $4.2$ & $0.27$ & $4.9$ & $5.7$ & $4.6$ \\
& Ours & $13 (+8\%)$ & $4.7 (+12\%)$ & $0.37 (+37\%)$ & $5.4 (+10\%)$ & $6.2 (+9\%)$ & $5.0 (+9\%)$ \\
\bottomrule
\end{tabular}
\end{footnotesize}
\end{sc}
\end{center}
\end{table}

\subsubsection{Effect of Parallelism}
We conclude the empirical evaluation of our approximate algorithm
by repeating the throughput-accuracy experiments with a different
number of CPUs. In our implementation, we take advantage of
access to multiple processors by parallelizing the computation
of inner product between queries and centroids (in Algorithm~\ref{algorithm:retrieval})
for each query, in addition to distributing the queries themselves to the
available CPUs. As a result of this concurrent paradigm, we expect that,
by reducing the number of CPUs available to the algorithm, throughput
will be more heavily affected in low-recall regions (when $\ell$ is small).

\begin{figure}[t]
\begin{center}
\centerline{
\subfloat[\splade{}]{
\includegraphics[width=0.45\linewidth]{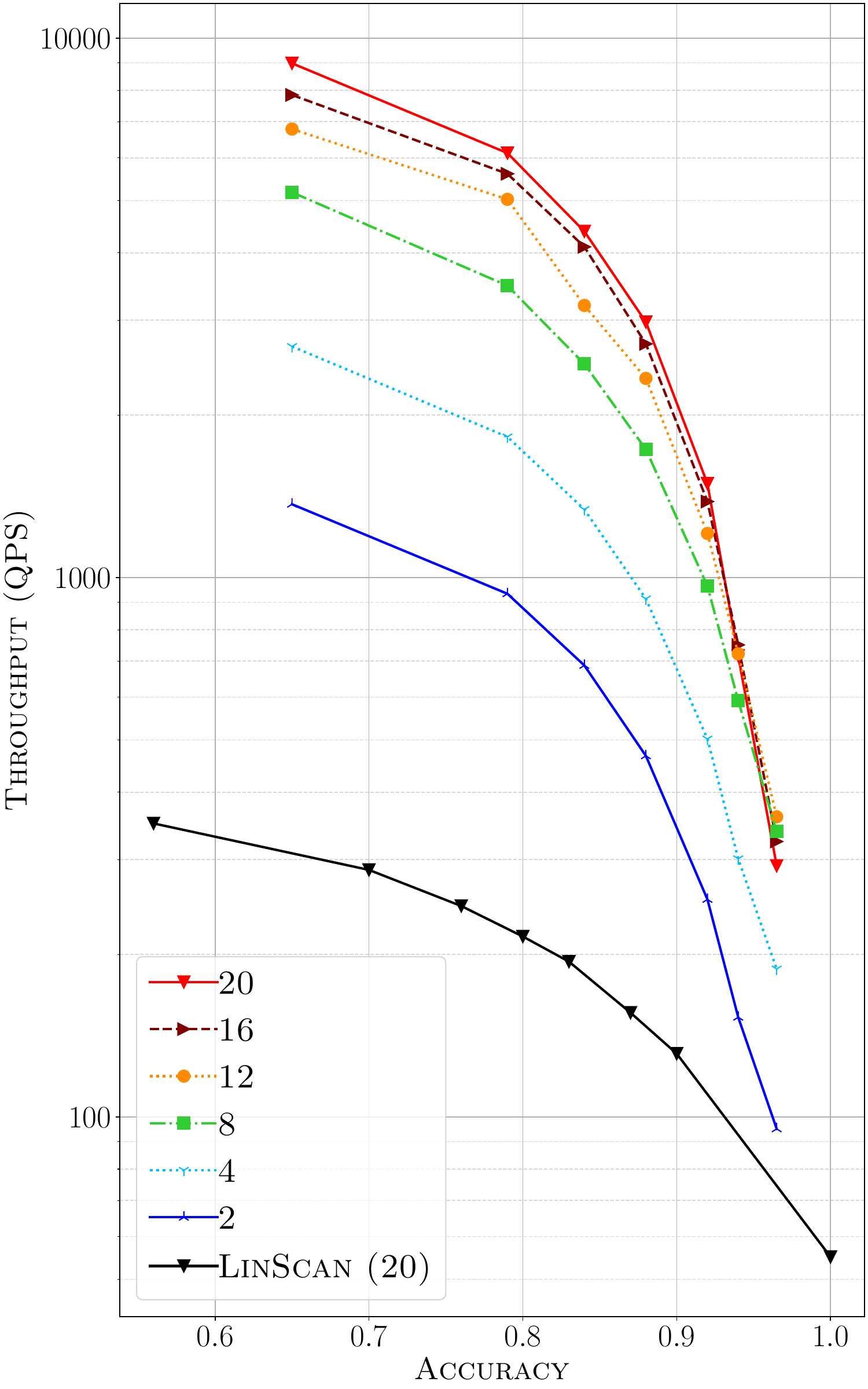}
}
\subfloat[\esplade{}]{
\includegraphics[width=0.45\linewidth]{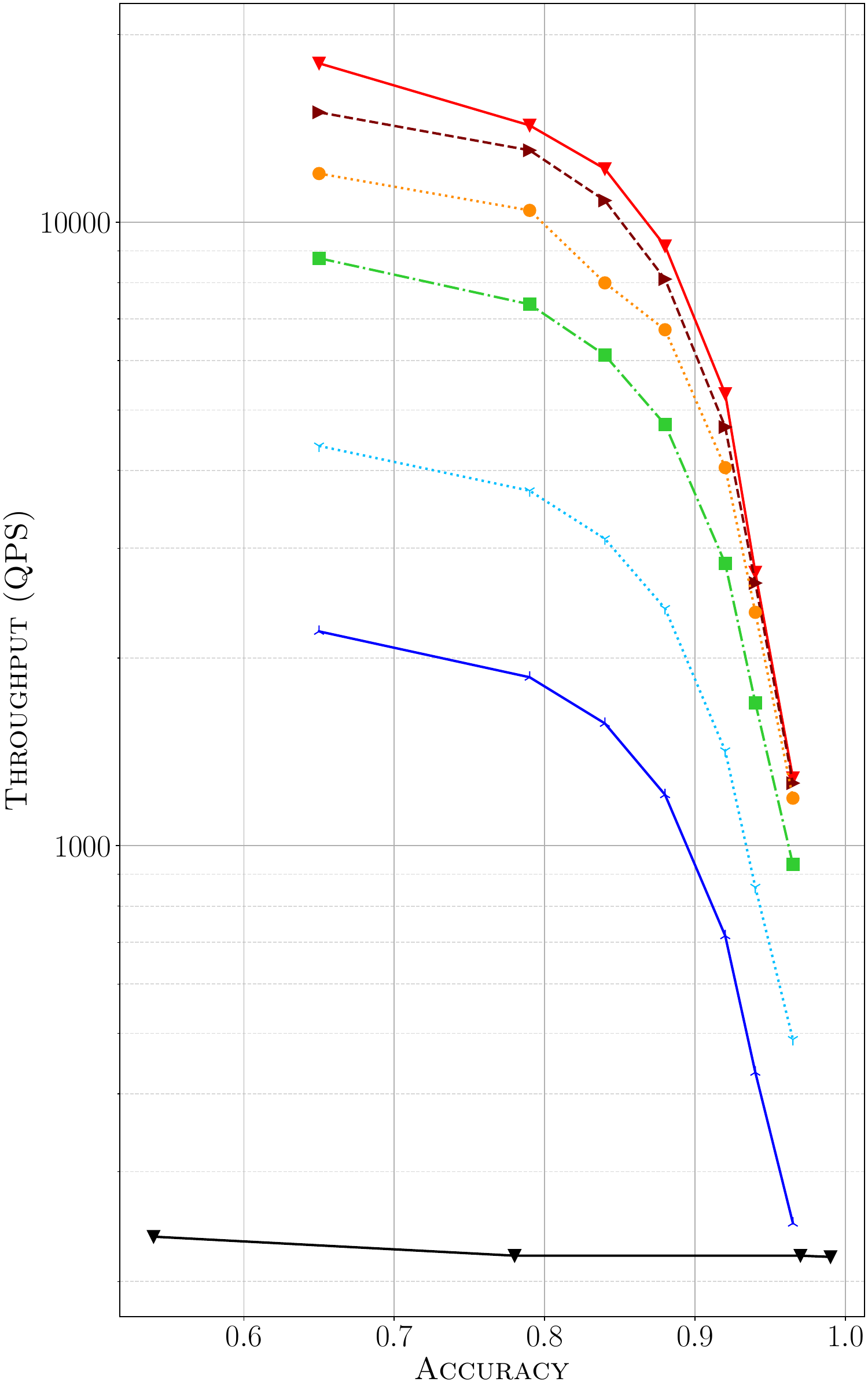}
}}
\vspace{0.2cm}
\caption{Effect of changing the number of CPUs on throughput. The figures
illustrate these measurements for \textsc{MS Marco}, and a particular configuration
of our algorithm that uses spherical KMeans over \weaksinnamon{} sketches.
We include \linscan{} executed on $20$ CPUs from Figure~\ref{figure:qps-accuracy:splade}
and~\ref{figure:qps-accuracy:esplade} as a point of reference.}
\label{figure:parallelism}
\end{center}
\end{figure}

Figure~\ref{figure:parallelism} shows the results of these experiments on the \splade{}-
and \esplade{}-encoded \textsc{MS Marco} dataset. The figures only include
a configuration of our algorithms with spherical KMeans and \weaksinnamon{}.
It is easy to confirm that our hypothesis from above holds: In low-recall
regions where computation is heavily dominated by the cost of computing inner product
with centroids, throughput decreases considerably as we reduce the number of CPUs.

\section{Towards a Unified Framework for MIPS}
\label{section:unified}

Sections~\ref{section:sketching} through~\ref{section:dynamic-pruning} presented a complete
instance of Algorithm~\ref{algorithm:retrieval} for IVF-based MIPS over \emph{sparse} vectors.
But, recall that, we borrowed the idea of IVF-based search from the dense MIPS literature.
So it is only natural to pose the following question: Now that we have an arbitrarily-accurate
IVF algorithm for sparse vectors, can we extend it to \emph{hybrid} vectors in $\mathbb{R}^{m+N}$?
In this section, we unpack that question superficially and investigate possible directions
at a high level to explore the feasibility and benefits of such an approach.
First, however, let us motivate this question.

\subsection{Motivation}
\label{section:unified:motivation}

We described the changing landscape of retrieval in Section~\ref{section:introduction}.
From lexical-semantic search to multi-modal retrieval, for many emerging applications
the ability to conduct MIPS over hybrid vectors efficiently and effectively is a requisite.
One viable approach to searching over a collection of hybrid vectors $\mathcal{X}$
is to simply decompose the process into separate MIPS questions,
one over the dense subspace $\mathcal{X}^d$ and the other over the sparse one $\mathcal{X}^s$,
followed by an aggregation of the retrieved sets.
Indeed this approach has become the \emph{de facto} solution
for hybrid vector retrieval~\cite{bruch2023fusion,chen2022ecir}.

The two-stage retrieval system works as follows: When a hybrid query vector $q \in \mathbb{R}^{m+N}$
arrives and the retrieval system is expected to return the top $k$ documents,
commonly, $q^d$ is sent to the dense MIPS system with a request
for the top $k^\prime \geq k$ vectors, and $q^s$ to
the sparse retrieval component with a similar request.
Documents in the union of the two sets are subsequently scored and reranked
to produce an approximate set of top-$k$ vectors, $\mathcal{\tilde{S}}$:
\begin{align}
&\mathcal{\tilde{S}} = \argmax^{(k)}_{x \in \mathcal{S}^d \cup \mathcal{S}^s} \; \langle q , x \rangle, \\
&\mathcal{S}^d = \argmax^{(k^\prime)}_{x \in \mathcal{X}} \; \langle q^d, x^d \rangle \; \text{ and, }
\mathcal{S}^s = \argmax^{(k^\prime)}_{x \in \mathcal{X}} \; \langle q^s, x^s \rangle.
\end{align}

Let us set aside the effectiveness of the setup above for a moment and consider its complexity
from a systems standpoint. It is clear that, both for researchers and practitioners,
studying and creating two disconnected, incompatible systems adds unwanted costs.
For example, systems developers must take care to keep all documents
in sync between the two indexes at all times.
Reasoning about the (mis)behavior of the retrieval system, as another example,
requires investigating one layer of indirection and understanding the processes
leading to two separate retrieved sets.
These collectively pose a challenge to systems researchers,
and add difficulty to operations in production.
Furthermore, it is easy to see that the least scalable
of the two systems dictates or shapes the overall latency and
throughput capacity.

Even if we accepted the cost of studying two separate systems or deemed it negligible,
and further decided scalability is not a concern,
it is not difficult to show that such a heterogeneous design may
prove wasteful or outright ineffective in the general case.
More concretely, depending on how the $\ell 2$ mass of the query and document
vectors is split between the dense subspace and the sparse subspace,
the two sub-systems involved may have to resort to a large $k^\prime$
in order to ensure an accurate final retrieved set at rank $k$.

\begin{figure}[t]
    \centering
    \includegraphics[width=0.6\linewidth]{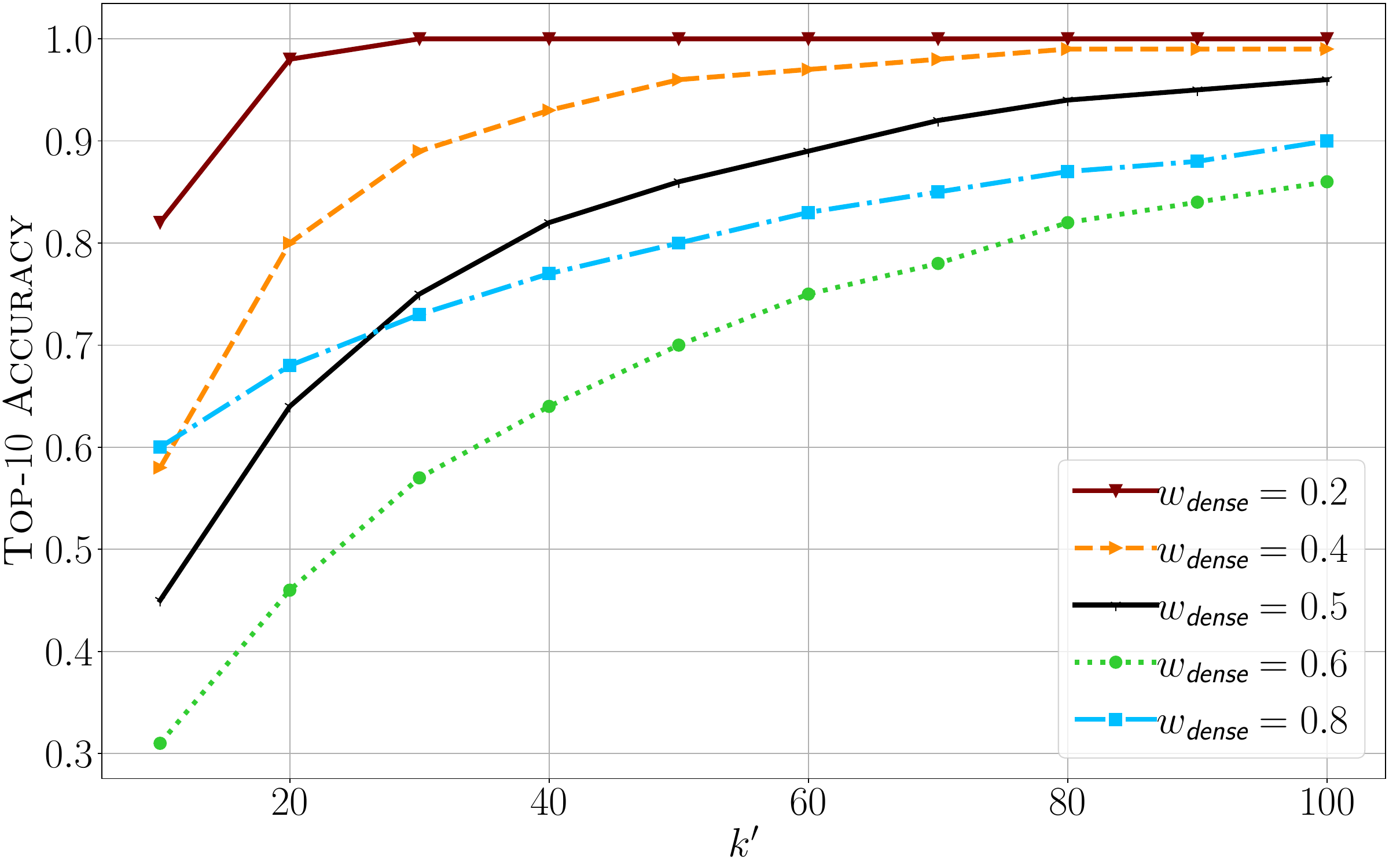}
    \caption{Top-$10$ accuracy of the two-stage retrieval system for hybrid vectors.
    We retrieve $k^\prime$ candidates from each
    sub-system and rerank them to find the top-$10$ set.
    We prepare the hybrid vectors by first normalizing the dense and sparse parts separately,
    then constructing query vectors as follows: $q = w_\textit{dense} q^d + (1 - w_\textit{dense}) q^s$,
    where $q^d$ and $q^s$ are sampled from the data distribution.
    In effect, $w_\textit{dense}$ shifts the $\ell 2$ mass from the sparse to the dense subspace,
    giving more importance to one subspace over the other during retrieval.}
    \label{figure:motivation-example}
\end{figure}

While the phenomenon above is provable, we demonstrate its effect
by a simple (though contrived) experiment.
We generate a collection of $100{,}000$ documents and $1{,}000$ queries.
Each vector is a hybrid of a dense and a sparse vector.
The dense vectors are in $\mathbb{R}^{64}$,  with each coordinate
drawing its value from the exponential distribution (with scale $0.5$).
The sparse vectors are in $\mathbb{R}^{1000}$ with an average of $\psi=16$ non-zero
coordinates, where non-zero values are drawn from the exponential distribution
(scale $0.5$). We use different seeds for the pseudo-random generator when creating
document and query vectors.

In order to study how the ratio of $\ell 2$ mass between dense and sparse subspaces
affects retrieval quality, we first normalize the generated dense and sparse vectors
separately. During retrieval, we amplify the dense part of the query vector by a weight
between $0$ and $1$ and multiply the sparse part by one minus that weight. In the end,
we are performing retrieval for a query vector $q$ that can be written as $w_\textit{dense} q^d + (1 - w_\textit{dense}) q^s$.
By letting $w_\textit{dense}$ sweep the unit interval, we simulate a shift of the $\ell 2$ mass
of the hybrid vector from the sparse to the dense subspace.

Over the generated collection, we conduct \emph{exact} retrieval using exhaustive
search and obtain the top $k=10$ vectors for each query by maximizing the inner product.
We then use the two-stage design by asking each sub-system to return the (exact) top
$k^\prime$ vectors for $k^\prime \in [100]$,
and reranking the union set to obtain the final top $k=10$ documents.
We then measure the top-$k$ accuracy of the two-stage architecture.

Figure~\ref{figure:motivation-example} plots accuracy versus $k^\prime$
for different values of $w_\textit{dense}$.
It is easy to see that, as one subspace becomes more important than the other,
the retrieval quality too changes. Importantly, a larger $k^\prime$ is often
required to attain a high accuracy.

The factors identified in this section---systems complexity, scalability bottleneck,
and the sub-optimality of retrieval quality---nudge
us in the direction of a unified framework for MIPS.

\subsection{IVF MIPS for Hybrid Vectors}
We present a simple extension of the IVF indexing and retrieval duo of
Algorithms~\ref{algorithm:indexing} and~\ref{algorithm:retrieval} to generalize
the logic to hybrid vectors. This is shown in Algorithms~\ref{algorithm:indexing-hybrid}
and~\ref{algorithm:retrieval-hybrid}, where the only two differences with the original
algorithms are that (a) sketching is applied only to the sparse
portion of vectors to form new vectors in $\mathbb{R}^{m+n}$ instead of $\mathbb{R}^{m+N}$,
and (b) that the sub-algorithm $\mathcal{R}$ is assumed to carry out top-$k$ retrieval
over hybrid vectors from a given set of partitions.

\begin{algorithm}[!t]
\SetAlgoLined
{\bf Input: }{Collection $\mathcal{X}$ of hybrid vectors in $\mathbb{R}^{m+N}$;
Number of clusters, $P$;
Random projector, $\phi: \mathbb{R}^N \rightarrow \mathbb{R}^n$ where $n \ll N$;
Clustering algorithm $\textsc{Cluster}$ that returns partitions
of input data and their representatives.}\\
\KwResult{Cluster assignments $\mathcal{P}_i = \{ j \;|\; x^{(j)} \in \text{ Partition } i \}$
and cluster representatives $\mathcal{C}_i$'s.}

\begin{algorithmic}[1]
    \STATE $\tilde{\mathcal{X}} \leftarrow \{ x^d \oplus \phi(x^s) \;|\; x^d \oplus x^s \in \mathcal{X} \}$
    \STATE $\textsc{Partitions}, \textsc{Representatives} \leftarrow \textsc{Cluster}(\tilde{\mathcal{X}}; P)$
    \STATE $\mathcal{P}_i \leftarrow \{ j \;|\; \tilde{x}^{(j)} \in \textsc{Partitions}[i] \}, \quad \forall 1 \leq i \leq P$
    \STATE $\mathcal{C}_i \leftarrow \textsc{Representatives}[i], \quad \forall 1 \leq i \leq P$
    \RETURN $\mathcal{P}$ and $\mathcal{C}$
 \end{algorithmic}
 \caption{Indexing of hybrid vectors}
\label{algorithm:indexing-hybrid}
\end{algorithm}

\begin{algorithm}[!t]
\SetAlgoLined
{\bf Input: }{Hybrid query vector, $q \in \mathbb{R}^{m+N}$;
Clusters and representatives,
$\mathcal{P}$, $\mathcal{C}$ obtained from Algorithm~\ref{algorithm:indexing-hybrid};
random projector $\phi: \mathbb{R}^N \rightarrow \mathbb{R}^n$;
Number of data points to examine, $\ell \leq |\mathcal{X}|$ where
$\lvert \mathcal{X} \rvert$ denotes the size of the collection;
hybrid MIPS sub-algorithm $\mathcal{R}$.}\\
\KwResult{Approximate set of top $k$ vectors that maximize inner product with $q$.}

\begin{algorithmic}[1]
    \STATE $\tilde{q} \leftarrow q^d \oplus \phi(q^s)$
    \STATE \textsc{SortedClusters} $\leftarrow \textbf{SortDescending}(\mathcal{P} \text{ by } \langle \tilde{q}, \mathcal{C}_i \rangle)$

    \STATE $\textsc{TotalSize} \leftarrow 0$
    \STATE $\mathcal{I} \leftarrow \emptyset$ \Comment*[r]{Records the index of the partitions $\mathcal{R}$ should probe.}
    \FOR{$\mathcal{P}_{\pi_i} \in $ \textsc{SortedClusters}}
    \STATE $\mathcal{I} \leftarrow \pi_i$
    \STATE $\textsc{TotalSize} \leftarrow \textsc{TotalSize} + \lvert \mathcal{P}_{\pi_i} \rvert$
    \STATE \textbf{break if} $\textsc{TotalSize} \geq \ell$
    \ENDFOR
    
    \RETURN Top $k$ vectors from partitions $\mathcal{P}_\mathcal{I} \triangleq \{ \mathcal{P}_i \;|\; i \in \mathcal{I} \}$
    w.r.t $\langle q, \cdot \rangle$ using $\mathcal{R}$
 \end{algorithmic}
 \caption{Retrieval of hybrid vectors}
\label{algorithm:retrieval-hybrid}
\end{algorithm}

In this section, we only verify the viability of the extended algorithms
and leave an in-depth investigation of the proposal to future work.
As such, we use exhaustive search as the sub-algorithm $\mathcal{R}$
and acknowledge that any observations made using such an algorithm only
speaks to the effectiveness of the method and not its efficiency.

\subsection{Empirical Evaluation}
Let us repeat the experiment from Section~\ref{section:unified:motivation} on synthetic
vectors and compare the two-stage retrieval process with the unified framework
in terms of retrieval accuracy. To that end, we design the following protocol.

First, we perform exact MIPS using exhaustive search over the hybrid collection of vectors.
The set of top-$k$ documents obtained in this way make up the ground-truth for
each query.

Next, we consider the two-stage system.
We retrieve through exhaustive search the exact set of top-$k^\prime$
(for a large $k^\prime$) documents according to their sparse inner product,
and another (possibly overlapping) set by their dense inner product.
From the two ranked lists, we accumulate enough documents from the
top such that the size of the resulting set is roughly equal to $k$.
In this way, we can measure the top-$k$ accuracy of the two-stage system
against the ground-truth.

Finally, we turn to the unified framework.
We use the JL transform to reduce the dimensionality
of sparse vectors, and spherical KMeans to partition the vectors. We then
proceed as usual and measure top-$k$ accuracy for different values of $\ell$.

\begin{figure}[t]
\begin{center}
\centerline{
\subfloat[$w_\textit{dense}=0.2$]{
\includegraphics[width=0.32\linewidth]{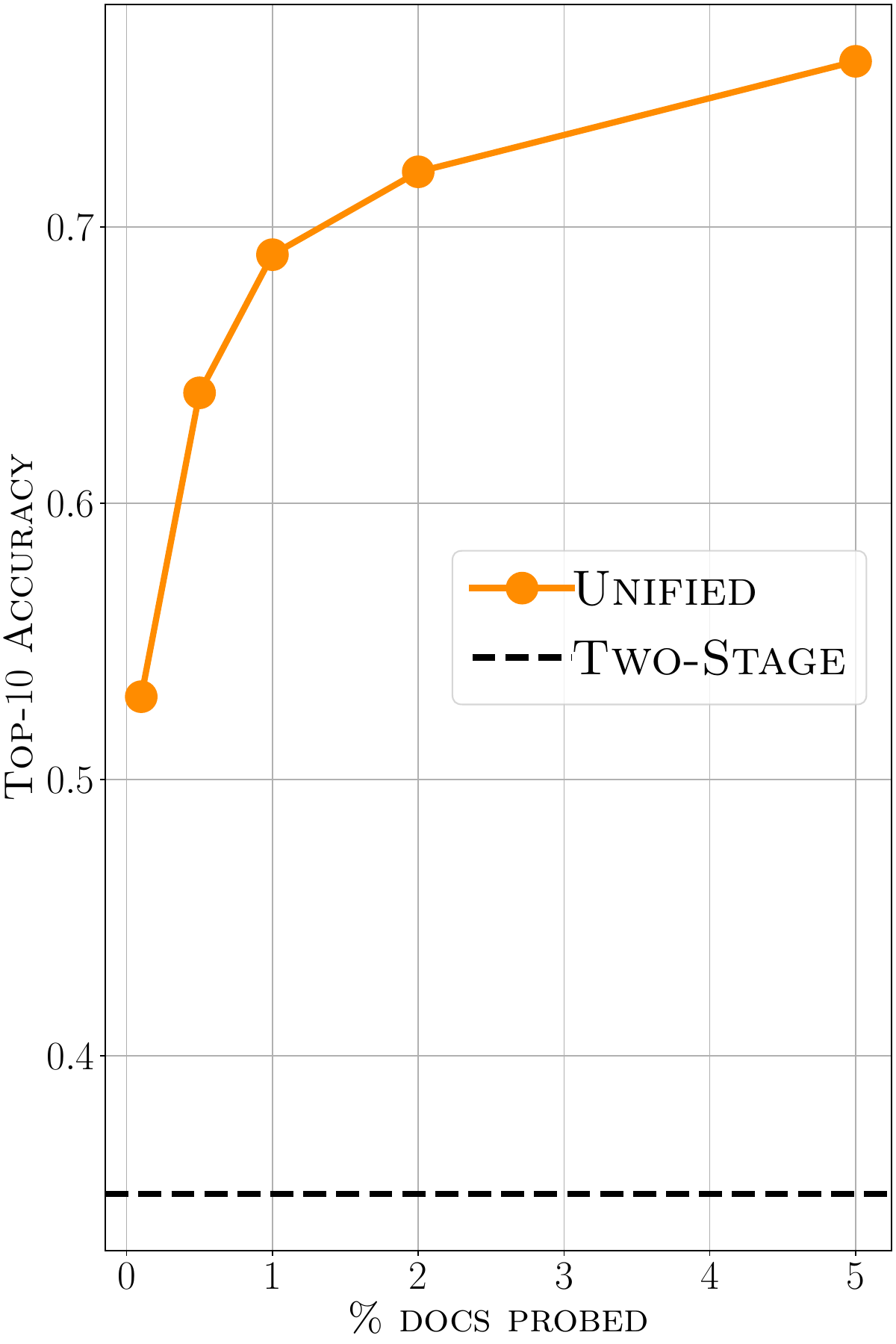}
}
\subfloat[$w_\textit{dense}=0.5$]{
\includegraphics[width=0.32\linewidth]{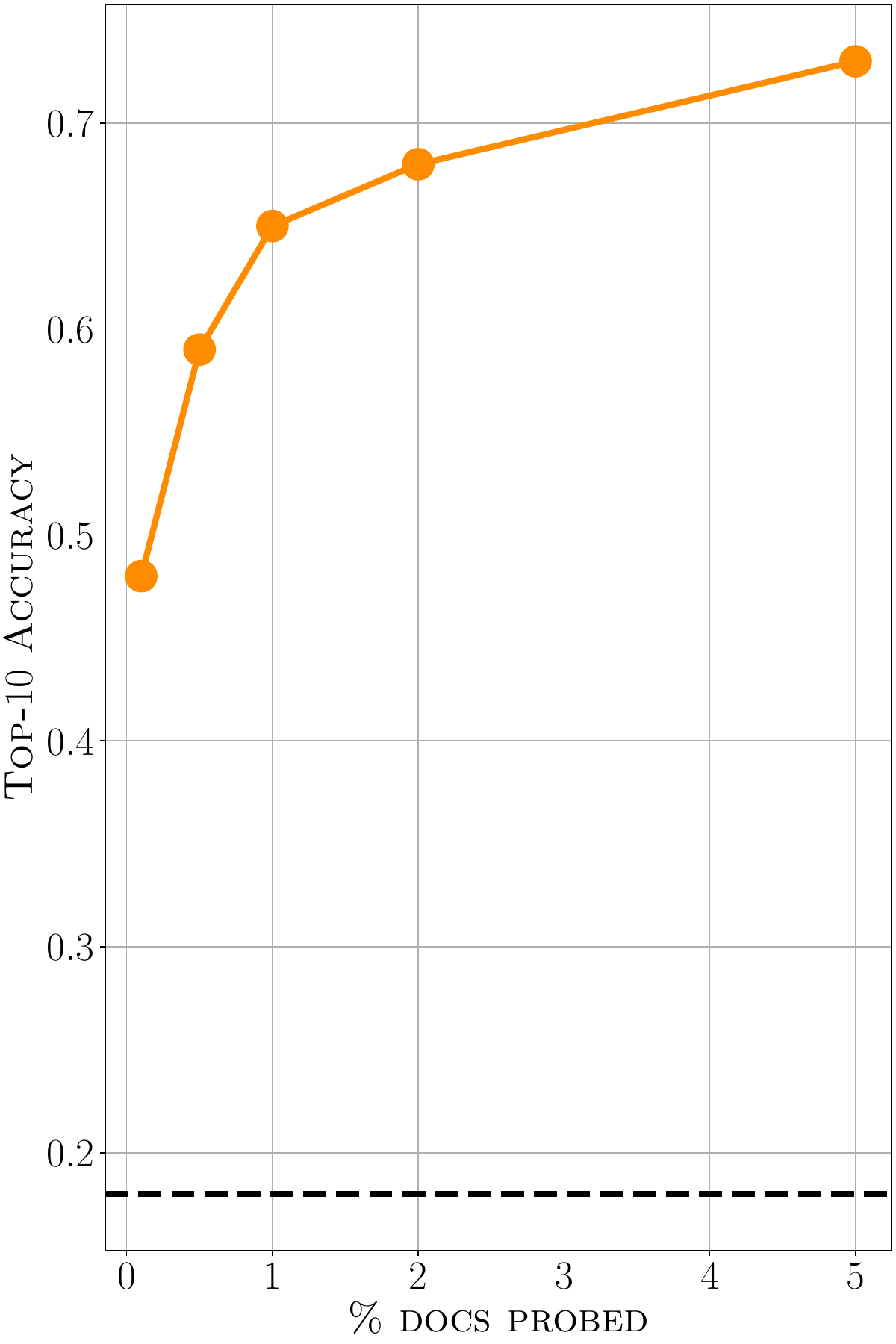}
}
\subfloat[$w_\textit{dense}=0.8$]{
\includegraphics[width=0.32\linewidth]{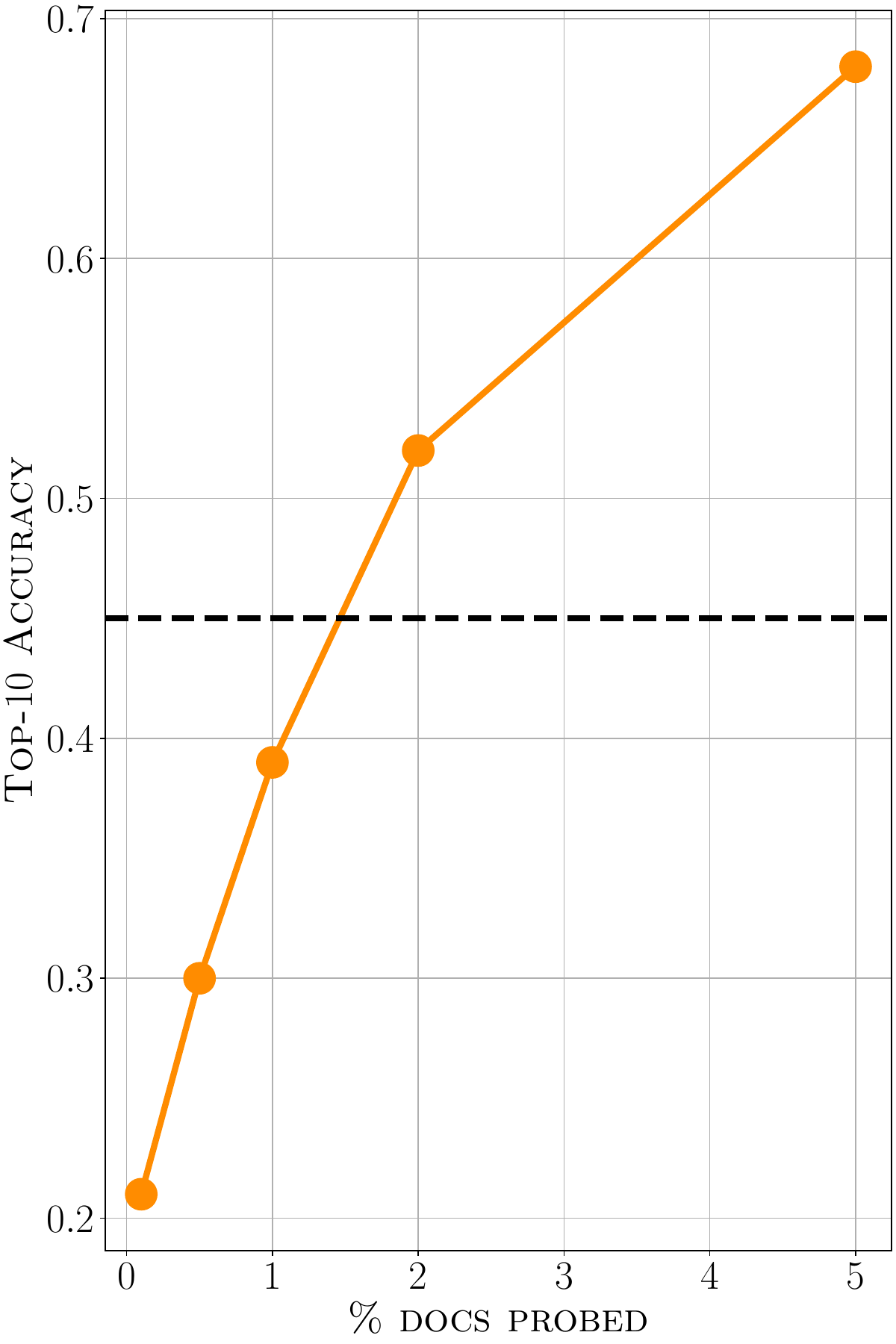}
}}
\vspace{0.2cm}
\caption{Top-$10$ accuracy over hybrid vectors as a function of the percentage of documents probed.
$w_\textit{dense}$ controls how much of the $\ell 2$ mass of a hybrid vector is concentrated in its
dense subspace. We also plot the performance of the two-stage system where each system returns 
the set of top-$k^\prime$ documents according to sparse or dense inner product scores, such that
the size of the union of the two sets is roughly $k$.}
\label{figure:hybrid:unified}
\end{center}
\end{figure}

From these experiments, we wish to understand
whether and when the accuracy of the unified framework exceeds the accuracy of
the two-stage setup. If the unified system is able to surpass the accuracy of the
two-stage system by examining a relatively small portion of the
collection---a quantity controlled through $\ell$---then that is indicative
of the viability of the proposal.
Indeed, as Figure~\ref{figure:hybrid:unified} shows, the unified system almost always reaches
a top-$10$ accuracy that is higher than the two-stage system's by evaluating less than $2\%$
of the collection.

\section{Discussion and Conclusion}
\label{section:conclusion}

We began this research with a simple question: Can we apply dense MIPS algorithms
to sparse vectors? That led us to investigate different dimensionality reduction
techniques for sparse vectors as a way to contain the curse of dimensionality.
We showed, for example, that the JL transform and \sinnamon{} behave differently
on sparse vectors and can preserve inner product to different degrees.
We also thoroughly evaluated the effect of clustering on sparse MIPS
in the context of an IVF-based retrieval system. Coupling dimensionality reduction
with clustering realized an effective IVF system for sparse vectors,
summarized in Algorithms~\ref{algorithm:indexing} and ~\ref{algorithm:retrieval}.

The protocol is easy to describe and is as follows. We sketch sparse vectors into
a lower-dimensional (dense or sparse) subspace in a first step. We then
apply clustering to the sketches and partition the data into a predetermined
number of clusters, each identified by a representative (e.g., a centroid).
When the system is presented with a query, we sketch the query (asymmetrically)
and identify the top partitions by taking inner product between the query and cluster
representatives. We then execute a secondary sub-algorithm to perform MIPS on the restricted
subset of document vectors.

In our presentation of the material above, we observed a strong, natural connection
between clustering for IVF and dynamic pruning methods for inverted indexes.
We developed that insight into an inverted index-based algorithm that could serve
as the sub-algorithm in the above search procedure. Importantly, the algorithm
organizes documents within an inverted list by partition identifier---rather than the
conventional arrangement by document identifier or impact score. Such an organization,
coupled with skip pointers, enables the algorithm to only search over the subset of
documents that belong to the top partitions determined by the IVF method. Crucially,
the algorithm is agnostic to the vector distribution and admits real-valued vectors.

Finally, we discussed how our proposal leads to a unified retrieval framework for
hybrid vectors. By sketching the sparse sub-vectors and constructing an IVF index
for the transformed hybrid vectors, we showed that it is possible to achieve
better recall than a two-stage system, where dense and sparse sub-vectors are
handled separately. The added advantage of the unified approach is that its
accuracy remains robust under different vector distributions, where the mass
shifts from the dense to the sparse subspace.

We limited our discussion of hybrid MIPS to synthetic vectors as
we were only interested in the viability of this byproduct of our primary research question.
We acknowledge that we have only scratched the surface of retrieval over hybrid vectors.
There are a multitude of open questions
within the unified regime that warrant further investigation,
including many minor but practical aspects of the framework that we conveniently
ignored in our high-level description. We leave those as future work.

We believe our investigation of MIPS for sparse (and hybrid vectors)
provides many opportunities for information retrieval researchers.
One line of research most immediately affected by our proposal is
sparse representation learning. Models
such as \splade{} are not only competitive on in- and out-of-domain tasks,
they also produce inherently-interpretable representations of text---a desirable behavior in many production systems.
However, sparse embeddings have, by and large, been tailored to existing retrieval regimes.
For example, \esplade{} learns sparser queries for better latency.
\textsc{uniCoil}~\cite{unicoil} collapses term representations of \textsc{Coil}~\cite{coil}
to a scalar for compatibility with inverted indexes.
We claim that our proposed regime is a step toward removing such
constraints, enabling researchers to explore sparse representations without much restraint,
leading to a potentially different behavior. As we observe in
Figures~\ref{figure:clustering-quality:splade} and~\ref{figure:clustering-quality:esplade},
for example, \splade{} vectors are more amenable to clustering
than \esplade{}, and may even prove more efficient within the new framework.
That is good news as there is evidence suggesting that \splade{}
is more effective than its other variant on out-of-domain data~\cite{lassance2022sigir}.

Another related area of research that can benefit from our proposed regime
is multi-modal and multimedia retrieval.
Because our framework is agnostic to the distribution of the hybrid vectors,
it is entirely plausible to formulate the multi-modal problem as MIPS over hybrid vectors,
especially when one of the modes involves textual data,
is data that is partially sparse, or where one may need to engineer (sparse)
features to augment dense embeddings.

\bibliographystyle{ACM-Reference-Format}
\bibliography{main}

\appendix

\section{Proof of Theorem~\ref{theorem:jl-variance-fixed-vectors}}
\label{appendix:jl-variance-fixed-vectors}

Fix two vectors $u$ and $v \in \mathbb{R}^N$.
Define $Z_\textsc{Sketch} = \langle \phi(u), \phi(v) \rangle$ as the random
variable representing the inner product of sketches of size $n$,
prepared using the projection $\phi(u) = R u$, with $R \in \{-1/\sqrt{n}, 1/\sqrt{n}\}^{n \times N}$.
$Z_\textsc{Sketch}$ is an unbiased estimator of $\langle u, v \rangle$.
Its distribution tends to a Gaussian with variance:
\begin{equation*}
\frac{1}{n} \big( \lVert u \rVert_2^2 \lVert v \rVert_2^2 + \langle u, v \rangle^2 - 2 \sum_i u_i^2 v_i^2  \big).
\end{equation*}

\begin{proof}
Consider the random variable $Z=\big( \sum_j R_j u_j \big)\big( \sum_k R_k v_k \big)$,
where $R_i$'s are Rademacher random variables. It is clear that $nZ$ is the product
of the sketch coordinate $i$ (for any $i$): $\phi(u)_i \phi(v)_i$.

We can expand the expected value of $Z$ as follows:
\begin{align*}
    \mathbb{E}[Z] &= \mathbb{E}\big[ \big( \sum_j R_j u_j \big) \big( \sum_k R_k v_k \big) \big] \\
    &= \mathbb{E}[\sum_i R_i^2 u_i v_i] + \mathbb{E}[ \sum_{j \neq k} R_j R_k u_j v_k ] \\
    & = \sum_i u_i v_i \underbrace{\mathbb{E}[R_i^2]}_1 + \sum_{j \neq k} u_j v_k \underbrace{\mathbb{E}[R_j R_k ]}_0 \\
    &= \langle u, v \rangle.
\end{align*}

The variance of $Z$ can be expressed as follows:
\begin{equation*}
    \mathit{Var}(Z) = \mathbb{E}[Z^2] - \mathbb{E}[Z]^2 =
    \mathbb{E}[\big( \sum_j R_j u_j \big)^2\big( \sum_k R_k v_k \big)^2] - \langle u, v\rangle^2.
\end{equation*}

We have the following:
\begin{align}
    \mathbb{E}&[\big( \sum_j R_j u_j \big)^2\big( \sum_k R_k v_k \big)^2] = \mathbb{E}\big[
        \big( \sum_i u_i^2 + \sum_{i \neq j} R_i R_j u_i u_j \big)
        \big( \sum_k v_k^2 + \sum_{k \neq l} R_k R_l v_k v_l \big)
    \big] \\
        &= \lVert u \rVert_2^2 \lVert v \rVert_2^2  +
            \underbrace{\mathbb{E}[\sum_i u_i^2 \sum_{k \neq l} R_k R_l v_k v_l]}_0 +
            \underbrace{\mathbb{E}[\sum_k v_k^2 \sum_{i \neq j} R_i R_j u_i u_j]}_0 +
            \mathbb{E}[\sum_{i \neq j} R_i R_j u_i u_j \sum_{k \neq l} R_k R_l v_k v_l].
            \label{equation:jl-variance-fixed-vectors:variance}
\end{align}
The last term can be decomposed as follows:
\begin{align*}
    \mathbb{E}&\big[ \sum_{i \neq j \neq k \neq l} R_i R_j R_k R_l u_i u_j v_k v_l \big] + \\
    &\mathbb{E}\big[ \sum_{i = k, j \neq l \lor i \neq k, j = l} R_i R_j R_k R_l u_i u_j v_k v_l \big] + \\
    & \mathbb{E}\big[ \sum_{i \neq j, i=k, j=l \lor i \neq j, i=l, j=k} R_i R_j R_k R_l u_i u_j v_k v_l \big].
\end{align*}
The first two terms are $0$ and the last term can be rewritten as follows:
\begin{equation}
\label{equation:jl-variance-fixed-vectors:reduction}
    2 \mathbb{E}\big[ \sum_i u_i v_i \big( \sum_j u_j v_j - u_i v_i \big) \big] = 2 \langle u, v \rangle^2 - 2 \sum_i u_i^2 v_i^2.
\end{equation}

We now substitute the last term in Equation~(\ref{equation:jl-variance-fixed-vectors:variance}) with
Equation~(\ref{equation:jl-variance-fixed-vectors:reduction}) to obtain:
\begin{equation}
\mathit{Var}(Z) = \lVert u \rVert_2^2 \lVert v \rVert_2^2 + \langle u, v \rangle^2 - 2 \sum_i u_i^2 v_i^2.
\end{equation}

Observe that $Z_\textsc{Sketch} = 1/n \sum_i \phi(u)_i \phi(v)_i$ is
the sum of independent, identically distributed random variables.
Furthermore, for bounded vectors $u$ and $v$, the variance is finite.
By the application of the Central Limit Theorem, we can deduce that
the distribution of $Z_\textsc{Sketch}$ tends to a normal distribution
with the stated expected value. Noting that
$\mathit{Var}(Z_\textsc{Sketch}) = 1/n^2 \sum_i \mathit{Var}(Z)$
gives the desired variance.
\end{proof}

\section{Proof of Theorem~\ref{theorem:jl-variance-fixed-query}}
\label{appendix:jl-variance-fixed-query}

Fix a query vector $q \in \mathbb{R}^N$ and let $X$ be
a random vector drawn according to the following probabilistic model.
Coordinate $i$, $X_i$, is non-zero with probability $p_i > 0$ and,
if it is non-zero, draws its value from a distribution with mean $\mu$
and variance $\sigma^2$. $Z_\textsc{Sketch} = \langle \phi(q), \phi(X) \rangle$,
with $\phi(u) = R u$ and $R \in \{-1/\sqrt{n}, 1/\sqrt{n}\}^{n \times N}$,
has expected value $\mu \sum_i p_i q_i$ and variance:
\begin{equation*}
    \frac{1}{n} \big[
    (\mu^2 + \sigma^2)\big( \lVert q \rVert_2^2 \sum_i p_i - \sum_i p_i q_i^2 \big) +
    \mu^2 \big( (\sum_i q_i p_i)^2 - \sum_i (q_i p_i)^2 \big)
    \big]
\end{equation*}

\begin{proof}
It is easy to see that:
\begin{equation*}
    \mathbb{E}[Z_\textsc{Sketch}] = \sum_i q_i \mathbb{E}[X_i] = \mu \sum_i p_i q_i.
\end{equation*}

As for variance, we start from Theorem~\ref{theorem:jl-variance-fixed-vectors}
and arrive at the following expression:
\begin{equation}
\label{equation:jl-variance-fixed-query:variance}
\frac{1}{n} \big( \lVert q \rVert_2^2 \mathbb{E}[\lVert X \rVert_2^2] + \mathbb{E}[\langle q, X \rangle^2] - 2 \sum_i q_i^2 \mathbb{E}[X_i^2]  \big),
\end{equation}
where the expectation is with respect to $X$. Let us consider the terms inside the
parentheses one by one. The first term becomes:
\begin{align*}
    \lVert q \rVert_2^2 \mathbb{E}[\lVert X \rVert_2^2] &= \lVert q \rVert_2^2 \sum_i\mathbb{E}[X_i^2] \\
    &= \lVert q \rVert_2^2 (\mu^2 + \sigma^2) \sum_i p_i.
\end{align*}
The second term reduces to:
\begin{align*}
\mathbb{E}[\langle q, X \rangle^2] &= \mathbb{E}\big[ \langle q, X \rangle \big]^2 +
    \mathit{Var}\big[ \langle q, X \rangle \big] + \\
&= \mu^2 (\sum_i q_i p_i)^2 + \sum q_i^2 \big[ (\mu^2 + \sigma^2) p_i - \mu^2 p_i^2 \big] \\
&= \mu^2 \big( (\sum_i q_i p_i)^2 - \sum_i q_i^2 p_i^2 \big) + \sum_i q_i^2 p_i (\mu^2 + \sigma^2).
\end{align*}
Finally, the last term breaks down to:
\begin{align*}
- 2 \sum_i q_i^2 \mathbb{E}[X_i^2] &= -2 \sum_i q_i^2 (\mu^2 + \sigma^2) p_i \\
&= -2 (\mu^2 + \sigma^2) \sum_i q_i^2 p_i.
\end{align*}

Putting all these terms back into Equation~(\ref{equation:jl-variance-fixed-query:variance}) yields
the desired expression for variance.
\end{proof}

\section{Proof of Theorem~\ref{theorem:sinnamon:upper-bound-sketch-error}}
\label{appendix:sinnamon:upper-bound-sketch-error}

Let $X$ be a random vector drawn according to the following probabilistic model.
Coordinate $i$, $X_i$, is non-zero with probability $p_i > 0$ and,
if it is non-zero, draws its value from a distribution with PDF $\phi$ and CDF $\Phi$.
Then:
\begin{equation*}
    \mathbb{P}[\overline{X}_{\pi(i)} - X_i \leq \delta] \approx
        (1 - p_i) \big( e^{-\frac{1}{m} (1 - \Phi(\delta)) \sum_{j \neq i} p_j} \big) +
        p_i \int e^{-\frac{1}{m} (1 - \Phi(\alpha + \delta)) \sum_{j \neq i} p_j} \phi(\alpha) d \alpha
\end{equation*}

\begin{proof}
Decomposing the probability of the event by conditioning on whether $X_i$ is
``active'' (i.e., its value is drawn from the distribution with PDF $\phi$) or
``inactive'' (i.e., it is $0$), we arrive at:
\begin{equation*}
    \mathbb{P}[\overline{X}_{\pi(i)} - X_i \leq \delta ] = 
    p_i \mathbb{P}[\overline{X}_{\pi(i)} - X_i \leq \delta \;|\; X_i \textit{ is active}] +
    (1 - p_i) \mathbb{P}[\overline{X}_{\pi(i)} \leq \delta \;|\; X_i \textit{ is inactive}].
\end{equation*}
The term conditioned on $X_i$ being active is given by Theorem~5.4 of~\cite{bruch2023sinnamon}.
The other event involving an inactive $X_i$ happens when all values that collide with $\overline{X}_{\pi(i)}$
are less than or equal to $\delta$. This event is equivalent to the event that every active coordinate
whose value is greater than $\delta$ maps to any sketch coordinate except $i$.
Using this alternative event, we can write the conditional probability as follows:
\begin{equation*}
    (1 - \frac{1}{m})^{(1 - \Phi(\delta)) \sum_{j \neq i} p_j} \approx
    e^{-\frac{1}{m} (1 - \Phi(\delta)) \sum_{j \neq i} p_j},
\end{equation*}
where we used $e^{-1} \approx (1 - 1/m)^m$. That completes the proof.
\end{proof}

\end{document}